\documentclass[fleqn,3p,onecolumn,11pt,nofootinbib,showkeys,showpacs]{revtex4-1}

\usepackage[small,justification=centerlast]{caption}
\usepackage{graphicx}
\usepackage{bm}
\usepackage{mathrsfs}
\usepackage[latin1]{inputenc}
\usepackage{array}
\usepackage{psfrag}
\usepackage{dsfont}
\usepackage{epsfig}
\usepackage{titletoc}
\usepackage{float}   
\usepackage{wrapfig} 
\usepackage{amsmath}\usepackage{amssymb,stmaryrd}
\usepackage{array}
\usepackage{epsfig}
\usepackage{cancel}
\usepackage[dvips]{feynmp}
\usepackage{upgreek}
\usepackage{subfig}
\usepackage{tabularx}
\usepackage{xcolor}
\usepackage{units}
\usepackage{textcomp}
\usepackage{wasysym}
\usepackage{amsthm}

\def\ring#1{{\mathaccent'27 #1}}

\usepackage{textfit} 

\usepackage{hyperref}

\renewcommand{\eqref}[1]{\mbox{Eq.~(\ref{#1})}}
\newcommand{\tabref}[1]{\mbox{Tab.~\ref{#1}}}
\newcommand{\figref}[1]{\mbox{Fig.~\ref{#1}}}
\newcommand{\secref}[1]{\mbox{Sec.~\ref{#1}}}

\newcommand{\appref}[1]{\mbox{App.~\ref{#1}}}

\begin{document}

\begin{fmffile}{diagrams}

\title{Vacuum Cherenkov radiation for Lorentz-violating fermions}

\author{M. Schreck} \email{marco.schreck@ufma.br}
\affiliation{Departamento de F\'{i}sica, Universidade Federal do Maranh\~{a}o \\
65080-805, S\~{a}o Lu\'{i}s, Maranh\~{a}o, Brazil}

\begin{abstract}
The current work focuses on the process of vacuum Cherenkov radiation for Lorentz-violating fermions that are described by the minimal Standard-Model Extension
(SME). To date, most considerations of this important hypothetical process have been restricted to Lorentz-violating photons, as the necessary theoretical tools for
the SME fermion sector have not been available. With their development in a very recent paper, we are now in a position to compute the decay rates based on a
modified Dirac theory. Two realizations of the Cherenkov process are studied. In the first scenario, the spin projection of the incoming fermion is assumed to be
conserved, and in the second, the spin projection is allowed to flip. The first type of process is shown to be still forbidden for the dimensionful $a$ and $b$
coefficients where there are strong indications that it is energetically disallowed for the $H$ coefficients, as well. However, it is rendered possible for the
dimensionless $c$, $d$, $e$, $f$, and $g$ coefficients. For large initial fermion energies, the decay rates for the $c$ and $d$ coefficients were found to grow
linearly with momentum
and to be linearly suppressed by the smallness of the Lorentz-violating coefficient where for the $e$, $f$, and $g$ coefficients this suppression is even quadratic.
The decay rates vanish in the vicinity of the threshold, as expected. The decay including a fermion spin flip plays a role for the spin-nondegenerate operators and
it was found to occur for the dimensionful $b$ and $H$ coefficients as well as for the dimensionless $d$ and $g$. The characteristics of this process differ much
from the properties of the spin-conserving one, e.g., there is no threshold. Based on experimental data of ultra-high-energy cosmic rays, new constraints on
Lorentz violation in the quark sector are obtained from the thresholds. However, it does not seem to be possible to derive bounds from the spin-flip decays. This
work reveals the usefulness of the quantum field theoretic methods recently developed to study the phenomenology of high-energy fermions within the framework of
the~SME.
\end{abstract}
\keywords{Lorentz violation; Dirac equation; Quantum field theory; Cosmic rays}
\pacs{11.30.Cp, 03.65.Pm, 03.70.+k, 26.40.+r}

\maketitle

\newpage
\setcounter{equation}{0}
\setcounter{section}{0}
\renewcommand{\theequation}{\arabic{section}.\arabic{equation}}

\section{Introduction}
\setcounter{equation}{0}

The search for signals of physics at the Planck scale has become more and more prominent over the past years. Since Planck-scale physics
is likely to modify the fundamental structure of spacetime and of particles, one of the most promising signals may be violations of
the fundamental symmetries of Lorentz and {\em CPT} invariance. Such effects were demonstrated to be present in certain string-theory models
\cite{Kostelecky:1988zi,Kostelecky:1989jp,Kostelecky:1989jw,Kostelecky:1991ak,Kostelecky:1994rn}, in loop quantum gravity
\cite{Gambini:1998it,Bojowald:2004bb}, in noncommutative theories \cite{AmelinoCamelia:1999pm,Carroll:2001ws}, spacetime foam models
\cite{Klinkhamer:2003ec,Bernadotte:2006ya,Hossenfelder:2014hha}, spacetimes with nontrivial topologies \cite{Klinkhamer:1998fa,Klinkhamer:1999zh},
and Ho\v{r}ava-Lifshitz gravity~\cite{Horava:2009uw}.\footnote{In this context it shall be briefly mentioned that diffeomorphism invariance in
Ho\v{r}ava-Lifshitz gravity is broken explicitly, which is known to be incompatible with the geometrical structure of the curved spacetime
manifold under consideration \cite{Kostelecky:2003fs}. Recently, an extension of Ho\v{r}ava-Lifshitz gravity has been constructed where diffeomorphism invariance is
broken dynamically in the ultraviolet regime \cite{Cognola:2016gjy}.}

The Standard-Model Extension (SME) provides a general field theory framework incorporating all Lorentz-violating operators that are compatible
with the gauge group of the Standard Model and with coordinate invariance \cite{Colladay:1996iz,Colladay:1998fq,Kostelecky:2003fs}. All operators
are decomposed into field operators of a particular mass dimension and controlling coefficients that govern Lorentz violation. The latter can be
interpreted as background fields
giving rise to preferred spacetime directions. The minimal SME contains all such field operators that are power-counting renormalizable, which
is a finite number. The nonminimal SME additionally includes the infinite number of higher-dimensional field operators
\cite{Kostelecky:2009zp,Kostelecky:2011gq,Kostelecky:2013rta}. The fundamental result of \cite{Greenberg:2002uu} that links Lorentz violation to
{\em CPT} violation in effective field theory ensures that all {\em CPT}-violating operators are contained in the SME automatically. For over
fifteen years, various experimental tests within particle physics and gravity have been carried out producing a high number of constraints on
{\em CPT} and Lorentz violation~\cite{Kostelecky:2008ts}. In particular, Lorentz-violating nonminimal models have also been proposed to look for
unusual electromagnetic interactions between fermions and photons \cite{Belich:2004ng,Casana:2012yj,Araujo:2015zsa,Araujo:2016hsn,Ding:2016lwt}
and altered interactions in the electroweak sector~\cite{Mouchrek-Santos:2016upa}.

Modified particle decays primarily provide excellent tests of Lorentz invariance, which was pointed out even before the SME existed
\cite{Beall:1970rw,Coleman:1998ti,Coleman:1997xq}. One of the most prominent of these processes is vacuum Cherenkov radiation. This is a
Cherenkov-type process of an electrically charged, massive particle that can occur in a Lorentz-violating vacuum without the presence of external electromagnetic
fields. It resembles ordinary Cherenkov radiation that takes place for a charged particle propagating through an optical medium when the particle
moves with a velocity exceeding the phase velocity of light in that medium. Under such conditions, a large number of molecules is polarized nearby the particle
trajectory, which then radiate coherently. A similar process can occur in the presence of a Lorentz-violating background field that permeates
the vacuum and turns it into an optical medium with a nontrivial refractive index.

Vacuum Cherenkov radiation has already been studied in several papers. In the context of the SME, it was investigated predominantly based
on Lorentz violation in the photon sector. A~primary interest was to understand vacuum Cherenkov radiation in the {\em CPT}-odd
photon sector, i.e., within, the (spacelike) Maxwell-Chern-Simons (MCS) theory. In the classical regime, the process was studied in
\cite{Lehnert:2004be,Lehnert:2004hq} where these works were complemented by investigations at the quantum level in \cite{Kaufhold:2005vj,Kaufhold:2007qd}.
After having established a profound understanding of this sector, the interest changed to the {\em CPT}-even modified Maxwell theory. In one of
the first papers \cite{Altschul:2006zz}, the classical regime was considered with quantum-theoretical calculations following in
\cite{Hohensee:2008xz,Klinkhamer:2008ky}. Recent interest has grown towards vacuum Cherenkov radiation for timelike MCS theory. Studies
thereof had been avoided before due to instabilities of that framework \cite{oai:arXiv.org:hep-ph/0101087}. The calculation in
\cite{Altschul:2014bba} was restricted to classical electrodynamics showing that there is no contribution to the decay rate up to second order in
the velocity of the radiating particle. This result was complemented by the finding of \cite{Schober:2015rya} leading to the conclusion that
a charged particle does not radiate photons at all for timelike MCS theory.
The recent article \cite{Colladay:2016rmy} now also includes the behavior in the quantum regime.
In \cite{Colladay:2016rsf} MCS-theory is quantized consistently for a general background vector and a general class of covariant gauges by introducing
a photon mass that lies well beyond observational limits. This procedure paves the way to performing phenomenology and to considering quantum corrections in such
frameworks.

Furthermore, \cite{Altschul:2016ces} extends our knowledge of vacuum Cherenkov radiation of pions producing a novel lower bound
of $-7\times 10^{-13}$ for Lorentz violation in the pion sector. Another recent article \cite{Diaz:2015hxa} even generalizes the calculations
carried out for pointlike particles in \cite{Klinkhamer:2008ky} by taking into account a partonic description of these particles. Last but not least,
gravitational Cherenkov radiation was focused on in \cite{Kostelecky:2015dpa}. This process is the counterpart in gravity that occurs when the
velocity of a massive particle or a photon exceeds the phase velocity of gravity. Under such circumstances, the particle loses energy by emitting
gravitational waves. The needed requirement can be fulfilled in a Lorentz-violating gravitational framework.

Also, it is reasonable to observe that a subset of Lorentz violation in the fermion sector can be converted into a subset of coefficients
in the photon sector \cite{Altschul:2006zz,Altschul:2007kr} to perform the calculation of the decay rate in the latter sector. Hence, a subset
of the results obtained in the photon sector can be interpreted to correspond to equivalent results in the fermion sector.

It shall be mentioned, as well, that vacuum Cherenkov radiation has been studied beyond the SME. The authors of \cite{Carmona:2014lqa} do not
work in the context of effective field theory. Instead, they impose a modified energy-momentum conservation law, which they apply to vacuum
Cherenkov radiation. A different point of view is taken in \cite{Martinez-Huerta:2016odc,Martinez-Huerta:2016azo}. In their papers, the authors
follow a philosophy that differs from the basic principles of the SME. Their intention is to find a modified dispersion relation directly that
yields decay rates for vacuum Cherenkov radiation and photon decay that have the
same characteristics as results based on the SME. To do so, they introduce a generic class of isotropic modifications of photon dispersion relations,
and they compute the decay rates solely from this modified kinematics. The result is compared to the decay rates obtained for the
isotropic {\em CPT}-even modified photon theory and the spacelike case of MCS theory. Furthermore, limits are set on the scale where Lorentz
violation is supposed to break down.

Note also that there is a framework that is called the ``Lorentz-violating extended standard model'' by its founding father Anselmi
\cite{Anselmi:2008bt}. Despite its name, it should not be confused with the SME. This framework comprises an energy scale $\Lambda_L$ that is
associated with Lorentz invariance breaking. It introduces Lorentz-violating contributions into the Standard Model by splitting spacetime
into two submanifolds with a subgroup of the Lorentz group operating as a symmetry group on one of these submanifolds only.
Within Anselmi's framework, vacuum Cherenkov radiation is considered for particular Lorentz-violating contributions
that include higher-dimensional operators \cite{Anselmi:2011ae}. The authors find that cosmic-ray data are consistent with a scale $\Lambda_L$ that is much smaller than
the Planck scale. Finally, the author of \cite{Zloshchastiev:2009aw,Zloshchastiev:2010uu} uses a logarithmic deformation of the Schr\"{o}dinger
equation that is supposed to follow from a nontrivial vacuum structure. Based on this modified quantum mechanics, a modified refractive index of
the vacuum is derived. The latter is then employed to make certain observations with regards to the properties of vacuum Cherenkov radiation.

What is still missing to date is an extended study of vacuum Cherenkov radiation (and other modified particle processes) in theories with Lorentz
invariance broken in the fermion sector of the SME. The current article is supposed to improve our understanding of this physical
scenario, which deals with the phenomenology of ultra-high-energy cosmic rays (UHECRs). We are now ready to reach that achievement as the methodology
to studying quantum processes of particles with Lorentz-violating fermions was developed in the very recent work \cite{Reis:2016hzu}.\footnote{see
\cite{Lehnert:2004ri} for similar considerations in the modified Dirac theory based on the minimal SME}
We primarily consider spin-nondegenerate frameworks in the fermion sector including the $b$, $d$, $g$, and $H$
coefficients. However, to perform individual cross checks and to complete the picture our studies also involve most of the remaining families
of coefficients.

The paper is organized as follows. In \secref{sec:kinematics} we will review the basic kinematics of the process. In that context, a couple of
statements will be made with regards to the Ward identity when Lorentz-violating fermions are involved. In \secref{eq:isotropic-frameworks}
we will continue investigating the process within isotropic frameworks. These studies cover the isotropic $b$, $c$, $d$, $e$, $f$, and $g$
coefficients. The next step will be to apply the developed methods to anisotropic background fields, which is carried out
in \secref{sec:anisotropic-frameworks}. Thereby, the $a$ coefficients and particular anisotropic cases of the $b$, $c$, $d$,
$e$, $f$, $g$, and $H$ coefficients will be covered. The basis of the Cherenkov process considered to that point will be
a conserved projection of particle spin of the incoming fermion. Therefore, \secref{sec:helicity-processes} will be devoted to investigating
helicity decays where the fermion spin projection can flip. All results will be compiled in \secref{sec:constraints}, followed by a set of new constraints
on certain combinations of controlling coefficients obtained from cosmic-ray data. Last but not least, the most important findings will be
summarized in \secref{eq:conclusions}. Calculational details are relegated to \appref{sec:propagators}. We use natural units with
$\hbar=c=1$ unless otherwise stated. Momentum components are always part of the contravariant momentum although their indices
are lower ones due to typesetting reasons.

\section{Basics of the process}
\setcounter{equation}{0}
\label{sec:kinematics}

An important statement with regards to vacuum Cherenkov radiation was made in \cite{Colladay:2016rsf,Colladay:2016eaw}. According to these references, an
occurrence of vacuum Cherenkov radiation can be considered from two points of view. If vacuum Cherenkov radiation is allowed in a certain
Lorentz-violating framework this points out that such a theory has instabilities in the form of negative-energy states in concordant observer
frames~\cite{Kostelecky:2000mm}. Such states were also shown to be related to spacelike particle momenta. Under this condition, photons with spacelike momenta
could be emitted by fermions traveling fast enough. From the standpoint that instabilities are a result of restricting the SME to its minimal version,
the process would be forbidden in the full framework, as instabilities could be remedied by taking into account higher-dimensional operators.
Even if Lorentz violation existed in nature, indeed, we could look for vacuum Cherenkov radiation forever without ever observing the process.
The second point of view is that the minimal SME describes Lorentz violation in nature completely where the operators from the nonminimal SME
do not play a role. As a result, vacuum Cherenkov radiation would be a physical process that can happen in nature, in fact. The absence of this
process then places constraints on Lorentz violation. In
this paper, the second opinion is followed, since all calculations will be restricted to the minimal SME anyhow. However, the first opinion
has its justification, as well, and it could be elaborated on in a future work by taking into account higher-dimensional operators.

Now, we consider a modified quantum electrodynamics (QED) with standard photons minimally coupled to Lorentz-violating spin-1/2
fermions. The Lagrange density can be conveniently written in the following form \cite{Kostelecky:2001jc}:
\begin{subequations}
\begin{align}
\mathcal{L}_{\mathrm{modQED}}[e,m_{\psi},X]&=\mathcal{L}_{\mathrm{photon}}+\mathcal{L}_{\mathrm{Dirac}}[e,m_{\psi},X]\,, \displaybreak[0]\\[2ex]
\mathcal{L}_{\mathrm{photon}}&=-\frac{1}{4}F_{\mu\nu}F^{\mu\nu}-\frac{1}{2}(\partial_{\mu}A^{\mu})^2\,, \displaybreak[0]\\[2ex]
\mathcal{L}_{\mathrm{Dirac}}[e,m_{\psi},X]&=\overline{\psi}\left[\Gamma^{\nu}\left(\frac{\mathrm{i}}{2}\overleftrightarrow{\partial_{\nu}}-eA_{\nu}\right)-M\right]\psi\,, \displaybreak[0]\\[2ex]
\label{eq:definition-quantity-gamma}
\Gamma^{\nu}&\equiv \gamma^{\nu}+c^{(4)\mu\nu}\gamma_{\mu}+d^{(4)\mu\nu}\gamma^5\gamma_{\mu}+e^{(4)\nu}\mathds{1}_4+\mathrm{i}f^{(4)\nu}\gamma^5+\frac{1}{2}g^{(4)\mu\varrho\nu}\sigma_{\mu\varrho}\,, \displaybreak[0]\\[2ex]
M&\equiv m_{\psi}+a^{(3)\mu}\gamma_{\mu}+b^{(3)\mu}\gamma^5\gamma_{\mu}+\frac{1}{2}H^{(3)\mu\varrho}\sigma_{\mu\varrho}\,, \displaybreak[0]\\[2ex]
F_{\mu\nu}&\equiv \partial_{\mu}A_{\nu}-\partial_{\nu}A_{\mu}\,, \displaybreak[0]\\[2ex]
A\overleftrightarrow{\partial_{\mu}}B&\equiv A\partial_{\mu}B-(\partial_{\mu}A)B\,, \displaybreak[0]\\[2ex]
\label{eq:set-controlling-coefficients}
X&\equiv \{a^{(3)\mu},b^{(3)\mu},c^{(4)\mu\nu},d^{(4)\mu\nu},e^{(4)\nu},f^{(4)\nu},g^{(4)\mu\varrho\nu},H^{(3)\mu\varrho}\}\,.
\end{align}
\end{subequations}
Here $\psi$ is the Dirac field, $\overline{\psi}\equiv \psi^{\dagger}\gamma^0$ the Dirac conjugate, $A_{\mu}$ is the photon field,
and $F_{\mu\nu}$ the electromagnetic field strength tensor. All fields are defined in Minkowski spacetime with the metric $\eta_{\mu\nu}$
where we use the signature $(+,-,-,-)$. The Dirac matrices are standard and they obey the Clifford algebra
$\{\gamma^{\mu},\gamma^{\nu}\}=2\eta^{\mu\nu}\mathds{1}_4$ with the unit matrix $\mathds{1}_4$ in spinor space. Furthermore, the
chiral Dirac matrix is defined by $\gamma_5=\gamma^5\equiv\mathrm{i}\gamma^0\gamma^1\gamma^2\gamma^3$ and the commutator of two
Dirac matrices is $\sigma^{\mu\nu}\equiv (\mathrm{i}/2)[\gamma^{\mu},\gamma^{\nu}]$. The elementary charge is $e$, $m_{\psi}$
is the fermion mass, and $X$ is the set of minimal controlling coefficients for fermions. All of these are encoded in the quantities
$\Gamma^{\nu}$ and $M$. The first contains all dimensionless coefficients where the latter is comprised by the dimensionful ones.
For completeness, a Feynman-'t Hooft gauge fixing term has been added to the photon sector Lagrange density.

Energy-momentum conservation with regards to the external states must be taken into consideration to study the kinematics of the vacuum
Cherenkov process. It is not difficult to implement three-momentum conservation as the latter is performed by choosing the three-momentum of the
outgoing fermion as the difference of the three-momenta of the incoming fermion and the outgoing photon. There are two possibilities for
the energy balance of the radiation process:
\begin{subequations}
\begin{align}
\label{eq:spin-conserving-process}
\Delta E^{(\pm)}\Big|_{\substack{\text{without} \\ \text{spin flip}}}=E^{(\pm)}(\mathbf{q})-|\mathbf{k}|-E^{(\pm)}(\mathbf{q}-\mathbf{k})\,, \\[2ex]
\label{eq:helicity-decay}
\Delta E^{(\pm)}\Big|_{\substack{\text{with} \\ \text{spin flip}}}=E^{(\pm)}(\mathbf{q})-|\mathbf{k}|-E^{(\mp)}(\mathbf{q}-\mathbf{k})\,.
\end{align}
\end{subequations}
Here we distinguish between the two fermion dispersion relations that may arise due to a possible spin nondegeneracy of the Lorentz-violating
operators. The first energy balance describes a spin-conserving process and the second a process including a spin flip of the fermion. Energy
conservation is enforced by claiming $\Delta E^{(\pm)}=0$, which renders the process energetically allowed.
\begin{figure}[t]
\centering
\includegraphics{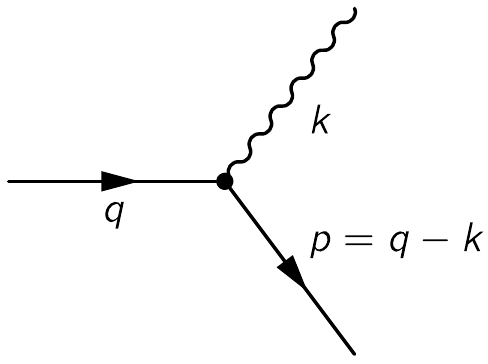}
\caption{Tree-level Feynman diagram for vacuum Cherenkov radiation. The incoming fermion is supposed to have four-momentum $q$ where
the outgoing photon and fermion have four-momenta $k$ and $p=q-k$, respectively.}
\label{fig:diagram-vacuum-cherenkov-radiation}
\end{figure}

To compute the decay rate we have to work within quantum field theory. The only contribution to the process at tree-level is represented by the
Feynman diagram in \figref{fig:diagram-vacuum-cherenkov-radiation}. The Feynman rules of the modified QED at first order in Lorentz violation can
be found in \cite{Kostelecky:2001jc}. First of all, free fermions, i.e., the spinor solutions of the Dirac equation are modified by Lorentz violation.
Second, there is a modified propagator. Third, all dimensionless coefficients contained in the quantity $\Gamma^{\nu}$ contribute to an altered
interaction between the fermion and the photon, cf.~Eqs.~(A7), (A8) in the latter reference. This interaction originates from the minimal coupling
of the photon to the fermion.

We intend to carry out the calculation at all orders in Lorentz violation. So besides the modified vertex, the modified particle energies and the spinor
matrices formed from a spinor and its Dirac conjugate are needed, i.e., expressions of the form $u\overline{u}$. The particle energies are obtained from
the condition of a nonvanishing determinant of the Dirac operator. The spinors can be computed directly by solving the Dirac equation with $p^0$
replaced by the particle energies. Since the explicit spinors will not be needed in the matrix element squared, it suffices to compute the spinor matrices
$u\overline{u}$ based on the validity of the optical theorem at tree-level according to the procedure described in \cite{Reis:2016hzu}.

At this point, a critical remark will be made on the matrix element. Let $\mathcal{M}=\varepsilon_{\mu}\mathcal{M}^{\mu}(k)$ be the amplitude for a QED process with external
on-shell particles and an external photon with four-momentum $k^{\mu}$ that is described by the polarization vector $\varepsilon_{\mu}(k)$. In standard QED, the
well-known Ward identity holds. It guarantees that $k_{\mu}\mathcal{M}^{\mu}(k)=0$ when the polarization vector is replaced by the photon four-momentum
\cite{Peskin:1995}. The Ward identity itself is a direct consequence of gauge invariance, and it follows from the more general Ward-Takahashi identity for
correlation functions involving external lines that are not on-shell. Hence, for $k_{\mu}\mathcal{M}^{\mu}(k)=0$ to be valid for the vacuum
Cherenkov amplitude, in particular, the external fermions must be on-shell. For the case of an incoming fermion line with momentum $q^{\mu}$ and an outgoing
fermion line with momentum $(q+k)^{\mu}$ meeting at a modified vertex, this can be demonstrated by rewriting the contraction $k_{\mu}\Gamma^{\mu}$
in the form
\begin{equation}
k_{\mu}\Gamma^{\mu}=\left[\Gamma^{\mu}(q+k)_{\mu}-M\right]-(\Gamma^{\mu}q_{\mu}-M)\,.
\end{equation}
Sandwiching this expression by the spinors $u(q)$ from the right and $\overline{u}(q+k)$ from the left, leads to the statement
\begin{equation}
\label{eq:ward-identity-proof-2}
k_{\mu}\mathcal{M}^{\mu}=0\,,\quad \mathcal{M}^{\mu}=\overline{u}(q+k)\Gamma^{\mu}u(q)\,,
\end{equation}
for on-shell momenta. This generalization of the Lorentz-invariant relationship can be derived by employing the conjugated Dirac equation for the
first expression and the Dirac equation for the second. Furthermore, the hermiticity conditions $\gamma^0(\Gamma^{\mu})^{\dagger}\gamma^0=\Gamma^{\mu}$
and $\gamma^0M^{\dagger}\gamma^0=M$ must be taken into account. So the Ward identity as a diagrammatical relationship is still valid with the Feynman
rules appropriately adapted, as the SME preserves gauge invariance.

However, care has to be taken when treating additional time derivatives in $\Gamma^{\mu}$. Such derivatives lead to an unconventional time
evolution of the asymptotic fermion states \cite{Colladay:2001wk}. Since we consider a process with external, on-shell particles, it is paramount to resolve that
problem. Fortunately, there exists a remedy, which relies on constructing a transformation in spinor space that can absorb the additional time derivatives. By
introducing a novel spinor $\chi$ in the Dirac operator according to $\psi=A\chi$, the time derivative can be removed when $A$ satisfies the condition
$A^{\dagger}\gamma^0\Gamma^0A=\mathds{1}_4$. Hence, both the Dirac operator and the spinor solutions are affected by this transformation. Let
$\chi=A^{-1}\psi$ be the transformed spinor and
\begin{equation}
\mathcal{L}=\frac{1}{2}\overline{\chi}S'^{-1}\chi+\text{H.c.}\,,\quad S'^{-1}=\gamma^0A^{\dagger}\gamma^0S^{-1}A\,,
\end{equation}
the transformed Lagrange density with the Dirac operator $S^{-1}$. The transformed Dirac equation and its conjugate can be obtained from the
Euler-Lagrange equations:
\begin{subequations}
\begin{align}
0&=S^{-1}A\chi\,, \\[2ex]
0&=\overline{\chi}(\gamma^0A^{\dagger}\gamma^0)S^{-1}\,.
\end{align}
\end{subequations}
Via \eqref{eq:ward-identity-proof-2} it was shown that the validity of the Ward identity at tree-level rests on the Dirac equation and its conjugate. Hence, to make this proof
of the Ward identity still work, we have to replace the modified interaction vertex by the appropriately transformed vertex:
\begin{equation}
-\mathrm{i}e\Gamma^{\mu} \mapsto -\mathrm{i}e(\gamma^0A^{\dagger}\gamma^0)\Gamma^{\mu}A\,.
\end{equation}
The Ward identity is essential when computing the amplitude corresponding to \eqref{fig:diagram-vacuum-cherenkov-radiation}. The amplitude
squared involves the sum over the photon polarization tensors $\varepsilon_{\mu}^{(\lambda)}(k)\varepsilon_{\nu}^{(\lambda)}(k)$ formed from the polarization
vectors of the physical polarization states $\lambda=1,2$. In standard QED, the sum over the polarization tensors is usually replaced by $-\eta_{\mu\nu}$ in
amplitudes, which corresponds to dropping all terms that involve at least one photon four-momentum $k^{\mu}$. The Ward identity renders this procedure
possible and we can follow it here, as well.

The matrix element squared can be computed such as in the standard case with the usual spinors replaced by the modified ones.
Furthermore, the modified interaction vertex is employed but the photon polarization vectors are taken as standard. Without explicitly evaluating
the trace in spinor space, the matrix element squared for a process with initial fermion spin $s$ and final fermion spin $s'$ can be written as follows:
\begin{subequations}
\label{eq:matrix-element-square}
\begin{align}
|\mathcal{M}^{(s,s')}|^2&=4\pi\alpha\,\mathrm{Tr}[\Lambda^{(s)}(q)\Gamma^{\mu}\Lambda^{(s')}(q-k)\Gamma^{\nu}]\Pi_{\mu\nu}(k)\,, \displaybreak[0]\\[2ex]
\Pi_{\mu\nu}(k)&\equiv \sum_{\lambda=1,2} \varepsilon_{\mu}^{(\lambda)}(k)\varepsilon_{\nu}^{(\lambda)}(k)\,,
\end{align}
\end{subequations}
with the electromagnetic fine-structure constant $\alpha=e^2/(4\pi)$. Furthermore, $\Gamma^{\mu}$ is the quantity defined in \eqref{eq:definition-quantity-gamma},
which involves the dimensionless controlling coefficients. The photon polarization sum is encoded in $\Pi_{\mu\nu}$ where $\varepsilon^{(\lambda)}_{\mu}$
is a standard (real) photon polarization vector associated with the polarization mode $\lambda$. The summation runs over the two physical (transverse)
polarizations. The objects $\Lambda^{(s)}$ contain the spinor matrices formed from the spinors $u^{(s)}$ of spin projection $s$ that contribute
to the process:
\begin{equation}
\Lambda_{ab}^{(s)}(p)\equiv u_a^{(s)}(p)\bar{u}_b^{(s)}(p)\,,
\end{equation}
and similarly for $s'$.
For spin-nondegenerate operators the different fermion spin states can be distinguished from each other. In principle, there are processes of the form $\oplus\rightarrow\oplus$, $\ominus\rightarrow\ominus$, $\oplus\rightarrow\ominus$, and $\ominus\rightarrow\oplus$ with the possible spin projections $\oplus$, $\ominus$. So the usual summation over particle spins is not implemented and the matrix element squared for a process with definite initial and final spin will be denoted as $|\mathcal{M}|^2\equiv |\mathcal{M}^{(s,s')}|^2$ for brevity. However, once the process is energetically allowed for a spin-degenerate operator, both fermion modes cannot be distinguished from each other, which is why we average over the initial and sum over the final fermion modes to obtain the matrix element squared:
\begin{equation}
|\mathcal{M}|^2=\frac{1}{2}\sum_{s,s'=\pm1/2} |\mathcal{M}^{(s,s')}|^2\,.
\end{equation}

\section{Isotropic frameworks}
\setcounter{equation}{0}
\label{eq:isotropic-frameworks}

The first part of the paper is dedicated to spin-conserving processes, whereas vacuum Cherenkov radiation including a spin flip of the fermion will be
discussed in \secref{sec:helicity-processes}. Isotropic frameworks shall be examined first because these are usually the simplest from a calculational
perspective. Furthermore, in general, it is more challenging to obtain experimental constraints for isotropic Lorentz violation because Earth-based experiments
often rely on sidereal variations, which only occur when there are spacelike preferred directions.

For an isotropic Lorentz-violating framework it is convenient to describe the phase space of the outgoing photon in spherical coordinates
$(k,\vartheta,\varphi)$ with the magnitude of the three-momentum $|\mathbf{k}|\equiv k$, the polar angle (colatitude) $\vartheta$, and the azimuthal angle $\varphi$.
The basis vectors $\{\hat{\mathbf{e}}_{\varrho},\hat{\mathbf{e}}_{\vartheta},\hat{\mathbf{e}}_{\varphi}\}$ are chosen as usual. The photon three-momentum
points radially away from the origin of the coordinate system, i.e., along the basis vector~$\hat{\mathbf{e}}_{\varrho}$. The physical photon polarization
vectors are orthogonal to the photon momentum, which is why each points along one of the two remaining basis vectors. Explicitly, the four-momentum
and polarization vectors are chosen as
\begin{equation}
\label{eq:polarization-vectors-isotropic}
k^{\mu}=k\begin{pmatrix}
1 \\
\sin\vartheta\cos\varphi \\
\sin\vartheta\sin\varphi \\
\cos\vartheta \\
\end{pmatrix}^{\mu}\,,\quad \varepsilon^{(1)\mu}=\begin{pmatrix}
0 \\
\cos\vartheta\cos\varphi \\
\cos\vartheta\sin\varphi \\
-\sin\vartheta \\
\end{pmatrix}^{\mu}\,,\quad \varepsilon^{(2)\mu}=\begin{pmatrix}
0 \\
-\sin\varphi \\
\cos\varphi \\
0 \\
\end{pmatrix}^{\mu}\,.
\end{equation}
The polarization sum can be decomposed as follows:
\begin{equation}
\label{eq:photon-polarization-sum-covariant}
\Pi^{\mu\nu}(k)=-\eta^{\mu\nu}-\frac{1}{\mathbf{k}^2}k^{\mu}k^{\nu}+\frac{1}{|\mathbf{k}|}(k^{\mu}n^{\nu}+n^{\mu}k^{\nu})\,.
\end{equation}
Here $n^{\mu}\equiv (1,0,0,0)^{\mu}$ is an auxiliary vector that is needed to write the polarization sum in a covariant way. However,
$n^{\mu}$ does not have any physical significance, as photons are not affected by Lorentz violation. The terms dependent on $k^{\mu}$
can be dropped due to the Ward identity.

The decay rate is then computed by integrating the matrix element squared of \eqref{eq:matrix-element-square} over the
two-particle phase space of the final particle state:
\begin{subequations}
\begin{align}
\Gamma&=\frac{1}{2E^{(\pm)}(\mathbf{q})}\gamma\,, \\[2ex]
\gamma&=\int \frac{\mathrm{d}^3k}{(2\pi)^3}\int \frac{\mathrm{d}^3p}{(2\pi)^3} \frac{1}{4\omega(\mathbf{k})E^{(\pm)}(\mathbf{p})}(2\pi)^4\delta^{(4)}(q-k-p)|\mathcal{M}|^2\,,
\end{align}
\end{subequations}
where $\omega(\mathbf{k})=|\mathbf{k}|$ is the standard dispersion relation of the photon.
The matrix element squared depends on the three-momenta $\mathbf{q}$, $\mathbf{k}$, and $\mathbf{p}$. The phase space integration
is carried out over all spatial momentum configurations of the final photon and fermion. The $\delta$ function in the integrand
encodes four-momentum conservation. The second of the two integrals can be computed quickly when three-momentum conservation
is employed. Furthermore, resorting to spherical coordinates for the photon momentum, the decay constant $\gamma$ results in:
\begin{equation}
\gamma=\frac{1}{8\pi} \int_0^{\infty} \mathrm{d}k\,k^2 \int_0^{\pi} \mathrm{d}\vartheta\,\frac{\sin\vartheta}{kE^{(\pm)}(\mathbf{q}-\mathbf{k})}\delta(\Delta E^{(\pm)})|\mathcal{M}|^2\,.
\end{equation}
Due to spatial isotropy, there is no dependence on the azimuthal angle $\varphi$, which is why the corresponding integration can
be performed immediately. Now the remaining integrals must be computed restricted to the configuration with $\Delta E^{(\pm)}=0$
where decays without a spin flip are considered. Note that the results can be conveniently adapted to the situation when a spin flip occurs.

The first reasonable step to do is to solve the energy balance equation with respect to $\cos\vartheta$ where $\vartheta$ is the
angle between the three-momenta of the incoming fermion and the outgoing photon. This delivers a solution $\vartheta_0=\vartheta_0(q,k,X_{\subset})$
dependent on a subset $X_{\subset}$ of controlling coefficients. The magnitude of the photon momentum itself is restricted by the requirement that
$\cos\vartheta_0 \in [-1,1]$. The latter is only possible for $k\in [0,k_{\mathrm{max}}]$, which gives
\begin{subequations}
\begin{align}
\label{eq:decay-rate-isotropic-intermediate}
\gamma&=\frac{1}{8\pi} \int_0^{k_{\mathrm{max}}} \mathrm{d}k\,\Pi(k)|\mathcal{M}|^2\Big|_{\vartheta=\vartheta_0}\,, \\[2ex]
\Pi(k)&=\frac{k\sin\vartheta}{E^{(\pm)}(\mathbf{q}-\mathbf{k})}\left|\frac{\partial \Delta E^{(\pm)}}{\partial\vartheta}\right|^{-1}\Bigg|_{\vartheta=\vartheta_0}\,.
\end{align}
\end{subequations}
The derivative of $\Delta E^{(\pm)}$ results from evaluating the $\delta$ function. Finally, we define the radiated-energy rate,
which gives the energy loss of a radiating fermion as a function of time:
\begin{equation}
\label{eq:radiated-energy-rate-isotropic}
\frac{\mathrm{d}W}{\mathrm{d}t}\equiv \frac{1}{8\pi} \int_0^{k_{\mathrm{max}}} \mathrm{d}k\,\Pi(k)|\mathcal{M}|^2\omega\,.
\end{equation}
The latter is obtained based on the definition of Eq.~(2.11) in \cite{Kaufhold:2007qd} with the phase-space integral reduced to the
final one-dimensional integral over the photon momentum. The remaining integrations will be performed when
considering special cases. However, due to the complexity of the integrands, an analytical integration will be either impractical
or it will produce complicated results that do not provide any further insight. For this reason, the final integration will be carried out
numerically with \textit{Mathematica} and the characteristic behavior of the decay rate as a function of the incoming fermion momentum will
be shown in a plot. However, analytical results will be stated for the high-energy behavior of the decay rates and the radiated-energy rates.
Finally, we introduce the preferred timelike direction $\ring{\lambda}^{\mu}\equiv (1,0,0,0)^{\mu}$, which will be useful at various places.
In this context, we refer to \cite{Kostelecky:2013rta} where all of the minimal isotropic coefficients and a couple of nonminimal ones were
identified and discussed.

\subsection{Isotropic $\boldsymbol{b}$ coefficients}
\label{sec:b-coefficients-isotropic}

There are four minimal {\em CPT}-odd component coefficients of mass dimension 1 that are comprised by the observer vector $b^{(3)\mu}$.
With the single isotropic coefficient $b^{(3)0}\equiv \ring{b}$ and all other coefficients set to zero, the corresponding modified fermion
energies can be written as follows:
\begin{equation}
E^{(\pm)}_{\ring{b}}=\sqrt{(|\mathbf{p}|\pm |\ring{b}|)^2+m_{\psi}^2}\,.
\end{equation}
To find out whether a process is energetically possible at all, it suffices to investigate a linear process with all momenta aligned along a single
line. This is reasonable, as additional energy available in the final state will be put into nonvanishing angles between particle momenta
\cite{Gagnon:2004xh}. Hence, considering a vanishing angle between the momenta of the incoming fermion and the outgoing photon, the energy balance
equation for the vacuum Cherenkov process is
\begin{subequations}
\begin{align}
\Delta E^{(\pm)}_{\ring{b}}&=\sqrt{\widetilde{q}^{\,(\pm)}+m_{\psi}^2}-k-\sqrt{(\widetilde{q}^{\,(\pm)}-k)^2+m_{\psi}^2}\,,\quad \widetilde{q}^{\,(\pm)}\equiv q\pm |\ring{b}|\,.
\end{align}
\end{subequations}
Therefore, it is possible to express the energy balance such as in the Lorentz-invariant case with the single controlling coefficient absorbed
into the initial fermion momentum. This works, since both quantities have the same mass dimension and it shows that the process remains
forbidden even for nonzero Lorentz violation.

\subsection{Isotropic $\boldsymbol{d}$ coefficients}
\label{sec:d-coefficients-isotropic}

The minimal $d$ coefficients are contained in a traceless observer two-tensor. The component coefficients themselves are
dimensionless and {\em CPT}-even but odd under a charge conjugation, which is why they change their signs for antiparticles.
The only isotropic framework for the $d$ coefficients is given by a diagonal, traceless matrix with a single nonvanishing controlling
coefficient:
\begin{equation}
\label{eq:definition-d-coefficients-isotropic}
d^{(4)\mu\nu}=\frac{\ring{d}}{3}\left[4\ring{\lambda}^{\mu}\ring{\lambda}^{\nu}-\eta^{\mu\nu}\right]=\ring{d}\times\mathrm{diag}\left(1,\frac{1}{3},\frac{1}{3},\frac{1}{3}\right)^{\mu\nu}\,,\quad \ring{d}\equiv d^{(4)00}\,.
\end{equation}
The particle energies for this isotropic sector can be cast into the form
\begin{align}
\label{eq:dispersion-relations-isotropic-d}
E^{(\pm)}_{\ring{d}}&=\frac{3}{|\ring{\mathfrak{y}}|}\sqrt{\ring{\mathfrak{X}}\mathbf{p}^2\pm 8\ring{d}|\mathbf{p}|\ring{\mathfrak{Y}}+\ring{\mathfrak{y}}m_{\psi}^2}=E_0\pm \frac{4}{3}\ring{d}|\mathbf{p}|+\dots\,, \displaybreak[0]\\[2ex]
\ring{\mathfrak{X}}&\equiv 9+22\ring{d}^2+\ring{d}^4\,, \displaybreak[0]\\[2ex]
\ring{\mathfrak{Y}}&\equiv \sqrt{\ring{\mathfrak{x}}^2\mathbf{p}^2+\ring{\mathfrak{y}}m_{\psi}^2}\,, \displaybreak[0]\\[2ex]
\label{eq:definitions-isotropic-x-y}
\ring{\mathfrak{x}}&\equiv 3+\ring{d}^2\,,\quad \ring{\mathfrak{y}}\equiv 9(1-\ring{d}^2)\,.
\end{align}
An important observation is made from the dispersion laws. They depend on the square of the controlling coefficient except
at the position of the two signs. In principle, this means that the latter sign is controlled completely by the sign of the controlling
coefficient. So we can restrict our calculation to $\ring{d}>0$ where for $\ring{d}<0$ both signs just have to be switched.

The coefficient $d^{(4)}_{00}$ is one of those that introduce additional time derivatives into the Lagrange density of the fermion
theory. Such derivatives lead to an unconventional time evolution of the asymptotic fermion states --- as mentioned in \secref{sec:kinematics}.
The solution is to introduce a novel spinor $\chi$
in the Dirac operator according to $\psi=A_{\ring{d}}\chi$. The time derivative can then be removed when $A_{\ring{d}}$ satisfies the condition
$A_{\ring{d}}^{\dagger}\gamma^0(\gamma^0+d^{(4)00}\gamma^5\gamma^0)A_{\ring{d}}=\mathds{1}_4$ \cite{Colladay:2001wk}.
This procedure defines both a new set of spinors and even a new propagator. The matrix $A_{\ring{d}}$ is given by
\eqref{eq:transformation-matrix-isotropic} and the modified propagator can be found in \eqref{eq:propagator-d-coefficients-isotropic}.
The altered propagator is then used in conjunction with Eq.~(4.11a) of \cite{Reis:2016hzu} to obtain the spinor matrices in this
framework (see Sec.~V~A in \cite{Reis:2016hzu}).\footnote{Note that in Eqs.~(4.11) of the published version of \cite{Reis:2016hzu},
the matrix $\gamma^5$ is missing in conjunction with the term involving $\Upsilon$.} So the spinor matrix corresponding to the mode with energy $E^{(+)}$ is given by
\begin{subequations}
\begin{align}
\Lambda_{\ring{d}}^{(+)}(\mathbf{p})&=\xi_{\ring{d}}^{\mu}\gamma_{\mu}+\Xi_{\ring{d}}\mathds{1}_4+\zeta_{\ring{d}}^{\mu}\gamma^5\gamma_{\mu}+\psi_{\ring{d}}^{\mu\nu}\sigma_{\mu\nu}\,, \displaybreak[0]\\[2ex]
\xi_{\ring{d}}^{\mu}&=\frac{1}{2\ring{\mathfrak{y}}\ring{\mathfrak{Y}}}\begin{pmatrix}
\ring{\mathfrak{y}}\ring{\mathfrak{Y}}E^{(+)} \\
3\ring{\mathfrak{x}}(\ring{\mathfrak{Y}}+4\ring{d}|\mathbf{p}|)\mathbf{p} \\
\end{pmatrix}^{\mu}\,, \displaybreak[0]\\[2ex]
\Xi_{\ring{d}}&=\frac{3}{2\sqrt{\ring{\mathfrak{y}}}}\frac{m_{\psi}}{\ring{\mathfrak{Y}}}(\ring{\mathfrak{Y}}+4\ring{d}|\mathbf{p}|)\,, \displaybreak[0]\\[2ex]
\zeta_{\ring{d}}^{\mu}&=\frac{|\mathbf{p}|}{2\ring{\mathfrak{Y}}\ring{\mathfrak{y}}}\begin{pmatrix}
\ring{\mathfrak{y}}\ring{\mathfrak{x}}E^{(+)} \\
3(\ring{\mathfrak{Y}}/\mathbf{p}^2)(\ring{\mathfrak{Y}}+4\ring{d}|\mathbf{p}|)\mathbf{p} \\
\end{pmatrix}^{\mu}\,, \displaybreak[0]\\[2ex]
\psi_{\ring{d}}^{\mu\nu}&=\frac{\sqrt{\ring{\mathfrak{y}}}E^{(+)}m_{\psi}}{4\ring{\mathfrak{Y}}|\mathbf{p}|}\begin{pmatrix}
0 & 0 & 0 & 0 \\
0 & 0 & p_3 & -p_2 \\
0 & -p_3 & 0 & p_1 \\
0 & p_2 & -p_1 & 0 \\
\end{pmatrix}^{\mu\nu}\,.
\end{align}
\end{subequations}
Analyzing the kinematics of vacuum Cherenkov radiation, there is no momentum configuration such that $\Delta E^{(-)}_{\ring{d}}=0$. However, for
the mode with energy $E^{(+)}_{\ring{d}}$, energy conservation can be fulfilled. Evaluating the energy balance condition restricts the
angle $\vartheta$ to a function $\vartheta_0$ depending on the initial fermion momentum $q$, the final photon momentum $k$, and
the controlling coefficient~$\ring{d}$:
\begin{subequations}
\label{eq:final-angle-isotropic-d}
\begin{align}
\cos\vartheta_0&=\frac{1}{2\ring{\mathfrak{z}}^2qk}\left[\ring{\mathfrak{z}}(\ring{\mathfrak{z}}q^2+9m_{\psi}^2)-9\ring{\mathfrak{X}}(E^{(+)}(\mathbf{q})-k)^2+\ring{\mathfrak{z}}^2k^2\right. \notag \\
&\phantom{{}={}\frac{1}{2\ring{\mathfrak{z}}^2qk}\Big[}\left.+\,72\ring{d}|E^{(+)}(\mathbf{q})-k|\sqrt{\ring{\mathfrak{x}}^2(E^{(+)}(\mathbf{q})-k)^2-\ring{\mathfrak{z}}m_{\psi}^2}\,\right]\,, \\[2ex]
\ring{\mathfrak{z}}&\equiv 9-\ring{d}^2\,.
\end{align}
\end{subequations}
The maximum photon momentum $k_{\mathrm{max}}$ is determined from the limiting condition that $\cos\vartheta_0=1$. It is
difficult to solve the corresponding equation analytically. Therefore, its solution is obtained numerically in conjunction with integrating
over the final particle phase space.
To compute the phase space factor, we need the derivative of the energy balance equation with respect to the angle in the final
particle state:
\begin{subequations}
\begin{equation}
\frac{\partial\Delta E_{\ring{d}}^{(+)}}{\partial\vartheta}=\left.-\frac{\partial E_{\ring{d}}^{(+)}(q)}{\partial q}\right|_{q=|\mathbf{q}-\mathbf{k}|}\frac{\partial |\mathbf{q}-\mathbf{k}|}{\partial\vartheta}=\left.-\frac{\partial E_{\ring{d}}^{(+)}(q)}{\partial q}\right|_{q=|\mathbf{q}-\mathbf{k}|}\frac{kq\sin\vartheta}{|\mathbf{q}-\mathbf{k}|}\,,
\end{equation}
where
\begin{align}
\frac{\partial E_{\ring{d}}^{(+)}(q)}{\partial q}&=\frac{1}{18qE_{\ring{d}}^{(+)}(q)}\frac{\ring{\mathfrak{z}}^2q^4-81\ring{\mathfrak{Z}}^2}{\ring{\mathfrak{y}}\ring{\mathfrak{Z}}+\ring{\mathfrak{X}}q^2}\,, \\[2ex]
\label{eq:definitions-isotropic-z}
\ring{\mathfrak{Z}}&\equiv m_{\psi}^2-\frac{\ring{\mathfrak{y}}}{9}[E_{\ring{d}}^{(+)}(q)]^2\,.
\end{align}
\end{subequations}
The phase space factor can now be obtained as follows:
\begin{subequations}
\begin{align}
\Pi_{\ring{d}}(k)&=\left.\frac{k\sin\vartheta}{E_{\ring{d}}^{(+)}(\mathbf{q}-\mathbf{k})}\left|\frac{\partial\Delta E_{\ring{d}}^{(+)}}{\partial\vartheta}\right|^{-1}\right|_{\vartheta=\vartheta_0}=\frac{18}{q}\left[\frac{\ring{\mathfrak{y}}\ring{\mathfrak{Z}}q^2+\ring{\mathfrak{X}}q^4}{\ring{\mathfrak{z}}^2q^4-81\ring{\mathfrak{Z}}^2}\right]_{q=|\mathbf{q}-\mathbf{k}|}\Bigg|_{\vartheta=\vartheta_0} \notag \\
&=\left.\frac{9}{\ring{\mathfrak{z}}^2q}\left[\ring{\mathfrak{X}}+\frac{4\ring{d}}{k-E^{(+)}}\frac{2\ring{\mathfrak{x}}^2(E_{\ring{d}}^{(+)}-k)^2-\ring{\mathfrak{z}}m_{\psi}^2}{\sqrt{\ring{\mathfrak{x}}^2(E_{\ring{d}}^{(+)}-k)^2-\ring{\mathfrak{z}}m_{\psi}^2}}\right]\right|_{\vartheta=\vartheta_0}\,.
\end{align}
\end{subequations}
Finally, when the vacuum Cherenkov process is allowed it has a (finite) threshold energy. Hence, it can only occur when the
incoming fermion energy is larger than a certain nonzero minimum value. The latter is expected to depend on the controlling
coefficient. The corresponding threshold momentum is the minimal incoming momentum to be delivered such that the process
is just about to be possible. It will then occur along a single line as the radiation of a photon with a nonvanishing angle $\vartheta$
always requires more energy. The same argument was used to show that the Cherenkov process never occurs for the isotropic
$b$ coefficient, which means that the threshold lies at infinity.
Thus, to determine the threshold momentum, we can just restrict the energy balance equation to $\vartheta=0$ and examine
its first-order expansion for small photon momenta. As the resulting equation is still involved we restrain Lorentz violation to
$\ring{d}\ll 1$, which is a reasonable assumption due to the large number of tight constraints already existing. The threshold
momentum is then determined to be
\begin{equation}
\label{eq:threshold-isotropic-d}
q^{\mathrm{th}}_{\ring{d}}=\frac{1}{2}\sqrt{\frac{3}{2}}\frac{m_{\psi}}{\sqrt{\ring{d}}}+\dots\,.
\end{equation}
The ellipses indicate higher-order corrections in the controlling coefficient. Several remarks are in order. First, due to dimensional
reasons, the threshold momentum linearly depends on the fermion mass. Second, it is inversely proportional to the
square-root of the controlling coefficient, which shows that it moves to infinity for vanishing Lorentz violation. This property is expected, as
the process is energetically forbidden in the Lorentz-invariant limit. Third, due to the square-root dependence,
\eqref{eq:threshold-isotropic-d} resembles the expression for the threshold energy of vacuum Cherenkov radiation in the isotropic
{\em CPT}-even modification of the photon sector, cf.~\cite{Klinkhamer:2008ky}.
Fourth, the particular dependence on the controlling coefficient is not
compulsory as we will see below when investigating the isotropic minimal $e$, $f$, and $g$ coefficients.

Now, in the matrix element squared of \eqref{eq:decay-rate-isotropic-intermediate} all occurrences of $\vartheta$ are replaced by
$\vartheta_0$ of \eqref{eq:final-angle-isotropic-d}, which followed from energy-momentum conservation. The resulting expression
then only depends on the incoming fermion momentum $q$, the outgoing photon momentum $k$, the fermion mass $m_{\psi}$, and the
controlling coefficient. The result of the numerical
integration is presented in \figref{fig:decay-rates-isotropic}. Several remarks are again in order. First, when the incoming fermion
momentum approaches the threshold, the decay rate goes to zero, as expected. Second, for large momenta $q$ of the incoming
fermion, i.e., for $q/m_{\psi}\gg 1$, the decay rate approaches the asymptote $\Gamma^{\infty}_{\ring{d}}=(16/9)\alpha\ring{g}q$.
Hence, for large momenta, the decay rate is a linear function of the fermion momentum and the controlling coefficient.

Recall that at the beginning the calculation was restricted to $\ring{d}>0$. All results can be carried over to $\ring{d}<0$ by changing
the sign before every $\ring{d}$, which means that the mode $E^{(-)}$ must be considered. The threshold energy is then given by
\eqref{eq:threshold-isotropic-d} with $\ring{d}$ replaced by $-\ring{d}$. Hence, it is possible to determine a two-sided bound on $\ring{d}$
from experimental data as shall be seen below.

\subsection{Isotropic $\boldsymbol{c}$ coefficients}
\label{sec:c-coefficients-isotropic}

To perform a cross check, an analog calculation of the decay rate for vacuum Cherenkov radiation will be carried out for the $c$
coefficients. These coefficients are {\em CPT}-even and they are spin-degenerate, which is why they exhibit a single modified dispersion
relation for particles only. Nevertheless, as we will see, the calculation has great parallels to that for the $d$ coefficients
with the only difference of being simpler. The isotropic case is characterized by the following choice:
\begin{equation}
c^{(4)\mu\nu}=\frac{\ring{c}}{3}\left[4\ring{\lambda}^{\mu}\ring{\lambda}^{\nu}-\eta^{\mu\nu}\right]=\ring{c}\times \mathrm{diag}\left(1,\frac{1}{3},\frac{1}{3},\frac{1}{3}\right)^{\mu\nu}\,,\quad \ring{c}\equiv c^{(4)00}\,,
\end{equation}
cf.~\eqref{eq:definition-d-coefficients-isotropic} for the $d$ coefficients. The dispersion relation for particles is simple and it can
be conveniently expressed in terms of two parameters involving the single nonzero controlling coefficient:
\begin{subequations}
\begin{align}
E_{\ring{c}}&=\frac{1}{\ring{\mathfrak{a}}}\sqrt{\ring{\mathfrak{b}}^2\mathbf{p}^2+m_{\psi}^2}\,, \displaybreak[0]\\[2ex]
\label{eq:isotropic-c-quantities-a-b}
\ring{\mathfrak{a}}&\equiv 1+\ring{c}\,,\quad \ring{\mathfrak{b}}\equiv 1-\frac{\ring{c}}{3}\,,
\end{align}
\end{subequations}
In principle, the global prefactor can be pulled under the square root to define a new parameter before $\mathbf{p}^2$ and to redefine the particle mass.
However, the particle energy will be kept as is. Since the coefficient matrix has been chosen such as for the $d$ coefficients, the
Lagrangian contains an additional time derivative, as well. To define proper asymptotic states, this time derivative is removed in
analogy to before by modifying the Dirac operator. That procedure is carried out with the matrix $A_{\ring{c}}$ given in \eqref{eq:transformation-matrices-a-coefficients}.

The general propagator for the $c$ coefficients can be found in Eq.~(B4) of \cite{Reis:2016hzu}. Transforming the Dirac operator
with the matrix $A_{\ring{c}}$, leads to an altered version of the propagator, which is stated in \eqref{eq:propagator-c-coefficients-isotropic}.
For spin-degenerate sectors, there is the peculiarity that both particle spins must be summed over since they will both contribute to the
vacuum Cherenkov process. So Eq.~(4.11a) of~\cite{Reis:2016hzu} is not directly applicable to derive the sum over the spinor matrices
for both fermion modes as the latter holds for spin-nondegenerate sectors only. However, it is not a great obstacle to derive an analog
result for the spin-degenerate coefficients, which is carried out in \appref{sec:spinor-matrices-spin-degenerate}. Finally, using
\eqref{eq:sum-spinor-matrices} allows for obtaining the sum over the spinor matrices:
\begin{subequations}
\label{eq:spinor-matrix-isotropic-c-coefficients}
\begin{align}
\Lambda_{\ring{c}}(\mathbf{p})&\equiv\sum_{s=\pm} u^{(s)}\overline{u}^{(s)}=\left[\xi_{\ring{c}}^{\mu}\gamma_{\mu}+\Xi_{\ring{c}}\mathds{1}_4\right]_{p_0=E_{\ring{c}}}\,, \displaybreak[0]\\[1ex]
\xi_{\ring{c}}^{\mu}&=\frac{\ring{\mathfrak{b}}}{\ring{\mathfrak{a}}}p^{\mu}+\frac{4}{3}\frac{\ring{c}}{\ring{\mathfrak{a}}}p_0\ring{\lambda}^{\mu}\,, \displaybreak[0]\\[2ex]
\Xi_{\ring{c}}&=\frac{m_{\psi}}{\ring{\mathfrak{a}}}\,.
\end{align}
\end{subequations}
To study the kinematics of the process, we introduce spherical coordinates in the momentum space of the photon. From the energy
balance equation, we can then compute the angle $\vartheta$ enclosed by the spatial momenta of the incoming fermion and the
outgoing photon. It is given by
\begin{equation}
\cos\vartheta=\frac{1}{\ring{\mathfrak{b}}^2q}\left(\ring{\mathfrak{a}}\sqrt{\ring{\mathfrak{b}}^2q^2+m_{\psi}^2}-\frac{2}{3}\ring{c}(\ring{\mathfrak{a}}+\ring{\mathfrak{b}})k\right)\,.
\end{equation}
The condition that $\cos\vartheta \in [-1,1]$ restricts the magnitude of the photon momentum to $[0,k_{\mathrm{max}}]$ with the
maximum value
\begin{equation}
k_{\mathrm{max}}=-\frac{3}{2}\frac{\ring{\mathfrak{b}}^2}{\ring{c}(\ring{\mathfrak{a}}+\ring{\mathfrak{b}})}q\left(1-\frac{\ring{\mathfrak{a}}}{\ring{\mathfrak{b}}^2q}\sqrt{\ring{\mathfrak{b}}^2q^2+m_{\psi}^2}\right)\,.
\end{equation}
The process takes place when a nonzero region in momentum space is accessible for the final particles. Thus, it is also possible to
obtain the threshold momentum from the condition that $k_{\mathrm{max}}=0$. The result at first order in Lorentz violation reads
\begin{equation}
\label{eq:threshold-isotropic-c}
q^{\mathrm{th}}_{\ring{c}}=\frac{1}{2}\sqrt{\frac{3}{2}}\frac{m_{\psi}}{\sqrt{-\ring{c}}}+\dots\,.
\end{equation}
From the latter threshold, it is evident that the radiation process is only rendered possible insofar $\ring{c}<0$. This is in contrast to the isotropic
$d$ coefficients that exhibit two distinct fermion dispersion relations allowing both signs of $\ring{d}$. We will come back to the latter point below.
Finally, the phase space element is computed as before leading to a compact result:
\begin{subequations}
\begin{align}
\frac{\partial\Delta E_{\ring{c}}}{\partial\vartheta}&=-\left(\frac{\ring{\mathfrak{b}}}{\ring{\mathfrak{a}}}\right)^2\frac{kq\sin\vartheta}{E_{\ring{c}}(\mathbf{q}-\mathbf{k})}\,, \displaybreak[0]\\[2ex]
\Pi_{\ring{c}}(k)&=\frac{k\sin\vartheta}{E_{\ring{c}}(\mathbf{q}-\mathbf{k})}\left|\frac{\partial\Delta E_{\ring{c}}}{\partial\vartheta}\right|^{-1}\bigg|_{\vartheta=\vartheta_0}=\frac{k\sin\vartheta}{E_{\ring{c}}(\mathbf{q}-\mathbf{k})}\left(\frac{\ring{\mathfrak{a}}}{\ring{\mathfrak{b}}}\right)^2\frac{E_{\ring{c}}(\mathbf{q}-\mathbf{k})}{kq\sin\vartheta}\bigg|_{\vartheta=\vartheta_0}\notag \displaybreak[0]\\
&=\frac{1}{q}\left(\frac{\ring{\mathfrak{a}}}{\ring{\mathfrak{b}}}\right)^2\,,
\end{align}
\end{subequations}
where any dependence on the angle $\vartheta$ drops out in the final expression. Now all ingredients are available to compute the
decay rate and the numerical result is shown in \figref{fig:decay-rates-isotropic}. Qualitatively, it behaves in the same way as the
decay rate for the $d$ coefficients for small enough Lorentz violation. The asymptotic decay rates for $q\gg m_{\psi}$ correspond to
each other due to the averaging over the initial spins and as the rates for processes with a spin flip are heavily suppressed by
Lorentz violation. This observation will be made in \secref{sec:helicity-processes} below.

There exists a coordinate transformation \cite{Altschul:2006zz} that maps the $c$ coefficients in the fermion sector to the nonbirefringent
coefficients $\overline{k}^{(4)}_F$ of the {\em CPT}-even modification of the photon sector. The latter modification is governed by an observer
four-tensor that is suitably contracted with two electromagnetic field strength tensors \cite{Kostelecky:2002hh,Bailey:2004na,Kostelecky:2009zp}.
Recall that $\overline{k}^{(4)}_F$ appears in the parameterization of the sector that produces nonbirefringent photon dispersion
laws at first order in Lorentz violation~\cite{Altschul:2006zz}. The Lagrange density and the decomposition of the observer four-tensor are given
by:
\begin{subequations}
\begin{align}
\mathcal{L}_{\mathrm{modMax}}&=-\frac{1}{4}F_{\mu\nu}F^{\mu\nu}-\frac{1}{4}k_F^{(4)\mu\nu\varrho\sigma}F_{\mu\nu}F_{\varrho\sigma}\,, \\[2ex]
\label{eq:nonbirefringent-ansatz}
k_F^{(4)\mu\nu\varrho\sigma}&=\frac{1}{2}\left(\eta^{\mu\varrho}\overline{k}^{(4)\nu\sigma}_F-\eta^{\mu\sigma}\overline{k}^{(4)\nu\varrho}_F-\eta^{\nu\varrho}\overline{k}^{(4)\mu\sigma}_F+\eta^{\nu\sigma}\overline{k}^{(4)\mu\varrho}_F\right)\,.
\end{align}
\end{subequations}
The coordinate transformation that moves Lorentz violation from the fermion to the photon sector and vice versa works at first order in
Lorentz violation, as well. It allows for moving nonbirefringent
photon coefficients to the fermion sector and vice versa. Hence, the decay rate of any allowed particle-physics process can be computed
for Lorentz violation either sitting in the $c$ coefficient matrix of the fermion sector or in the nonbirefringent modified Maxwell
theory. The leading-order terms of both results should correspond to each other.

Vacuum Cherenkov radiation and photon decay were considered in \cite{Klinkhamer:2008ky} for the isotropic part of modified Maxwell
theory, which is characterized by the single controlling coefficient $\widetilde{\kappa}_{\mathrm{tr}}$. Due to the coordinate
transformation discovered in \cite{Altschul:2006zz}, $\widetilde{\kappa}_{\mathrm{tr}}$ corresponds to the currently studied coefficient
$\ring{c}$ in the fermion sector where $\widetilde{\kappa}_{\mathrm{tr}}=-(4/3)\ring{c}$. The opposite sign can be explained when keeping
in mind that, in principle, Lorentz-violating contributions are moved from one side of an equation to the other. Hence, a positive $\widetilde{\kappa}_{\mathrm{tr}}$ translates to a negative $\ring{c}$ in the fermion sector. Since vacuum Cherenkov
radiation was found to be possible for $\widetilde{\kappa}_{\mathrm{tr}}>0$, it should also be possible for $\ring{c}<0$. Comparing the leading-order terms in the expansions of the threshold momentum and the decay rate for large momenta and
small Lorentz-violating coefficients reveals that they are equal. This is an excellent cross check for both results as both calculations were
performed independently from each other.

\subsection{Isotropic $\boldsymbol{e}$ coefficients}
\label{sec:e-coefficients-isotropic}

The minimal $e$ coefficients are dimensionless and comprised by an observer four-vector. There is a single isotropic coefficient that
corresponds to the zeroth component of this vector, i.e., $\ring{e}\equiv e^{(4)0}$. The isotropic dispersion law can be conveniently
cast into the form
\begin{subequations}
\label{eq:dispersion-relation-isotropic-e}
\begin{align}
E_{\ring{e}}(\mathbf{p})&=\frac{1}{\mathfrak{r}_{\ring{e}}^2}\left[\mathfrak{R}(\mathbf{p})-\ring{e}m_{\psi}\right]\,, \displaybreak[0]\\[2ex]
\mathfrak{R}(\mathbf{p})&=\sqrt{(\mathfrak{r}_{\ring{e}}\mathbf{p})^2+m_{\psi}^2}\,, \displaybreak[0]\\[2ex]
\mathfrak{r}_{\ring{e}}&=\sqrt{1-\ring{e}^2}\,.
\end{align}
\end{subequations}
The Lagrange density for this case is plagued by an additional time derivative, as well. Hence, we have to find a matrix $A_{\ring{e}}$ such that
$A_{\ring{e}}^{\dagger}\gamma^0[\gamma^0+\ring{e}\mathds{1}_4]A_{\ring{e}}=\mathds{1}_4$. Making a particular \textit{Ansatz} for $A_{\ring{e}}$
containing a subset of the 16 Dirac bilinears enables us to find the matrix, which is stated in \eqref{eq:transformation-matrix-e-coefficients}.
Using Eq.~(B4) of \cite{Reis:2016hzu}, the propagator of the Dirac operator transformed with the matrix $A_{\ring{e}}$ can be computed
and it is given in \eqref{eq:propagator-e-coefficients-isotropic}. From this propagator, the sum over the spinor matrices is obtained
with \eqref{eq:sum-spinor-matrices}:
\begin{subequations}
\begin{align}
\Lambda_{\ring{e}}(\mathbf{p})&\equiv\sum_{s=\pm} u^{(s)}\overline{u}^{(s)}=\frac{1}{\mathfrak{r}_{\ring{e}}^2}\left(1-\frac{\ring{e}m_{\psi}}{\mathfrak{R}(\mathbf{p})}\right)\left[\xi^{\mu}_{\ring{e}}\gamma_{\mu}+\Xi_{\ring{e}}\mathds{1}_4\right]_{p_0=E_{\ring{e}}}\,, \displaybreak[0]\\[2ex]
\xi^{\mu}_{\ring{e}}&=\mathfrak{r}_{\ring{e}}p^{\mu}+\ring{e}\left(m_{\psi}-\frac{\ring{e}E_{\ring{e}}}{1+\mathfrak{r}_{\ring{e}}^{-1}}\right)\ring{\lambda}^{\mu}\,, \displaybreak[0]\\[2ex]
\Xi_{\ring{e}}&=-m_{\psi}\,.
\end{align}
\end{subequations}
Note that there is now a global nontrivial prefactor that did not
appear in \eqref{eq:spinor-matrix-isotropic-c-coefficients} for the isotropic $c$ coefficients. The kinematics of the process is not difficult to
evaluate. Solving the energy balance equation for the angle between the three-momenta of the incoming fermion and the outgoing photon gives
\begin{equation}
\cos\vartheta=\frac{1}{q}\left[\frac{\ring{e}^2}{2}k+\mathfrak{R}(\mathbf{q})\right]\,.
\end{equation}
The condition for the angle to lie within the interval $[0,\pi]$ restricts the magnitude of the outgoing photon momentum as follows:
\begin{equation}
k\in [0,k_{\mathrm{max}}]\,,\quad k_{\mathrm{max}}=\frac{2}{\ring{e}^2}\left[q-\mathfrak{R}(\mathbf{q})\right]\,.
\end{equation}
The process takes place only when there is a nonvanishing region in the phase space of the final particles where energy-momentum
conservation is fulfilled. The latter is only possible when $k_{\mathrm{max}}>0$ where the condition $k_{\mathrm{max}}=0$ solved for
$q$ leads to the threshold momentum. The equation has two solutions:
\begin{equation}
\label{eq:threshold-momenta-isotropic-e}
q^{\mathrm{th}}_{\ring{e}}=\pm \frac{m_{\psi}}{\ring{e}}\,,
\end{equation}
whereby both signs are permissible. First of all, there is a critical difference to the threshold momenta obtained previously; compare to
Eqs.~(\ref{eq:threshold-isotropic-d}), (\ref{eq:threshold-isotropic-c}). Here, the threshold is not inversely proportional to the
square root of the coefficient, but to the coefficient itself. Let us assume that the order of magnitude for $\ring{c}$ is the same as for
$\ring{e}$. If the initial fermion propagates through a background field connected to $\ring{c}$ it will start radiating photons at a much smaller
energy compared to the case when it moves through a background field generated by $\ring{e}$. This behavior of the threshold momentum
has a direct implication on the decay rate, as we shall see.

Furthermore, the controlling coefficient can be negative, which would render the second threshold momentum positive. Thus, for
the isotropic $e$ coefficient, vacuum Cherenkov radiation is possible not only for a fixed sign of the controlling coefficient. This particular
behavior is in contrast to the characteristics of the isotropic $c$ coefficient, but the isotropic $d$ coefficient behaves in a similar manner. However, the
latter has two distinct dispersion relations, whereas the currently considered case has a single one only. A calculation in the classical regime
confirms the result, though. Computing the group velocity of the incoming particle, which corresponds to its physical propagation velocity, leads to
\begin{equation}
v_{\mathrm{gr}}=\left|\frac{\partial E}{\partial\mathbf{q}}\right|=\frac{|\mathbf{q}|}{\mathfrak{R}(\mathbf{q})}\,.
\end{equation}
The vacuum Cherenkov process is possible when the phase velocity of light is smaller than the maximum attainable velocity of massive particles.
Since the photon sector is Lorentz-invariant, the classical condition is $v_{\mathrm{gr}}\geq 1$. Solving $v_{\mathrm{gr}}=1$ for $|\mathbf{q}|$ leads to the
same threshold momenta that are given in \eqref{eq:threshold-momenta-isotropic-e}. With regards to phenomenology, this behavior allows for
obtaining a two-sided constraint on the controlling coefficient $\ring{e}$ such as for $\ring{d}$.

With these results at hand, we compute the phase space factor, which has a quite simple form, as well:
\begin{subequations}
\begin{align}
\Pi_{\ring{e}}(k)&=\frac{k\sin\vartheta}{E_{\ring{e}}(\mathbf{q}-\mathbf{k})}\left|\frac{\partial\Delta E_{\ring{e}}}{\partial\vartheta}\right|^{-1}\bigg|_{\vartheta=\vartheta_0}=\frac{(1-\ring{e}^2)\mathfrak{S}}{(\mathfrak{S}-\ring{e}m_{\psi})q}\,, \\[2ex]
\mathfrak{S}&=\mathfrak{R}(\mathbf{q})-(1-\ring{e}^2)k\,.
\end{align}
\end{subequations}
The final step is to perform the numerical integration producing the decay rate shown in \figref{fig:decay-rates-isotropic}. The latter differs
crucially from the previous two results for the $c$ and $d$ coefficients. For high momenta it approaches the straight line that is given by
$\Gamma^{\infty}_{\ring{e}}=(2/3)\alpha\ring{e}^2q$, i.e., it is suppressed by the square of the controlling coefficient.  When the momentum
decreases and approaches the threshold, the decay rate starts deviating from the asymptote and it goes to zero, as expected. However, the
threshold is several orders of magnitude larger than that of the $c$ and $d$ coefficients since it depends on the inverse of the controlling
coefficient instead of the inverse square root.

\subsection{Isotropic $\boldsymbol{f}$ coefficients}
\label{sec:f-coefficients-isotropic}

The $f$ coefficients have a structure very similar to that of the $e$ coefficients although there is a crucial difference, as we will see below.
These coefficients are contained in an observer four-vector, and the isotropic component is given by $\ring{f}\equiv f^{(4)0}$.
The modified fermion dispersion relation depends on the square of the latter coefficient only:
\begin{subequations}
\begin{align}
E_{\ring{f}}&=\frac{1}{\mathfrak{r}_{\ring{f}}}\sqrt{\mathbf{p}^2+m_{\psi}^2}\,, \\[2ex]
\label{eq:quantity-rf}
\mathfrak{r}_{\ring{f}}&\equiv\sqrt{1-\ring{f}^2}\,.
\end{align}
\end{subequations}
Note the similarities in the quantities $\mathfrak{r}_{\ring{e}}$ and $\mathfrak{r}_{\ring{f}}$. Again, we have to find a matrix $A_{\ring{f}}$ to
remove the additional time derivative in the Lagrange density, i.e., $A_{\ring{f}}^{\dagger}\gamma^0[\gamma^0+\mathrm{i}\ring{f}\gamma^5]A_{\ring{f}}=\mathds{1}_4$.
Making a similar \textit{Ansatz} as before, we compute the transformation described by \eqref{eq:transformation-matrix-f-coefficients-2}. However, the
latter matrix has a quite complicated structure. A simpler transformation matrix is obtained from an observation that was made in \cite{Altschul:2006ts}.
In the latter paper, a transformation was found to map the $f$ coefficients onto the $c$ coefficients. This result can be employed to arrive at the
more suitable $A_{\ring{f}}$ stated in \eqref{eq:transformation-matrix-f-coefficients}. The procedure to do so will be described in \appref{sec:scalar-f-coefficients}
in more detail. The propagator is taken from Eq.~(C4) of \cite{Reis:2016hzu} and it is transformed with the matrix $A_{\ring{f}}$
leading to \eqref{eq:propagator-e-coefficients-isotropic}. Based on \eqref{eq:sum-spinor-matrices}, the sum of the spinor matrices reads
\begin{subequations}
\begin{align}
\Lambda_{\ring{f}}(\mathbf{p})&=\sum_{s=\pm} u^{(s)}\overline{u}^{(s)}=\frac{1}{\mathfrak{r}_{\ring{f}}^2}[\xi^{\mu}_{\ring{f}}\gamma_{\mu}+\Xi_{\ring{f}}\mathds{1}_4]_{p_0=E_{\ring{f}}}\,, \\[2ex]
\label{eq:xi-spinor-sum-matrices-anisotropic-f}
\xi^{\mu}_{\ring{f}}&=\mathfrak{r}_{\ring{f}}\left[p^{\mu}+E_{\ring{f}}(\mathfrak{r}_{\ring{f}}-1)\lambda^{\mu}\right]\,, \\[2ex]
\Xi_{\ring{f}}&=\mathfrak{r}_{\ring{f}}m_{\psi}\,.
\end{align}
\end{subequations}
The kinematics can be evaluated as before. The angle between the momenta of the incoming fermion and the outgoing photon is
\begin{equation}
\cos\vartheta=\frac{1}{2q}\left[\ring{f}^2k+2\mathfrak{r}_{\ring{f}}\sqrt{q^2+m_{\psi}^2}\,\right]\,,
\end{equation}
and the magnitude of the photon momentum is restricted to
\begin{equation}
k\in [0,k_{\mathrm{max}}]\,,\quad k_{\mathrm{max}}=\frac{2}{\ring{f}^2}\left[q-\mathfrak{r}_{\ring{f}}\sqrt{q^2+m_{\psi}^2}\,\right]\,.
\end{equation}
From the condition that $k_{\mathrm{max}}=0$ for a nonvanishing phase space volume, the threshold momentum is obtained to be
\begin{equation}
\label{eq:threshold-isotropic-f}
q^{\mathrm{th}}_{\ring{f}}=\pm \frac{m_{\psi}}{\ring{f}}+\dots\,,
\end{equation}
which is a result analog to \eqref{eq:threshold-momenta-isotropic-e}. The only difference is that there are higher-order contributions in
the controlling coefficient. Both signs are permissible as the sign of the coefficient is not fixed, in principle. The phase space element
is extraordinarily simple:
\begin{equation}
\Pi_{\ring{f}}(k)=\frac{\mathfrak{r}_{\ring{f}}^2}{q}\,.
\end{equation}
Now we are in a position to compute the phase space integral over the matrix element squared. The decay rate is plotted in
\figref{fig:decay-rates-isotropic} and it is congruent with the result for $\ring{e}$.
Further important observations can be made when considering the map between the $c$ and the $f$ coefficients \cite{Altschul:2006ts},
which was referred to above. Explicitly, it is given by:
\begin{subequations}
\begin{align}
\label{eq:transformation-c-f-general}
c^{(4)\mu\nu}&=\frac{f^{(4)\mu}f^{(4)\nu}}{(f^{(4)})^2}\left[\sqrt{1-(f^{(4)})^2}-1\right]\,, \\[2ex]
\label{eq:transformation-c-f-timelike}
\ring{c}&=\mathfrak{r}_{\ring{f}}-1=-\frac{1}{2}\ring{f}^2+\dots\,.
\end{align}
\end{subequations}
In the latter map, the $f$ coefficients appear quadratically only. A nonvanishing isotropic coefficient $\ring{f}$ induces an isotropic
part of $c^{(4)\mu\nu}$. However, only the traceless part of $c^{(4)\mu\nu}$ contributes to physical observables and the latter is given by:
\begin{equation}
\label{eq:isotropic-c-traceless-part}
\mathrm{diag}(\ring{c},0,0,0)^{\mu\nu}-\frac{\ring{c}}{4}\eta^{\mu\nu}=\frac{3}{4}\ring{c}\times \mathrm{diag}\left(1,\frac{1}{3},\frac{1}{3},\frac{1}{3}\right)^{\mu\nu}\,.
\end{equation}
Hence, when mapping the $\ring{f}$ onto the $\ring{c}$ coefficient, due to $\ring{c}=-\ring{f}^2/2$ at first order in Lorentz violation,
the combination $\ring{c}'\equiv (3/4)\ring{c}=-(3/8)\ring{f}^2$ is expected to contribute to observables. This is exactly what
we observe. First of all, the threshold momentum is then given by
\begin{equation}
q^{\mathrm{th}}_{\ring{c}'}=\frac{1}{2}\sqrt{\frac{3}{2}}\frac{m_{\psi}}{\sqrt{-\ring{c}'}}+\dots=\frac{m_{\psi}}{|\ring{f}|}+\dots\,,
\end{equation}
which corresponds to the result of \eqref{eq:threshold-isotropic-f} when the absolute-value bars are interpreted as the two different signs.
A similar correspondence exists between the decay rates for a large initial fermion momentum, which we will see below. According to our
deductions in \secref{sec:c-coefficients-isotropic}, the process is only possible for $\ring{c}<0$. Since the square of $\ring{f}$
flows into $\ring{c}$, both signs of $\ring{f}$ are permissible.

\subsection{Isotropic $\boldsymbol{g}$ coefficients}
\label{sec:g-coefficients-isotropic}

The minimal $g$ coefficients are comprised by an observer three-tensor that is antisymmetric in the first two indices. The coefficients
are dimensionless, odd under {\em CPT}, but even under charge conjugation. Hence, they keep their signs for antiparticles. The single
isotropic sector is constructed by assigning the same $g^{(4)123}$ to the six coefficients whose index triples are permutations of
$\{1,2,3\}$ where the signs of the permutations are taken into account. All remaining coefficients shall be zero. Hence, any
component coefficient with at least one timelike index vanishes and the spacelike part is chosen to be totally antisymmetric:
\begin{equation}
g^{ijk}=\ring{g}\times \varepsilon^{ijk}\,,\quad \ring{g}\equiv g^{(4)123}\,.
\end{equation}
It is easier to evaluate the kinematics in comparison to the framework of isotropic $d$ coefficients. This can already be seen from the
dispersion relations
\begin{subequations}
\begin{align}
E^{(\pm)}_{\ring{g}}&=\sqrt{\mathbf{p}^2+(\mathfrak{r}_{\ring{g}}^{(\pm)})^2}\,, \\[2ex]
\mathfrak{r}^{(\pm)}_{\ring{g}}&\equiv m_{\psi}\pm \ring{g}|\mathbf{p}|\,,
\end{align}
\end{subequations}
which are evidently much simpler than \eqref{eq:dispersion-relations-isotropic-d}. Such as for the $d$ coefficients, the different signs
that are responsible for having two dispersion relations are directly linked to the coefficient $\ring{g}$. Therefore, without a restriction
of generality, we assume $\ring{g}>0$ and consider the mode described by $E^{(+)}_{\ring{g}}$. A negative controlling coefficient just
means that we switch the labels of the dispersion laws.

Although the $g$ coefficients are, in general, more involved than the $d$ coefficients, the Lagrange density of the isotropic $g$ sector
is simpler than the Lagrange density of the isotropic $d$ sector. The reason is that nonzero spatial coefficients $g^{(4)ijk}$ do not introduce
additional time derivatives into the Lagrange density. Thus, the previously mentioned issues with the unconventional time evolution of
asymptotic states does not occur here. Instead, Eq.~(4.11a) of \cite{Reis:2016hzu} can be applied directly to obtain the spinor matrices:
\begin{subequations}
\begin{align}
\Lambda_{\ring{g}}^{(+)}(\mathbf{p})&=\xi_{\ring{g}}^{\mu}\gamma_{\mu}+\Xi_{\ring{g}}\mathds{1}_4+\zeta_{\ring{g}}^{\mu}\gamma^5\gamma_{\mu}+\psi_{\ring{g}}^{\mu\nu}\sigma_{\mu\nu}\,, \displaybreak[0]\\[2ex]
\xi_{\ring{g}}^{\mu}&=\frac{1}{2}\begin{pmatrix}
E^{(+)} \\
\mathbf{p} \\
\end{pmatrix}^{\mu}\,, \displaybreak[0]\\[2ex]
\Xi_{\ring{g}}&=\frac{\mathfrak{r}_{\ring{g}}^{(+)}}{2}\,, \displaybreak[0]\\[2ex]
\zeta_{\ring{g}}^{\mu}&=\frac{1}{2|\mathbf{p}|}\begin{pmatrix}
\mathbf{p}^2 \\
E^{(+)}\mathbf{p} \\
\end{pmatrix}^{\mu}\,, \displaybreak[0]\\[2ex]
\psi_{\ring{g}}^{\mu\nu}&=\frac{\mathfrak{r}_{\ring{g}}^{(+)}}{4|\mathbf{p}|}\begin{pmatrix}
0 & 0 & 0 & 0 \\
0 & 0 & p_3 & -p_2 \\
0 & -p_3 & 0 & p_1 \\
0 & p_2 & -p_1 & 0 \\
\end{pmatrix}^{\mu\nu}\,.
\end{align}
\end{subequations}
The calculation of the decay rate for vacuum Cherenkov
radiation is performed in the same manner as for the $d$ coefficients. However, there are also crucial differences. First of all,
the energy balance equation is solved with respect to the angle $\vartheta$ assigned to the outgoing photon, which gives
\begin{align}
\cos\vartheta_0&=\frac{1}{2(1+\ring{g}^2)^2kq}\left\{(1+\ring{g}^2)\left[\ring{g}^2k^2+2kE^{(+)}(\mathbf{q})-2\ring{g}m_{\psi}q\right]\right. \notag \\
&\phantom{{}={}\frac{1}{2(1+\ring{g}^2)^2kq}\Big\}}\left.+\,2\ring{g}m_{\psi}\left[\sqrt{(1+\ring{g}^2)[k-E^{(+)}(\mathbf{q})]^2-m_{\psi}^2}-\ring{g}m_{\psi}\right]\right\}\,.
\end{align}
Since $\vartheta_0\in [0,\pi]$ the outgoing photon momentum $k$ is restricted to $[0,k_{\mathrm{max}}]$. However, the equation to be
solved involves a third-order polynomial in $k$. Its solutions are lengthy and not transparent, which is why they will not be stated. The
threshold momentum follows from the energy balance of the linear process, which is expanded in $k$ and $\ring{g}$. At first order in Lorentz
violation, it is given by
\begin{equation}
\label{eq:threshold-isotropic-g}
q^{\mathrm{th}}_{\ring{g}}=\frac{m_{\psi}}{\ring{g}}+\dots\,.
\end{equation}
Note the similarity to Eqs.~(\ref{eq:threshold-momenta-isotropic-e}), (\ref{eq:threshold-isotropic-f}) for the isotropic $e$ and $f$
coefficients, respectively. For $\ring{g}>0$, only one mode contributes to the process leading to only one sign in the latter threshold.
The phase space factor is computed in analogy to that of the isotropic $d$ coefficients. To do so, we need the derivative of the energy
balance with respect to the angle in the final particle state:
\begin{subequations}
\begin{align}
\frac{\partial\Delta E^{(+)}_{\ring{g}}}{\partial\vartheta}&=\left.-\frac{\partial E_{\ring{g}}^{(+)}(q)}{\partial q}\right|_{q=|\mathbf{q}-\mathbf{k}|}\frac{kq\sin\vartheta}{|\mathbf{q}-\mathbf{k}|}\,, \displaybreak[0]\\[2ex]
\frac{\partial E_{\ring{g}}^{(+)}(q)}{\partial q}&=\frac{(1+\ring{g}^2)q+\ring{g}m_{\psi}}{E_{\ring{g}}^{(+)}(q)}\,.
\end{align}
\end{subequations}
The following relation is useful giving the magnitude of the difference between the initial fermion momentum and the photon momentum evaluated at
the allowed angle $\vartheta_0$:
\begin{equation}
|\mathbf{q}-\mathbf{k}|\Big|_{\vartheta=\vartheta_0}=\frac{1}{2\ring{g}m_{\psi}}\left\{(E_{\ring{g}}^{(+)}-k)^2-\left[(1+\ring{g}^2)(\mathbf{q}-\mathbf{k})^2+m_{\psi}^2\right]\right\}\Big|_{\vartheta=\vartheta_0}\,.
\end{equation}
Finally, with these ingredients, the phase space factor is obtained to have the following form:
\begin{align}
\Pi_{\ring{g}}(k)&=\frac{1}{q}\left[1+\ring{g}^2+\frac{\ring{g}m_{\psi}}{|\mathbf{q}-\mathbf{k}|}\right]^{-1}\bigg|_{\vartheta=\vartheta_0} \notag \displaybreak[0]\\
&=\left.\frac{1}{(1+\ring{g}^2)q}\left[1-\frac{\ring{g}m_{\psi}}{\sqrt{(1+\ring{g}^2)k(k-2E_{\ring{g}}^{(+)})+\left[(1+\ring{g}^2)q+\ring{g}m_{\psi}\right]^2}}\right]\right|_{\vartheta=\vartheta_0} \notag \displaybreak[0]\\
&=\left.\frac{1}{(1+\ring{g}^2)q}\left[1-\frac{\ring{g}m_{\psi}}{\sqrt{(1+\ring{g}^2)(E_{\ring{g}}^{(+)}-k)^2-m_{\psi}^2}}\right]\right|_{\vartheta=\vartheta_0}\,.
\end{align}
Although the kinematics for the isotropic $g$ coefficients is not as involved as that for the isotropic $d$ coefficients, the integrand of \eqref{eq:decay-rate-isotropic-intermediate}
is still too complicated to perform the integration analytically. Hence, the numerical result can be found in \figref{fig:decay-rates-isotropic}. The behavior of the rate is
similar to that for the $e$ and $f$ coefficients, i.e., it is suppressed by the square of the controlling coefficient. For large momenta, it corresponds to
$\Gamma^{\infty}_{\ring{e}}$ or $\Gamma^{\infty}_{\ring{f}}$, as for the spin-degenerate coefficients, the initial fermion spin is averaged over and the
processes with spin flip are heavily suppressed, cf.~\secref{sec:helicity-processes}. For $\ring{g}<0$, the process takes place for the mode labeled by
$(-)$ where the calculation yields analog results.

\subsection{Comparison of results and discussion}

\begin{figure}[t!]
\centering
\includegraphics[scale=0.5]{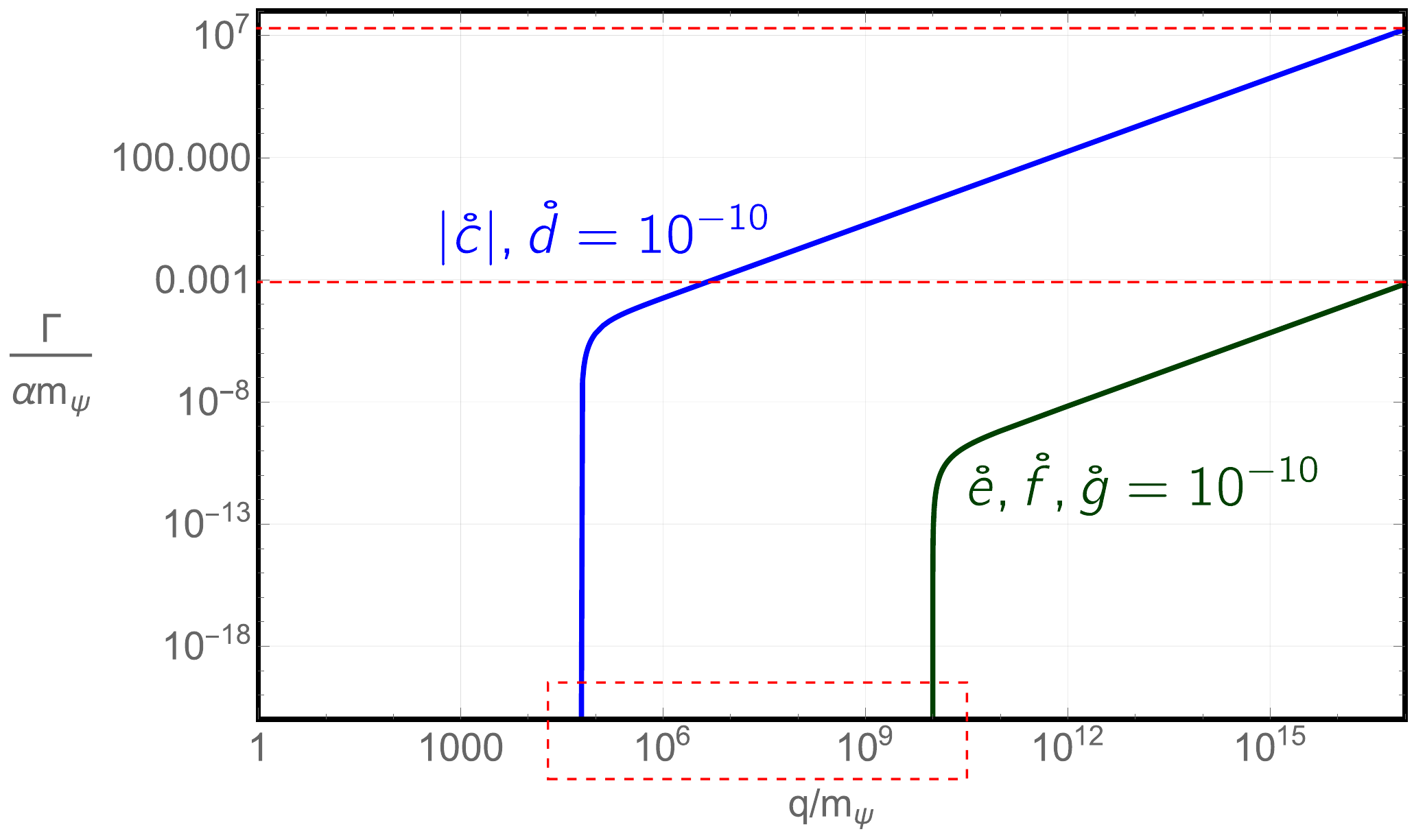}
\caption{Double-logarithmic plot of the decay rate $\Gamma/(\alpha m_{\psi})$ of vacuum Cherenkov radiation for the coefficients $\ring{c}$,
$\ring{d}$ (blue, plain), $\ring{e}$, $\ring{f}$, $\ring{g}$ (green, plain) as functions of the incoming particle
momentum $q/m_{\psi}$. The Lorentz-violating coefficients are chosen equally as $|\ring{c}|=\ring{d}=\ring{e}=\ring{f}=\ring{g}=10^{-10}$.
The dashed, red lines were added to guide the eye.}
\label{fig:decay-rates-isotropic}
\end{figure}
The decay rates for the isotropic $c$, $d$, $e$, $f$, and $g$ coefficients are shown in \figref{fig:decay-rates-isotropic}. For momenta that are much larger
than the thresholds, all curves grow linearly with the momentum where the decay rates decrease quickly when the momentum approaches the thresholds, as expected.
The very similar behavior of the decay rates for $\ring{c}$, $\ring{d}$ and for $\ring{e}$, $\ring{f}$, and $\ring{g}$ is indicative. The characteristics of
the decay rates for $\ring{e}$, $\ring{f}$, and $\ring{g}$ correspond to the properties of the curves for $\ring{c}$, $\ring{d}$ qualitatively. However, on
the one hand, the threshold momenta lie many orders of magnitude apart. On the other hand, the decay rates for $\ring{e}$,
$\ring{f}$, and $\ring{g}$ are suppressed by an additional power of the controlling coefficient, which renders the process inefficient for the latter coefficients.

\section{Anisotropic frameworks}
\setcounter{equation}{0}
\label{sec:anisotropic-frameworks}

After considering a couple of isotropic frameworks, our interest lies in gaining some understanding of the vacuum Cherenkov process in anisotropic theories.
Our studies will be restricted to such frameworks with a residual two-dimensional isotropy. They are characterized by a single preferred spacelike direction
$\bar{\lambda}^{\mu}$ and we choose a suitable observer frame where the latter is purely spacelike.\footnote{Choices of purely
timelike or spacelike preferred directions are always reasonable from a calculational point of view. Due to observer Lorentz invariance, not much is to be
gained from a calculation involving a spacelike preferred direction with nonvanishing time component. We will see that there are some interesting
cases that are covered by such choices of the preferred direction.}
For a system that is isotropic in a plane, it is reasonable to choose cylindrical coordinates of the form $(k_{\bot},\varphi,k_{\parallel})$.
Here the momentum component $k_{\bot}$ shall be perpendicular to the preferred direction and $k_{\parallel}$ is the momentum component parallel to the
preferred direction. Both $k_{\bot}$ and the azimuthal angle $\varphi$ parameterize the plane where the dispersion relation is isotropic. The photon four-momentum
and the physical polarization vectors are then chosen as follows:
\begin{equation}
\label{eq:momentum-polarization-anisotropic}
k^{\mu}=\begin{pmatrix}
|\mathbf{k}| \\
k_{\bot}\cos\varphi \\
k_{\bot}\sin\varphi \\
k_{\parallel} \\
\end{pmatrix}^{\mu}\,,\quad \varepsilon^{(1)\mu}=\begin{pmatrix}
0 \\
-\sin\varphi \\
\cos\varphi \\
0 \\
\end{pmatrix}^{\mu}\,,\quad \varepsilon^{(2)\mu}=\frac{1}{|\mathbf{k}|}\begin{pmatrix}
0 \\
k_{\parallel}\cos\varphi \\
k_{\parallel}\sin\varphi \\
-k_{\bot} \\
\end{pmatrix}^{\mu}\,.
\end{equation}
The polarization vectors are transverse, i.e., orthogonal to the momentum. Furthermore, they are also orthogonal to each other and properly normalized.
The sum over the polarization tensors can be expressed according to \eqref{eq:photon-polarization-sum-covariant} with the auxiliary vector $n^{\mu}$.

Coming to the decay rate, the phase space integral is cast into the form
\begin{subequations}
\begin{align}
\Gamma&=\frac{1}{2E^{(\pm)}(\mathbf{q})}\gamma\,, \\[2ex]
\gamma&=\frac{1}{16\pi^2} \int_0^{2\pi} \mathrm{d}\varphi \int_{-\infty}^{\infty} \mathrm{d}k_{\parallel}\int_0^{\infty} \mathrm{d}k_{\bot}\frac{k_{\bot}}{|\mathbf{k}|E^{(\pm)}(\mathbf{q}-\mathbf{k})}\delta(\Delta E^{(\pm)})|\mathcal{M}|^2\,.
\end{align}
\end{subequations}
The decay constant $\gamma$ is an observer Lorentz scalar, which is why it only depends on observer scalars themselves. For an anisotropic sector with the
single preferred direction $\bar{\lambda}^{\mu}$, this means that it can only depend on particle spin and the kinematic quantities
$q^2$, $q\cdot\bar{\lambda}$, and $\bar{\lambda}^2$ where
$q^{\mu}$ is the initial fermion momentum, which is on-shell. For now, we assume the generic modified fermion dispersion law to be of the form:
\begin{subequations}
\label{eq:generic-dispersion-relation-anisotropic}
\begin{align}
E^{(\pm)}(\mathbf{q})&=\sqrt{q_{\bot}^2+f^{(\pm)}(q_{\parallel},m_{\psi},X_{\subset})}\,, \\[2ex]
q_{\parallel}&\equiv\mathbf{q}\cdot \widehat{\boldsymbol{\lambda}}\,,\quad q_{\bot}\equiv\sqrt{\mathbf{q}^2-q_{\parallel}^2}\,,\quad \widehat{\boldsymbol{\lambda}}\equiv\frac{\boldsymbol{\bar{\lambda}}}{|\boldsymbol{\bar{\lambda}}|}\,,
\end{align}
\end{subequations}
with three-momentum components $q_{\bot}$ and $q_{\parallel}$ that are perpendicular and parallel, respectively, to the preferred direction. The
functions $f^{(\pm)}$ are generic and they shall depend only on the parallel momentum component, the fermion mass, and a subset $X_{\subset}$ of the
Lorentz-violating coefficients. A dependence on particle spin is linked to certain nonzero controlling coefficients and it is
indicated by the index~$(\pm)$. In the Lorentz-invariant case, it holds that $f^{(\pm)}=q_{\parallel}^2+m_{\psi}^2$, independently of particle spin.
Under these circumstances, the on-shell fermion momentum satisfies
\begin{equation}
q^2=E(\mathbf{q})^2-q_{\bot}^2-q_{\parallel}^2=f^{(\pm)}(q_{\parallel},m_{\psi},X_{\subset})-q_{\parallel}^2\,,
\end{equation}
where for vanishing Lorentz violation, $q^2=m_{\psi}^2$, as expected. This demonstrates that $\gamma$ cannot depend on the perpendicular momentum
component as long as the fermion dispersion relation is of the form indicated. We will see that this condition is, indeed, satisfied for the
anisotropic $c$, $d$, $e$, and $f$ coefficients, whereas the $g$ coefficients behave differently. Hence, setting $q_{\bot}=0$ for the first
four cases is permissible.
It is crucial that $\gamma$ does not depend on $q_{\bot}$ for calculational reasons because for $q_{\bot}\neq 0$, the kinematics of the process
involves third-order polynomials in the momentum, which are arduous to solve and make the calculation impractical.

With the assumption of $q_{\bot}=0$ at hand, energy-momentum conservation can be used to get rid of the integral over $k_{\bot}$. The $\delta$ function sets $k_{\bot}$ to the value
$k_{\bot,0}$ that is in accordance with energy-momentum conservation. Note that this also affects the integration over $k_{\parallel}$. As
$k_{\bot}\in [0,\infty)$, the parallel momentum component is restricted and we will see that $k_{\parallel}$ can only vary from zero to
a maximum value $k_{\parallel,\mathrm{max}}$. Since the energy-momentum relation does not depend on the azimuthal angle explicitly,
the limits of the integral over $\varphi$ remain unaffected. The decay rate is then given~by
\begin{subequations}
\begin{align}
\gamma&=\frac{1}{16\pi^2} \int_0^{2\pi}\mathrm{d}\varphi \int_0^{k_{\parallel,\mathrm{max}}} \mathrm{d}k_{\parallel}\,\Pi(\mathbf{k})|\mathcal{M}|^2\Big|_{k_{\bot}=k_{\bot,0}}\,, \displaybreak[0]\\[2ex]
\label{eq:phase-space-factor}
\Pi(\mathbf{k})&=\frac{k_{\bot}}{|\mathbf{k}|E^{(\pm)}(\mathbf{q}-\mathbf{k})}\left|\frac{\partial \Delta E^{(\pm)}}{\partial k_{\bot}}\right|^{-1}\bigg|_{k_{\bot}=k_{\bot,0}}\,.
\end{align}
\end{subequations}
The remaining integral over the azimuthal angle can be easily performed as long as the matrix element squared does not depend on $\varphi$. We will
see that this is the case for the anisotropic sector of $c$, $d$, $e$, and $f$ coefficients. However, for the $g$ coefficients the situation is
different and the nontrivial integral over $\varphi$ must be carried out numerically. Finally, we define the radiated-energy rate in analogy to
\eqref{eq:radiated-energy-rate-isotropic}:
\begin{equation}
\frac{\mathrm{d}W}{\mathrm{d}t}\equiv\frac{1}{16\pi^2} \int_0^{2\pi}\mathrm{d}\varphi \int_0^{k_{\parallel,\mathrm{max}}} \mathrm{d}k_{\parallel}\,\Pi(\mathbf{k})|\mathcal{M}|^2\omega\Big|_{k_{\bot}=k_{\bot,0}}\,.
\end{equation}
In all of the cases considered, the spacelike preferred direction will be chosen to point along the third spatial axis of the coordinate system, i.e.,
$\bar{\lambda}^{\mu}\equiv (0,\bar{\boldsymbol{\lambda}})^{\mu}$ with $\bar{\boldsymbol{\lambda}}=(0,0,1)^T$.

\subsection{Isotropic and anisotropic $\boldsymbol{a}$ coefficients}
\label{sec:a-coefficients}

The $a$ coefficients are {\em CPT}-odd and there are four minimal controlling coefficients that can be put into a Lorentz observer four-vector
$a^{(3)\mu}=(a^{(3)0},\mathbf{a})^{\mu}$.
We will not perform separate analyses for the isotropic and anisotropic parts of the $a$ coefficients as it is convenient to treat both sectors in one
go. Since the framework is spin-degenerate, there is a single fermion dispersion relation. It is reasonable to split the momentum vector into a part
$\mathbf{p}_{\bot}$ perpendicular to the preferred spatial direction $\mathbf{a}$ and into a part $\mathbf{p}_{\parallel}$ parallel to $\mathbf{a}$:
\begin{subequations}
\begin{align}
E_a&=a^{(3)0}+\sqrt{p_{\bot}^2+(p_{\parallel}-|\mathbf{a}|)^2+m_{\psi}^2}\,, \\[2ex]
p_{\parallel}&\equiv\mathbf{p}\cdot \widehat{\mathbf{a}}\,,\quad p_{\bot}\equiv\sqrt{\mathbf{p}^2-p_{\parallel}^2}\,,\quad \widehat{\mathbf{a}}\equiv\frac{\mathbf{a}}{|\mathbf{a}|}\,.
\end{align}
\end{subequations}
The $a$ coefficients have the peculiarity that the four-momentum is shifted by $-a^{(3)\mu}$. Note that the isotropic coefficient $a^{(3)0}$
is subtracted from the particle energy on the left-hand side of the dispersion relation and it has been brought to the right-hand side. To
investigate whether the process is allowed energetically it again suffices to consider a linear process \cite{Gagnon:2004xh}. Hence, the energy
balance for a vanishing perpendicular photon momentum component can be brought into the form
\begin{subequations}
\begin{align}
\Delta E_a&=\sqrt{\widetilde{q}^{\,2}+m_{\psi}^2}-k_{\parallel}-\sqrt{(\widetilde{q}-k_{\parallel})^2+m_{\psi}^2}\,, \\[2ex]
\widetilde{q}&\equiv q-|\mathbf{a}|\,.
\end{align}
\end{subequations}
It is clear that the energy balance for vacuum Cherenkov radiation has the standard shape when $|\mathbf{a}|$ is absorbed by the incoming fermion
momentum. The isotropic coefficient has dropped out of the equation completely. Hence, a background field generated by the minimal $a$ coefficients
does not render vacuum Cherenkov radiation possible.

\subsection{Anisotropic $\boldsymbol{b}$ coefficients}
\label{sec:b-coefficients-anisotropic}

The isotropic component of the $b$ coefficient vector was investigated in \secref{sec:b-coefficients-isotropic} with the result that it does
not allow a charged fermion to radiate photons in vacuo. The next step is to study the anisotropic $b$ coefficients that are contained in
the spatial vector $\mathbf{b}$ and the latter gives rise to a preferred spacelike direction. We will again define momentum components
perpendicular and parallel to the preferred direction. The modified particle energy has the form
\begin{subequations}
\begin{align}
E^{(\pm)}_{\bar{b}}&=\sqrt{p_{\bot}^2+\left(\sqrt{p_{\parallel}^2+m_{\psi}^2}\pm |\mathbf{b}|\right)^2}\,, \\[2ex]
p_{\parallel}&\equiv\mathbf{p}\cdot \widehat{\mathbf{b}}\,,\quad p_{\bot}\equiv\sqrt{\mathbf{p}^2-p_{\parallel}^2}\,,\quad \widehat{\mathbf{b}}\equiv\frac{\mathbf{b}}{|\mathbf{b}|}\,.
\end{align}
\end{subequations}
The kinematics of the vacuum Cherenkov process can again be studied conveniently by looking at the energy balance for a vanishing
perpendicular component of the photon momentum:
\begin{align}
\Delta E_{\bar{b}}^{(\pm)}&=\left|\sqrt{q^2+m_{\psi}^2}\pm |\mathbf{b}|\right|-k_{\parallel}-\left|\sqrt{(q-k_{\parallel})^2+m_{\psi}^2}\pm |\mathbf{b}|\right| \notag \\
&=E_0(q)-k_{\parallel}-E_0(q-k_{\parallel})\,.
\end{align}
It is suggestive that the outer square root can be completely eliminated considering absolute-value bars of the expressions contained.
As Lorentz violation is assumed to be perturbative, the expressions enclosed are positive and the bars can be dropped.
This results in the preferred direction $\mathbf{b}$ dropping out of the energy balance altogether leading to the standard result. Thus,
vacuum Cherenkov radiation is not possible in a background generated by the $b$ coefficients --- at least not without a spin flip.

\subsection{Anisotropic $\boldsymbol{H}$ coefficients}
\label{sec:H-coefficients-anisotropic}

The minimal $H$ coefficients form a set of six independent coefficients that have mass dimension~1. They are contained in an antisymmetric observer
Lorentz two-tensor. Three of these coefficients have one timelike and one spacelike index where the remaining three can be classified as purely
spacelike. To deal with the first class of coefficients, it is convenient to introduce a preferred direction $\mathbf{h}$ composed of the coefficients.
The two-tensor $H^{(3)\mu\nu}$ is then cast into the form
\begin{equation}
H^{(3)\mu\nu}=\begin{pmatrix}
0 & \mathbf{h}^T \\
-\mathbf{h} & \mathbf{0}\otimes \mathbf{0} \\
\end{pmatrix}^{\mu\nu}\,,\quad h^i=H^{(3)0i}\,.
\end{equation}
There are two distinct dispersion relations that are conveniently written as follows:
\begin{subequations}
\label{eq:dispersion-relation-mixed-h}
\begin{align}
E_{\bar{h}}^{(\pm)}(\mathbf{p})&=\sqrt{p_{\parallel}^2+(p_{\bot}\pm |\mathbf{h}|)^2+m_{\psi}^2}\,, \displaybreak[0]\\[2ex]
p_{\parallel}&\equiv\mathbf{p}\cdot \widehat{\mathbf{h}}\,,\quad p_{\bot}\equiv\sqrt{\mathbf{p}^2-p_{\parallel}^2}\,,\quad \widehat{\mathbf{h}}\equiv\frac{\mathbf{h}}{|\mathbf{h}|}\,.
\end{align}
\end{subequations}
Hence, the dispersion relation can be expressed in terms of momentum components parallel and orthogonal to the preferred direction. The framework
is isotropic in the plane perpendicular to $\mathbf{h}$. Considering a linear process, i.e., for a vanishing perpendicular photon momentum component
$k_{\bot}$, the energy balance equation is
\begin{equation}
\Delta E_{\bar{h}}^{(\pm)}=\sqrt{q^2+\widetilde{m}^2}-k_{\parallel}-\sqrt{(q-k_{\parallel})^2+\widetilde{m}^2}\,,\quad \widetilde{m}^2\equiv m_{\psi}^2+\mathbf{h}^2\,.
\end{equation}

The energy balance has the same form as in the Lorentz-invariant case when the controlling coefficients are absorbed by the fermion
mass. The sign before $|\mathbf{h}|$ does then not even matter anymore. For this reason, Lorentz violation caused by any of the three coefficients
$h^{(3)0i}$ does not render vacuum Cherenkov radiation possible.

Coming to the framework characterized by the three purely spacelike coefficients, we can proceed along a similar path. It again makes sense to introduce
a preferred direction $\widetilde{\mathbf{h}}$ whose components correspond to the three nonzero coefficients where the ordering has to be chosen
appropriately. Here, the two-tensor $H^{(3)\mu\nu}$ is decomposed as
\begin{equation}
H^{(3)\mu\nu}=\begin{pmatrix}
0 & \mathbf{0}^T \\
\mathbf{0} & (\varepsilon^{ijk}\widetilde{h}^k) \\
\end{pmatrix}^{\mu\nu}\,,\quad \widetilde{\mathbf{h}}\equiv\begin{pmatrix}
H^{(3)23} \\
-H^{(3)13} \\
H^{(3)12} \\
\end{pmatrix}\,,
\end{equation}
and the modified dispersion relations are given by
\begin{subequations}
\begin{align}
E_{\bar{\bar{h}}}^{(\pm)}(\mathbf{p})&=\sqrt{p_{\parallel}^2+\left(\sqrt{p_{\bot}^2+m_{\psi}^2}\pm |\widetilde{\mathbf{h}}|\right)^2}\,, \displaybreak[0]\\[2ex]
p_{\parallel}&\equiv\mathbf{p}\cdot \widehat{\widetilde{\mathbf{h}}}\,,\quad p_{\bot}\equiv\sqrt{\mathbf{p}^2-p_{\parallel}^2}\,,\quad \widehat{\widetilde{\mathbf{h}}}\equiv\frac{\widetilde{\mathbf{h}}}{|\widetilde{\mathbf{h}}|}\,.
\end{align}
\end{subequations}
It is again wise to decompose the momentum vector into parts parallel and perpendicular to the preferred direction. Although there is again a residual isotropy
in a plane, there is a crucial difference to the dispersion relation of \eqref{eq:dispersion-relation-mixed-h}. In the first, the fermion mass appears in combination
with $p_{\parallel}$ whereas, in the latter, it comes together with $p_{\bot}$. However, this difference does not change the energy balance for vacuum
Cherenkov radiation critically:
\begin{equation}
\Delta E^{(\pm)}_{\bar{\bar{h}}}=\sqrt{q^2+\widetilde{m}_{\pm}^2}-k_{\parallel}-\sqrt{(q-k_{\parallel})^2+\widetilde{m}_{\pm}^2}\,,\quad \widetilde{m}_{\pm}\equiv m_{\psi}\pm |\widetilde{\mathbf{h}}|\,.
\end{equation}
Just as before, all controlling coefficients can be absorbed into the fermion mass. So the structure of the energy balance equation is taken over from
the Lorentz-invariant case. Thus, nonzero minimal purely spacelike $H$ coefficients do not induce a vacuum Cherenkov process, as well.

The general dispersion relation for all six coefficients nonzero at the same time seems to be highly complicated. What can be treated,
though, are cases with just two nonzero coefficients $H^{(3)0i}\equiv h_{0i}$, $H^{(3)jk}\equiv h_{jk}$ where $\{i,j,k\}$ is a permutation of $\{1,2,3\}$.
The dispersion relations for such a case can be written in the form
\begin{equation}
E^{(\pm)}(\mathbf{p})=\sqrt{\mathbf{p}^2+m_{\psi}^2+h_{jk}^2+\widetilde{h}_{jk}^2\pm 2\sqrt{(p_j^2+p_k^2)(h_{jk}^2+\widetilde{h}_{jk}^2)+h_{jk}^2m_{\psi}^2}}\,.
\end{equation}
Here, $\widetilde{h}_{ij}$ are the spatial components of the dual tensor $\widetilde{H}^{(3)\mu\nu}\equiv (1/2)\varepsilon^{\mu\nu\varrho\sigma}H_{\varrho\sigma}^{(3)}$
with the four-dimensional Levi-Civita tensor $\varepsilon^{\mu\nu\varrho\sigma}$ where $\varepsilon^{0123}=1$. Since the momentum components
$p_j$, $p_k$ are multiplied by Lorentz-violating coefficients, it is suggestive to express the dispersion law as
\begin{subequations}
\begin{align}
E^{(\pm)}&=\sqrt{p_{\bot}^2+m_{\psi}^2-\delta m_{\psi}^2+(P_{\parallel}^{(\pm)})^2}\,, \\[2ex]
P_{\parallel}^{(\pm)}&\equiv \sqrt{p_{\parallel}^2+\delta m_{\psi}^2}\pm \sqrt{h_{jk}^2+\widetilde{h}_{jk}^2}\,, \\[2ex]
\delta m_{\psi}^2&\equiv \frac{h_{jk}^2}{h_{jk}^2+\widetilde{h}_{jk}^2}m_{\psi}^2\,,\quad p_{\parallel}\equiv\sqrt{p_j^2+p_k^2}\,,\quad p_{\bot}\equiv\sqrt{\mathbf{p}^2-p_{\parallel}^2}\,,
\end{align}
\end{subequations}
where we defined a new momentum $P_{\parallel}^{(\pm)}$. This is possible here, since in a first step, we can absorb $\delta m_{\psi}^2$ into
$p_{\parallel}^2$ to define the momentum $\widetilde{p}_{\parallel}^{\,2}$ and in a second step, we absorb $\pm(h_{jk}^2+\widetilde{h}_{jk}^2)^{1/2}$
into $\widetilde{p}_{\parallel}$. Both of these quantities do not depend on the momentum itself and they are just additive to $p_{\parallel}^2$ and
to $\widetilde{p}_{\parallel}$, respectively. Based on this result, the energy balance equation is considered for a linear decay. However, it does
not provide a solution for the outgoing photon momentum that is nonzero. This is clear as the modified dispersion relation expressed by the new
momentum $P_{\parallel}^{(\pm)}$ is formally equivalent to the dispersion relation of a standard fermion for each spin projection.

Different combinations of mixed and purely spacelike coefficients lead to highly complicated dispersion relations involving third roots, which are
impractical to study. The only possibility at reach for these cases is to plot the energy balance for randomly chosen values of the coefficients.
This two-dimensional surface was plotted three times as a function of two of the three spatial momentum components. Such plots strongly suggest that
$\Delta E^{(\pm)}$ cannot reach nonnegative values. Thus, vacuum Cherenkov radiation without a spin flip is most probably forbidden for the
entire set of minimal $H$ coefficients, but a rigorous proof is not available at this point.

\subsection{Anisotropic $\boldsymbol{d}$ coefficients}
\label{sec:d-coefficients-anisotropic}

We will investigate a case of anisotropic $d$ coefficients that is characterized by the following diagonal and traceless matrix containing
a single coefficient:

\begin{equation}
\label{eq:d-coefficients-anisotropic}
d^{(4)\mu\nu}=\frac{\bar{d}}{3}\left[4\bar{\lambda}^{\mu}\bar{\lambda}^{\nu}+\eta^{\mu\nu}\right]=\bar{d}\times\mathrm{diag}\left(\frac{1}{3},-\frac{1}{3},-\frac{1}{3},1\right)^{\mu\nu}\,,\quad \bar{d}\equiv d^{(4)33}\,.
\end{equation}
The dispersion relation has a form similar to \eqref{eq:dispersion-relations-isotropic-d}:
\begin{subequations}
\label{eq:dispersion-relations-anisotropic-d}
\begin{align}
E^{(\pm)}_{\bar{d}}&=\sqrt{p_{\bot}^2+f_{\bar{d}}(p_{\parallel})}=E_0\pm \frac{4}{3}\bar{d}\sqrt{p_{\parallel}^2+m_{\psi}^2}\frac{|p_{\parallel}|}{E_0}+\dots\,, \displaybreak[0]\\[2ex]
f_{\bar{d}}(p_{\parallel})&\equiv\frac{9}{\bar{\mathfrak{z}}^2}\left[\bar{\mathfrak{X}}p_{\parallel}^2\pm 8\bar{d}|p_{\parallel}|\bar{\mathfrak{Y}}+\bar{\mathfrak{z}}m_{\psi}^2\right]\,, \displaybreak[0]\\[2ex]
\label{eq:anisotropic-d-quantity-X}
\bar{\mathfrak{X}}&\equiv 9+22\bar{d}^2+\bar{d}^4\,, \displaybreak[0]\\[2ex]
\bar{\mathfrak{Y}}&\equiv\sqrt{\bar{\mathfrak{x}}^2p_{\parallel}^2+\bar{\mathfrak{z}}m_{\psi}^2}\,, \displaybreak[0]\\[2ex]
\label{eq:anisotropic-d-quantities-x-y}
\bar{\mathfrak{x}}&\equiv 3+\bar{d}^2\,,\quad \bar{\mathfrak{y}}\equiv 9(1-\bar{d}^2)\,,\quad \bar{\mathfrak{z}}\equiv 9-\bar{d}^2\,.
\end{align}
\end{subequations}
The third momentum component takes a special role in the fermion energy, since Lorentz violation only appears in the function $f_{\bar{d}}$ of the
third momentum component. This behavior can be traced back to the existence of a preferred
spacelike direction that points along the third spatial axis of the coordinate system. A crucial and helpful property for computing the decay rate of
vacuum Cherenkov radiation is that the dispersion relation is purely conventional in the plane perpendicular to the preferred direction, cf.
the description at the beginning of the current section.

Let $\bar{d}>0$ where the dispersion relation $E^{(+)}_{\bar{d}}$ is taken as a base.
For the choice of anisotropic coefficients of \eqref{eq:d-coefficients-anisotropic} there is a similar issue for the asymptotic states such as
for the isotropic sector, which was studied in \secref{sec:d-coefficients-isotropic}. Again, a matrix $A_{\bar{d}}$ has to be constructed such
that the additional time derivative in the Lagrange density can be removed. This means that the condition
$A_{\bar{d}}^{\dagger}\gamma^0[\gamma^0+(d^{(4)33}/3)\gamma^5\gamma^0]A_{\bar{d}}=\mathds{1}_4$ must be fulfilled. The matrix $A_{\bar{d}}$
is stated in \eqref{eq:transformation-matrix-anisotropic} and the modified propagator is given by \eqref{eq:propagator-d-coefficients-anisotropic}.
The spinor matrix needed for the decay rate is obtained from Eq.~(4.11a) of \cite{Reis:2016hzu} where the components of the previous propagator
are inserted. This leads to
\begin{subequations}
\begin{align}
\Lambda_{\bar{d}}^{(+)}(\mathbf{p})&=\xi_{\bar{d}}^{\mu}\gamma_{\mu}+\Xi_{\bar{d}}\mathds{1}_4+\zeta_{\bar{d}}^{\mu}\gamma^5\gamma_{\mu}+\psi_{\bar{d}}^{\mu\nu}\sigma_{\mu\nu}\,, \displaybreak[0]\\[2ex]
\xi_{\bar{d}}^{\mu}&=\frac{1}{2}\begin{pmatrix}
E^{(+)} \\
\mathbf{p} \\
\end{pmatrix}^{\mu}+\frac{1}{2}\left[\frac{3\bar{\mathfrak{x}}p_{\parallel}}{\bar{\mathfrak{z}}\bar{\mathfrak{Y}}}(\bar{\mathfrak{Y}}+4\bar{d}p_{\parallel})-p_{\parallel}\right]\bar{\lambda}^{\mu}\,, \displaybreak[0]\\[2ex]
\Xi_{\bar{d}}&=\frac{3m_{\psi}}{2\sqrt{\bar{\mathfrak{z}}}\bar{\mathfrak{Y}}}(\bar{\mathfrak{Y}}+4\bar{d}p_{\parallel})\,, \displaybreak[0]\\[2ex]
\zeta_{\bar{d}}^{\mu}&=\frac{\bar{\mathfrak{x}}p_{\parallel}}{2\bar{\mathfrak{Y}}}\begin{pmatrix}
E^{(+)} \\
\mathbf{p} \\
\end{pmatrix}^{\mu}+\left[\frac{3}{2\bar{\mathfrak{z}}}(\bar{\mathfrak{Y}}+4\bar{d}p_{\parallel})-\frac{\bar{\mathfrak{x}}p_{\parallel}^2}{2\bar{\mathfrak{Y}}}\right]\bar{\lambda}^{\mu}\,, \displaybreak[0]\\[2ex]
\psi_{\bar{d}}^{\mu\nu}&=\frac{m_{\psi}\sqrt{\bar{\mathfrak{z}}}}{4\bar{\mathfrak{Y}}}\begin{pmatrix}
0 & -p_2 & p_1 & 0 \\
p_2 & 0 & E^{(+)} & 0 \\
-p_1 & -E^{(+)} & 0 & 0 \\
0 & 0 & 0 & 0 \\
\end{pmatrix}^{\mu\nu}\,.
\end{align}
\end{subequations}
For dispersion relations of the special form of \eqref{eq:dispersion-relations-anisotropic-d}, the momentum component $k_{\bot,0}$
following from energy-momentum conservation is conveniently expressed via the K\"{a}ll\'{e}n function~$\Omega$:
\begin{subequations}
\label{eq:orthogonal-momentum-d-coefficients}
\begin{align}
k_{\bot,0}&=\frac{\sqrt{\Omega(f_{\bar{d}}(q),k_{\parallel}^2,f_{\bar{d}}(q-k_{\parallel}))}}{2E_{\bar{d}}^{(+)}(q)}\,, \\[2ex]
\label{eq:kallen-function}
\Omega(x,y,z)&\equiv x^2+y^2+z^2-2xy-2yz-2xz\,.
\end{align}
\end{subequations}
The parallel photon momentum $k_{\parallel}$ follows from the condition that $k_{\bot}=0$. However, the resulting equation is too complicated
to be solved analytically, which is why it is solved numerically for certain values of the incoming fermion momentum and controlling coefficient.
A first-order approximation for the threshold momentum can be obtained, nevertheless, from the energy balance of the linear process with a vanishing
perpendicular photon momentum component $k_{\bot}$. An expansion in $k_{\parallel}$ and $\bar{d}$ delivers
\begin{equation}
\label{eq:threshold-anisotropic-d}
q^{\mathrm{th}}_{\parallel,\bar{d}}=\frac{1}{2}\sqrt{\frac{3}{2}}\frac{m_{\psi}}{\sqrt{\bar{d}}}+\dots\,.
\end{equation}
Due to the structure of the dispersion relation, the phase space factor can be computed elegantly, as well. First of all, the derivative of the energy
balance equation with respect to $k_{\bot}$ is given by
\begin{equation}
\left.\frac{\partial\Delta E_{\bar{d}}^{(+)}}{\partial k_{\bot}}\right|_{k_{\bot}=k_{\bot,0}}=\left.-\left(\frac{k_{\bot}}{|\mathbf{k}|}+\frac{k_{\bot}}{E_{\bar{d}}^{(+)}(\mathbf{q}-\mathbf{k})}\right)\right|_{k_{\bot}=k_{\bot,0}}=\left.-\frac{E_{\bar{d}}^{(+)}(\mathbf{q})k_{\bot}}{|\mathbf{k}|E_{\bar{d}}^{(+)}(\mathbf{q}-\mathbf{k})}\right|_{k_{\bot}=k_{\bot,0}}\,.
\end{equation}
The phase space factor is especially simple, as it depends on the initial fermion energy only:
\begin{equation}
\label{eq:phase-space-d-coefficients-anisotropic}
\Gamma_{\bar{d}}(k)=\frac{k_{\bot}}{|\mathbf{k}|E_{\bar{d}}^{(+)}(\mathbf{q}-\mathbf{k})}\left|\frac{\partial\Delta E_{\bar{d}}^{(+)}}{\partial k_{\bot}}\right|^{-1}\Bigg|_{k_{\bot}=k_{\bot,0}}=\frac{1}{E_{\bar{d}}^{(+)}(\mathbf{q})}\,.
\end{equation}
Therefore, it can be pulled out of the phase space integral, finally giving:
\begin{equation}
\gamma=\frac{1}{8\pi E_{\bar{d}}^{(+)}(\mathbf{q})} \int_0^{k_{\parallel,\mathrm{max}}} \mathrm{d}k_{\parallel}\,|\mathcal{M}|^2\Big|_{k_{\bot}=k_{\bot,0}}
\end{equation}
Thus, in the matrix element squared, all $k_{\bot}$ have to be replaced by \eqref{eq:orthogonal-momentum-d-coefficients}. The resulting
expression must be integrated over $k_{\parallel}$. When the matrix element squared does itself not depend on the azimuthal angle, the integral
over $\varphi$ just gives a factor of $2\pi$. The numerical result for the decay rate as a function of the incoming fermion momentum is
shown in \figref{fig:decay-rates-anisotropic}. The curve has the typical characteristics of the decay rate for the corresponding isotropic
coefficient $\ring{d}$. The calculation can be carried out in an analog way for the second fermion mode when $\bar{d}<0$.

\subsection{Anisotropic $\boldsymbol{c}$ coefficients}
\label{sec:c-coefficients-anisotropic}

A proper cross check for calculations carried out in anisotropic frameworks with a remaining two-dimensional isotropy is performed within
a fermion theory modified by the following set of $c$ coefficients:
\begin{equation}
\label{eq:c-coefficients-anisotropic}
c^{(4)\mu\nu}=\frac{\bar{c}}{3}\left[4\bar{\lambda}^{\mu}\bar{\lambda}^{\nu}+\eta^{\mu\nu}\right]=\bar{c}\times\mathrm{diag}\left(\frac{1}{3},-\frac{1}{3},-\frac{1}{3},1\right)^{\mu\nu}\,,\quad \bar{c}\equiv c^{(4)33}\,.
\end{equation}
Note the similarity to \eqref{eq:d-coefficients-anisotropic}.
Hence, many of the following calculations are very similar to those presented in the previous section. The modified fermion energy can be
decomposed into the standard orthogonal part and a contribution dependent on the momentum component parallel to the preferred direction.
The latter contains all Lorentz violation and it is described by the function~$f_{\bar{c}}$:
\begin{subequations}
\label{eq:dispersion-relation-anisotropic-c}
\begin{align}
E_{\bar{c}}&=\sqrt{p_{\bot}^2+f_{\bar{c}}(p_{\parallel})}\,, \displaybreak[0]\\[2ex]
f_{\bar{c}}(p_{\parallel})&\equiv\frac{1}{\bar{\mathfrak{b}}^2}(\bar{\mathfrak{a}}^2p_{\parallel}^2+m_{\psi}^2)\,, \displaybreak[0]\\[2ex]
\label{eq:anisotropic-c-quantities-a-b}
\bar{\mathfrak{a}}&\equiv 1-\bar{c}\,,\quad \bar{\mathfrak{b}}\equiv 1+\frac{\bar{c}}{3}\,.
\end{align}
\end{subequations}
This particular framework induces an additional time derivative in the Lagrange density, as well. It is removed from the Dirac operator with
the matrix $A_{\bar{c}}$ stated in \eqref{eq:transformation-matrices-a-coefficients}. This results in the modified propagator that is given
by \eqref{eq:propagator-c-coefficients-anisotropic}. Summing the spinor matrices produces
\begin{subequations}
\begin{align}
\Lambda_{\bar{c}}(\mathbf{p})&\equiv\sum_{s=\pm} u^{(s)}\overline{u}^{(s)}=[\xi_{\bar{c}}^{\mu}\gamma_{\mu}+\Xi_{\bar{c}}\mathds{1}_4]_{p_0=E_{\bar{c}}}\,, \displaybreak[0]\\[2ex]
\xi_{\bar{c}}^{\mu}&=p^{\mu}-\frac{4}{3}\frac{\bar{c}}{\bar{\mathfrak{b}}}p_{\parallel}\bar{\lambda}^{\mu}\,, \displaybreak[0]\\[2ex]
\Xi_{\bar{c}}&=\frac{m_{\psi}}{\bar{\mathfrak{b}}}\,.
\end{align}
\end{subequations}
The next step is to analyze the kinematics of the process. The particular form of the dispersion relation again allows for expressing the
photon momentum component perpendicular to the preferred direction via the K\"{a}ll\'{e}n function, cf.~\eqref{eq:orthogonal-momentum-d-coefficients}.
Note that here the function $f_{\bar{c}}$ and the dispersion relation $E_{\bar{c}}$ have to be employed. The photon momentum component parallel to the
preferred direction is restricted by the requirement that $k_{\bot}\geq 0$. This leads to $k_{\parallel}\in [0,k_{\parallel,\mathrm{max}}]$
with
\begin{equation}
k_{\parallel,\mathrm{max}}=\frac{3}{2\bar{c}(\bar{\mathfrak{a}}+\bar{\mathfrak{b}})}\left(-\bar{\mathfrak{a}}^2q+\bar{\mathfrak{b}}\sqrt{\bar{\mathfrak{a}}^2q^2+m_{\psi}^2}\right)\,.
\end{equation}
From the condition $k_{\parallel,\mathrm{max}}=0$, we obtain the threshold momentum
\begin{equation}
\label{eq:threshold-anisotropic-c}
q^{\mathrm{th}}_{\parallel,\bar{c}}=\frac{1}{2}\sqrt{\frac{3}{2}}\frac{m_{\psi}}{\sqrt{-\bar{c}}}+\dots\,.
\end{equation}
Hence, the process only occurs if $\bar{c}<0$. This makes sense for the same reason that was outlined in \secref{sec:c-coefficients-isotropic}. The
phase space factor is computed in analogy to the previous section with the simple result of \eqref{eq:phase-space-d-coefficients-anisotropic} where
the suitable particle energy of \eqref{eq:dispersion-relation-anisotropic-c} has to be inserted. The decay rate is plotted in \figref{fig:decay-rates-anisotropic}
and it is similar to that for the $\bar{d}$ coefficient due to the same reasons stated for the isotropic cases.

\subsection{Anisotropic $\boldsymbol{e}$ coefficients}
\label{sec:e-coefficients-anisotropic}

The anisotropic case of $e$ coefficients is generated by nonvanishing spatial coefficients, i.e, $e^{(4)\mu}\equiv (0,\mathbf{e}^{(4)})^{\mu}$. It is
reasonable to choose the observer frame such that $\mathbf{e}^{(4)}$ points along the third axis of the coordinate system: $\mathbf{e}^{(4)}\equiv\bar{e}\,\bar{\boldsymbol{\lambda}}$
with $\bar{e}\equiv |\mathbf{e}^{(4)}|$. The fermion dispersion relation can then be written as follows:
\begin{subequations}
\label{eq:dispersion-relations-anisotropic-e}
\begin{align}
E_{\bar{e}}&=\sqrt{p_{\bot}^2+f_{\bar{e}}(p_{\parallel})}\,, \\[2ex]
f_{\bar{e}}(p_{\parallel})&=\mathfrak{r}_{\bar{e}}(p_{\parallel})^2+p_{\parallel}^2\,, \\[2ex]
\label{eq:definition-r}
\mathfrak{r}_{\bar{e}}(p_{\parallel})&\equiv \bar{e}p_{\parallel}+m_{\psi}\,.
\end{align}
\end{subequations}
Besides, it is convenient to introduce the quantity $\mathfrak{r}_{\bar{e}}$ that will appear at various places.
The anisotropic sector does not exhibit any additional time derivatives in its Lagrange density. Thus, the spinor solutions do not have
to be transformed and we obtain the sum over the spinor matrices directly from \eqref{eq:sum-spinor-matrices} of the current paper
and Eq.~(B4) of \cite{Reis:2016hzu}:
\begin{equation}
\Lambda_{\bar{e}}(\mathbf{p})\equiv \sum_{s=\pm} u^{(s)}\overline{u}^{(s)}=[\cancel{p}+\mathfrak{r}_{\bar{e}}\mathds{1}_4]_{p_0=E_{\bar{e}}}\,.
\end{equation}
The kinematics of the process is evaluated as usual for sectors with a single spacelike preferred direction. First, it makes sense to
determine the perpendicular photon momentum component from energy-momentum conservation. In the dispersion law,
Lorentz violation is only linked to the parallel momentum component. Hence, for the anisotropic $e$ coefficients, it is again
possible to express the perpendicular photon momentum component via the K\"{a}ll\'{e}n function, cf.~\eqref{eq:orthogonal-momentum-d-coefficients}.
Since the process only takes place as long as $k_{\bot}>0$, the parallel component $k_{\parallel}$ is restricted to the interval
$[0,k_{\parallel,\mathrm{max}}]$ with
\begin{equation}
k_{\parallel,\mathrm{max}}=\frac{2}{\bar{e}^2}\left[q_{\parallel}+\bar{e}\mathfrak{r}_{\bar{e}}(q_{\parallel})-\sqrt{q_{\parallel}^2+\mathfrak{r}_{\bar{e}}(q_{\parallel})^2}\,\right]\,.
\end{equation}
Note that here the quantity $\mathfrak{r}_{\bar{e}}$ of \eqref{eq:definition-r} is a function of $q_{\parallel}$.
From the condition $k_{\parallel,\mathrm{max}}=0$ we obtain the minimal initial fermion momentum that renders the process possible:
\begin{equation}
q^{\mathrm{th}}_{\parallel,\bar{e}}=\frac{m_{\psi}}{\bar{e}}+\dots\,.
\end{equation}
In contrast to the isotropic case of $e$ coefficients with the single nonvanishing $e^{(4)0}$, there is only a single solution for the
threshold energy because the magnitude of $\mathbf{e}^{(4)}$ is positive. Last but not least, the phase space factor is given by
\eqref{eq:phase-space-d-coefficients-anisotropic} with the particle energy taken from \eqref{eq:dispersion-relations-anisotropic-e}.
The numerical result for the decay rate is presented in \figref{fig:decay-rates-anisotropic}. Note the similarities to the decay rate for
the isotropic coefficient $\ring{e}$. For $q_{\parallel}\gg m_{\psi}$ it is suppressed by the square of the controlling coefficient and
its threshold momentum lies several orders of magnitude above the thresholds for $\bar{c}$ and $\bar{d}$.

\subsection{Anisotropic $\boldsymbol{f}$ coefficients}
\label{sec:f-coefficients-anisotropic}

The anisotropic case of $f$ coefficients is chosen in analogy to the anisotropic sector of the $e$ coefficients, i.e.,
$f^{(4)\mu}\equiv (0,\mathbf{f}^{(4)})^{\mu}$. A suitable observer frame makes $\mathbf{f}^{(4)}$ point along the third spatial axis
of the coordinate system: $\mathbf{f}^{(4)}\equiv\bar{f}\,\bar{\boldsymbol{\lambda}}$ with $\bar{f}\equiv |\mathbf{f}^{(4)}|$. The modified dispersion relation
then reads
\begin{subequations}
\label{eq:dispersion-relations-anisotropic-f}
\begin{align}
E_{\bar{f}}&=\sqrt{p_{\bot}^2+f_{\bar{f}}(p_{\parallel})}\,, \displaybreak[0]\\[2ex]
f_{\bar{f}}(p_{\parallel})&=\mathfrak{r}_{\bar{f}}^2p_{\parallel}^2+m_{\psi}^2\,, \displaybreak[0]\\[2ex]
\mathfrak{r}_{\bar{f}}&=\sqrt{1+{\bar{f}}^2}\,,
\end{align}
\end{subequations}
where the quantity $\mathfrak{r}_{\bar{f}}$ is independent of the momentum in contrast to $\mathfrak{r}_{\bar{e}}$ of \eqref{eq:definition-r}.
Since there are not any additional time derivatives in the Lagrange density the sum over the spinor matrices follows simply from
\eqref{eq:sum-spinor-matrices} of the current article and Eq.~(C4) of \cite{Reis:2016hzu}:
\begin{equation}
\Lambda_{\bar{f}}(\mathbf{p})\equiv \sum_{s=\pm} u^{(s)}\overline{u}^{(s)}=[\cancel{p}+m_{\psi}\mathds{1}_4-\mathrm{i}(\mathbf{p}\cdot\mathbf{f}^{(4)})\gamma^5]_{p_0=E_{\bar{f}}}\,.
\end{equation}
Energy-momentum conservation forces the perpendicular momentum component of the final photon to be given by \eqref{eq:orthogonal-momentum-d-coefficients}
with the function $f_{\bar{f}}$ inserted. The parallel momentum component is restricted to
\begin{equation}
k_{\parallel}\in [0,k_{\parallel,\mathrm{max}}]\,,\quad k_{\parallel,\mathrm{max}}=\frac{2}{\bar{f}^2}\left[\mathfrak{r}_{\bar{f}}^2q-\sqrt{\mathfrak{r}_{\bar{f}}^2q^2+m_{\psi}^2}\,\right]\,,
\end{equation}
which finally leads to the threshold momentum:
\begin{equation}
q^{\mathrm{th}}_{\bar{f}}=\frac{m_{\psi}}{\bar{f}}+\dots\,.
\end{equation}
As $\bar{f}>0$, the threshold momentum has a fixed sign only. The phase space factor simply corresponds to
\eqref{eq:phase-space-d-coefficients-anisotropic} with the fermion energy of \eqref{eq:dispersion-relations-anisotropic-f}. The decay rate is
plotted in \figref{fig:decay-rates-anisotropic} and the result is similar to the decay rate for the anisotropic $\bar{e}$ studied in the previous
section. Furthermore, the observation made in \cite{Altschul:2006ts} can be applied to the current case, as well. Thus, we try to map the
$f$ onto the $c$ coefficients where a nonvanishing $f_3^{(4)}$ will be considered only. We then obtain from the general transformation of
\eqref{eq:transformation-c-f-general}:
\begin{equation}
\bar{c}=1-\sqrt{1+(f^{(4)}_3)^2}=-\frac{1}{2}(f^{(4)}_3)^2+\dots\,.
\end{equation}
Furthermore, when transforming the $f^{(4)}_3$ into the $c$ coefficient sector, only the traceless part contributes:
\begin{equation}
\mathrm{diag}(0,0,0,\bar{c})^{\mu\nu}+\frac{\bar{c}}{4}\eta^{\mu\nu}=\frac{3}{4}\bar{c}\times \left(\frac{1}{3},-\frac{1}{3},-\frac{1}{3},1\right)^{\mu\nu}\,,
\end{equation}
cf.~\eqref{eq:isotropic-c-traceless-part}. The rest of the current consideration works in analogy to what we did for the isotropic $f$ coefficients
in \secref{sec:f-coefficients-isotropic}. This demonstrates the consistency between the results for the anisotropic $c$ and $f$ coefficients, and the
transformation obtained in \cite{Altschul:2006ts}.

\subsection{Anisotropic $\boldsymbol{g}$ coefficients}
\label{sec:g-coefficients-anisotropic}

Among the anisotropic choices of minimal $g$ coefficients, we choose
$g^{(4)\mu\nu\varrho}=\bar{g}(\ring{\lambda}^{\mu}\bar{\lambda}^{\nu}-\bar{\lambda}^{\mu}\ring{\lambda}^{\nu})\bar{\lambda}^{\varrho}$
with $g^{(4)033}\equiv \bar{g}$. The anisotropy is linked to the preferred direction pointing along the third spatial axis of the coordinate system.
The corresponding dispersion relations are
\begin{subequations}
\begin{align}
E^{(\pm)}_{\bar{g}}&=\sqrt{(\mathfrak{r}^{(\pm)}_{\bar{g}})^2+p_{\parallel}^2+m_{\psi}^2}\,, \\[2ex]
\mathfrak{r}^{(\pm)}_{\bar{g}}&\equiv p_{\bot}\pm \bar{g}|p_{\parallel}|\,.
\end{align}
\end{subequations}
Despite the isotropy in the plane perpendicular to the preferred direction, their form differs from the shape of the dispersion relations
for the anisotropic sectors of $c$, $d$, $e$, and $f$ stated in Eqs.~(\ref{eq:dispersion-relation-anisotropic-c}), (\ref{eq:dispersion-relations-anisotropic-d}),
(\ref{eq:dispersion-relations-anisotropic-e}), and (\ref{eq:dispersion-relations-anisotropic-f}). The perpendicular and parallel momentum
components cannot be separated from each other, but there is a product of both in the dispersion relation. The reason is
that $g^{(4)\mu\nu\varrho}$ is a third-rank tensor that is constructed from two preferred spacetime directions, in contrast to the vectors and
tensors studied so far.
Therefore, according to the
argument at the beginning of the current section, the decay rate is expected to depend on the perpendicular component $q_{\bot}$ of the
initial fermion momentum, as well. However, the calculation for a nonvanishing $q_{\bot}$ turns out to be impractical, which is why we
restrict it to the case of $q_{\bot}=0$, nevertheless. Therefore, in contrast to the $c$, $d$, $e$, and $f$ coefficients, the decay rate
for $q_{\bot}=0$ can only be considered as a special case, i.e., when the initial fermion propagates along the preferred direction. However,
deeper insights from studying the general case are not expected.

\begin{figure}[t!]
\centering
\includegraphics[scale=0.5]{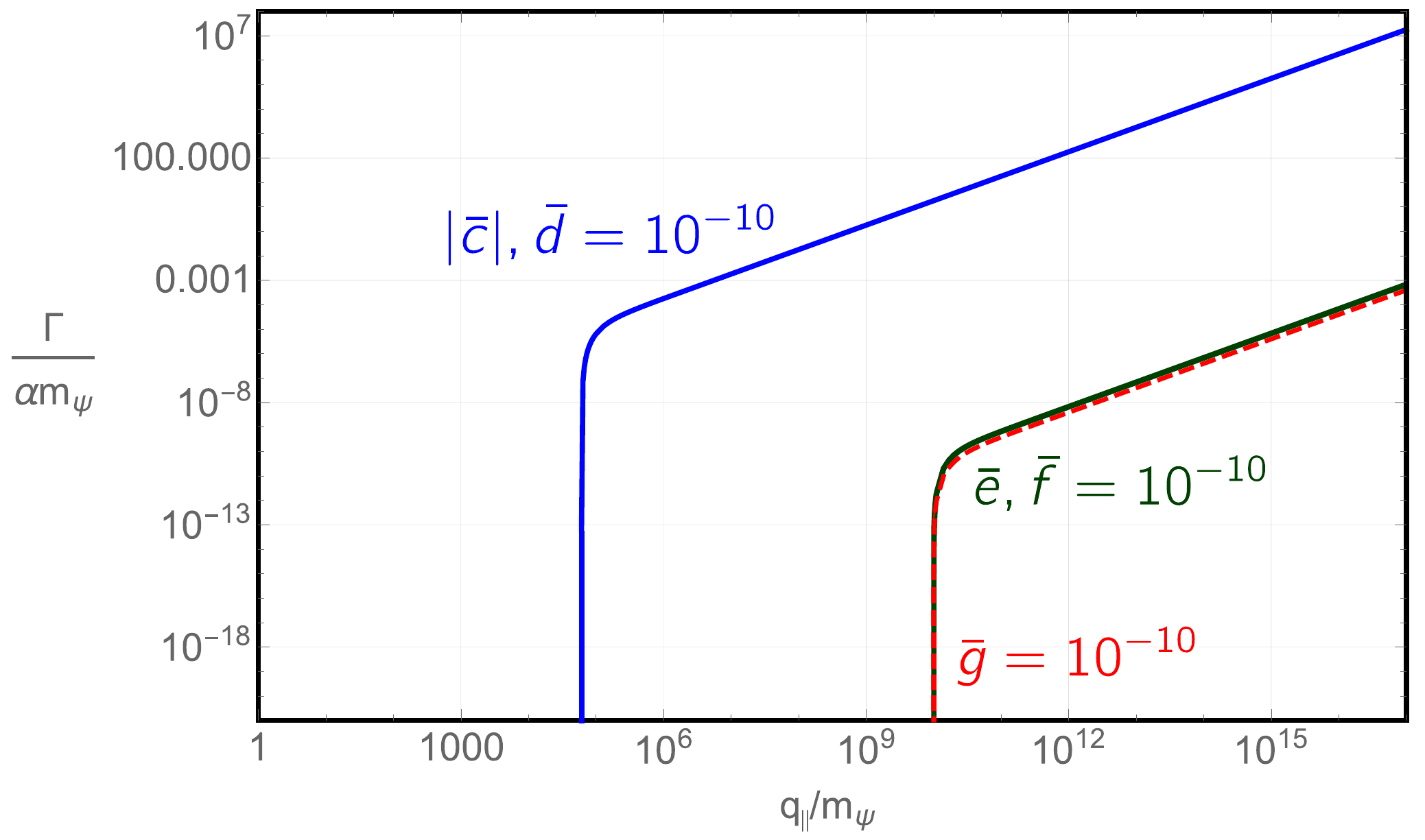}
\caption{Double-logarithmic plot of the decay rate $\Gamma/(\alpha m_{\psi})$ of vacuum Cherenkov radiation for the coefficients
$\bar{c}$, $\bar{d}$ (blue, plain), $\bar{e}$, $\bar{f}$ (green, plain), and $\bar{g}$ (red, dashed) as functions
of the incoming particle momentum $q_{\parallel}/m_{\psi}$. The Lorentz-violating coefficients are chosen equally as
$|\bar{c}|=\bar{d}=\bar{e}=\bar{f}=\bar{g}=10^{-10}$.}
\label{fig:decay-rates-anisotropic}
\end{figure}

We use the restriction $\bar{g}>0$ and consider the fermion mode described by $E^{(+)}_{\bar{g}}$.
As there are no additional time derivatives in the Lagrange density, the spinor matrix $u\bar{u}$ is obtained directly from Eq.~(4.11a) of \cite{Reis:2016hzu}:
\begin{subequations}
\begin{align}
\Lambda_{\bar{g}}^{(+)}(\mathbf{p})&=\xi_{\bar{g}}^{\mu}\gamma_{\mu}+\Xi_{\bar{g}}\mathds{1}_4+\zeta_{\bar{g}}^{\mu}\gamma^5\gamma_{\mu}+\psi_{\bar{g}}^{\mu\nu}\sigma_{\mu\nu}\,, \displaybreak[0]\\[2ex]
\xi_{\bar{g}}^{\mu}&=\frac{1}{2}\begin{pmatrix}
E^{(+)} \\
\mathbf{p} \\
\end{pmatrix}^{\mu}+\frac{\bar{g}|p_{\parallel}|}{2p_{\bot}}\begin{pmatrix}
0 \\
p_1 \\
p_2 \\
0 \\
\end{pmatrix}^{\mu}\,, \displaybreak[0]\\[2ex]
\Xi_{\bar{g}}&=\frac{m_{\psi}}{2}\,, \displaybreak[0]\\[2ex]
\zeta_{\bar{g}}^{\mu}&=\frac{m_{\psi}\mathrm{sgn}(p_{\parallel})}{2p_{\bot}}\begin{pmatrix}
0 \\
-p_2 \\
p_1 \\
0 \\
\end{pmatrix}^{\mu}\,, \displaybreak[0]\\[2ex]
\psi_{\bar{g}}^{\mu\nu}&=\frac{\mathrm{sgn}(p_{\parallel})}{4p_{\bot}}\begin{pmatrix}
0 & p_1p_{\parallel} & p_2p_{\parallel} & -p_{\bot}\mathfrak{r}_{\bar{g}}^{(+)} \\
-p_1p_{\parallel} & 0 & 0 & -p_1E^{(+)} \\
-p_2p_{\parallel} & 0 & 0 & -p_2E^{(+)} \\
p_{\bot}\mathfrak{r}_{\bar{g}}^{(+)} & p_1E^{(+)} & p_2E^{(+)} & 0 \\
\end{pmatrix}^{\mu\nu}\,.
\end{align}
\end{subequations}
Since the dispersion relation has a different form in comparison to all of the anisotropic cases previously studied, the perpendicular
momentum component, which is determined by energy-momentum conservation, cannot be elegantly expressed via the K\"{a}ll\'{e}n
function. Instead, we obtain a more complicated result that can be best written in terms of three functions as follows:
\begin{subequations}
\begin{align}
k_{\bot,0}&=\frac{1}{2g(q)}\left\{\bar{g}|q-k_{\parallel}|h(q)-\sqrt{f(q)\left[4k_{\parallel}^2g(q)+h^2(q)\right]}\right\}\,, \displaybreak[0]\\[2ex]
f(q)&\equiv (1+\bar{g}^2)q^2+m_{\psi}^2\,, \displaybreak[0]\\[2ex]
g(q)&\equiv \bar{g}^2(q-k_{\parallel})^2-f(q)\,, \displaybreak[0]\\[2ex]
h(q)&\equiv k_{\parallel}^2+f(q)-f(q-k_{\parallel})\,.
\end{align}
\end{subequations}
The condition $k_{\bot,0}\geq 0$ restricts the parallel component, with the maximum momentum being simple enough
to be expressed analytically:
\begin{equation}
k_{\parallel}\in [0,k_{\parallel,\mathrm{max}}]\,,\quad k_{\parallel,\mathrm{max}}=\frac{2}{\bar{g}^2}\left[(1+\bar{g}^2)q-\sqrt{f(q)}\right]\,.
\end{equation}
Finally, the condition for a nonvanishing accessible region in phase space, $k_{\parallel,\mathrm{max}}=0$, delivers the threshold momentum:
\begin{equation}
\label{eq:threshold-anisotropic-g}
q^{\mathrm{th}}_{\bar{g}}=\frac{m_{\psi}}{\bar{g}}+\dots\,.
\end{equation}
Compare the latter to \eqref{eq:threshold-isotropic-g}. Via the derivative of the energy balance equation,
\begin{equation}
\frac{\partial \Delta E_{\bar{g}}^{(+)}}{\partial k_{\bot}}=-\left[\frac{k_{\bot}}{|\mathbf{k}|}+\frac{k_{\bot}+\bar{g}|q-k_{\parallel}|}{E_{\bar{g}}^{(+)}(\mathbf{q}-\mathbf{k})}\right]\,,
\end{equation}
the phase space factor is stated as follows:
\begin{align}
\Pi_{\bar{g}}(k)&=\frac{k_{\bot}}{|\mathbf{k}|E_{\bar{g}}^{(+)}(\mathbf{q}-\mathbf{k})}\left|\frac{\partial \Delta E_{\bar{g}}^{(+)}}{\partial k_{\bot}}\right|^{-1}\Bigg|_{k_{\bot}=k_{\bot,0}} \notag \\
&=\frac{1}{E_{\bar{g}}^{(+)}(\mathbf{q})+\bar{g}\sqrt{1+k_{\parallel}^2/k_{\bot}^2}|q-k_{\parallel}|}\Bigg|_{k_{\bot}=k_{\bot,0}}\,.
\end{align}
The procedure of computing the decay rate numerically differs from that carried out for the coefficients $\bar{d}$, $\bar{c}$, $\bar{e}$, and $\bar{f}$
as the matrix element squared depends on the azimuthal angle $\varphi$. Therefore, $\varphi$ has to be integrated over numerically, as well.
The numerical result for the decay rate can be found in \figref{fig:decay-rates-anisotropic}. Note the similarities to the decay rates for
$\bar{e}$ and $\bar{f}$. However, it must
be kept in mind that this result is only a special case for $q_{\bot}=0$. As indicated within the current section, due to the particular form
of the dispersion relation for $\bar{g}$, a dependence on the perpendicular momentum component of the incoming fermion is expected. Furthermore,
an analog result is obtained by studying $\bar{g}<0$ in conjunction with the mode described by $E^{(-)}_{\bar{g}}$.

\subsection{Comparison of results and discussion}

The decay rates for the anisotropic sectors of the $c$, $d$, $e$, $f$, and $g$ coefficients are plotted in \figref{fig:decay-rates-anisotropic} as
functions of the parallel momentum component $q_{\parallel}$ of the incoming fermion. The behaviors of the curves are very similar
when they are compared to the isotropic results presented in \figref{fig:decay-rates-isotropic}. The linear behavior for momenta
$q_{\parallel}\gg m_{\psi}$ and the quick decrease near the threshold are clearly visible. The only caveat concerns the anisotropic
$g$ coefficients where the presented result presumably is just a special case when the initial fermion propagates along the preferred
direction. Due to the form of the dispersion relation, the decay rate is expected to depend on the perpendicular momentum component
$q_{\bot}$ additionally.

\section{Helicity processes}
\setcounter{equation}{0}
\label{sec:helicity-processes}

For the spin-nondegenerate operators there is a peculiarity that arises from the existence of two distinct dispersion relations for particles. In a
vacuum Cherenkov process, the spin projection of the incoming fermion can switch, in principle. As long as the fermion dispersion
relation does not depend on the spin projection, such spin-flip processes may deliver a partial contribution to a nonzero decay rate where the contribution
cannot be isolated from the spin-conserving processes. However, this changes when there are two different fermion dispersion laws, as now the fermion can
jump from one branch of the dispersion law to the other when its spin projection changes. Hence, the energy balance equation crucially differs from that of a
spin-conserving decay, which forces us to consider such particular decays separately. Spin-flip processes, which are sometimes called helicity processes,
can be important for the spin-nondegenerate operators and we will analyze their impact for the $b$ and $H$ coefficients, in particular. In this context, fermions
in a spin-up state will be denoted as $\oplus$ and fermions in a spin-down state as $\ominus$.

\subsection{Isotropic $\boldsymbol{b}$ coefficients}

\begin{figure}
\centering
\subfloat[]{\label{fig:helicity-decay-b-isotropic}\includegraphics[scale=0.45]{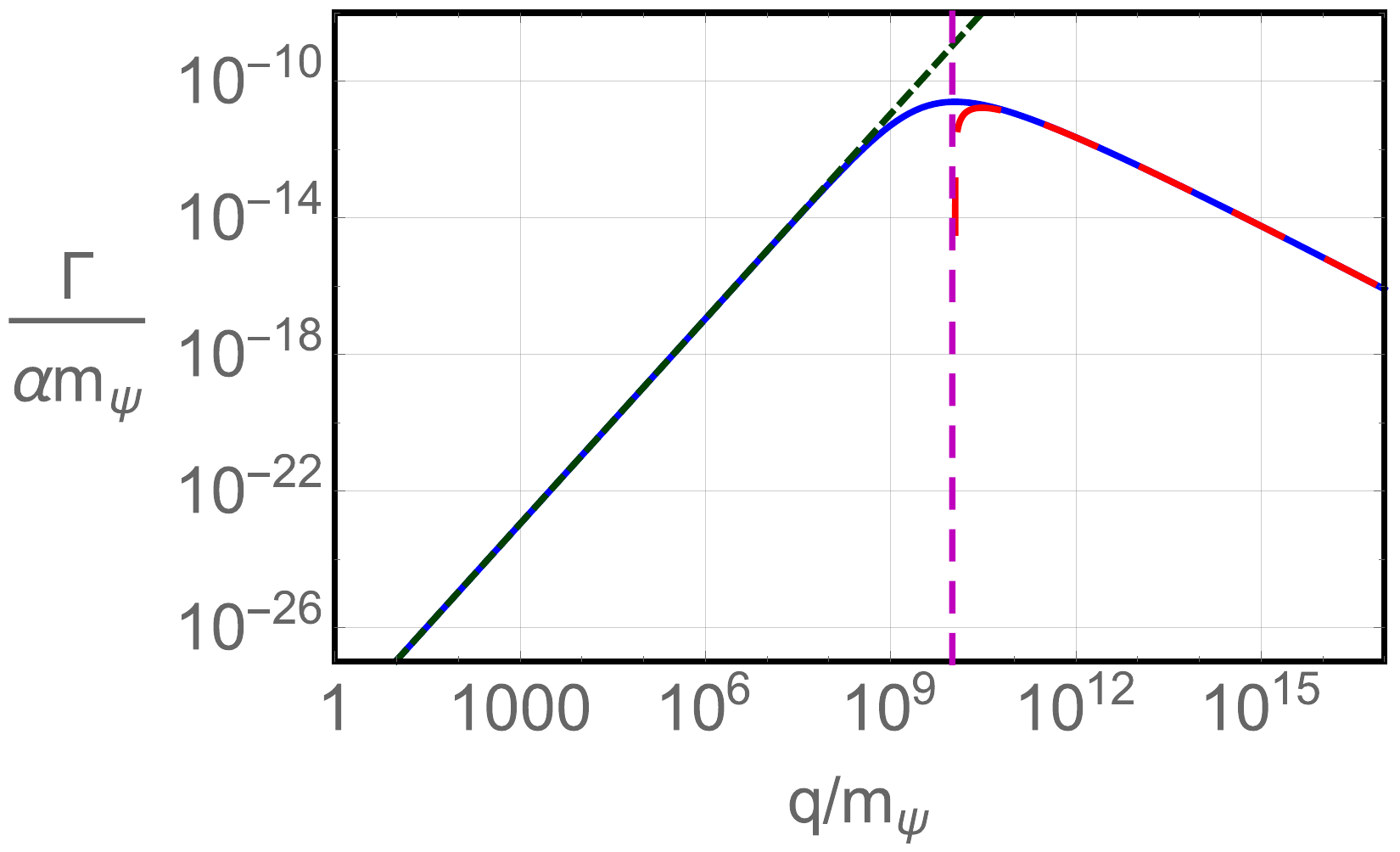}}\qquad
\subfloat[]{\label{fig:helicity-decay-b-anisotropic}\includegraphics[scale=0.45]{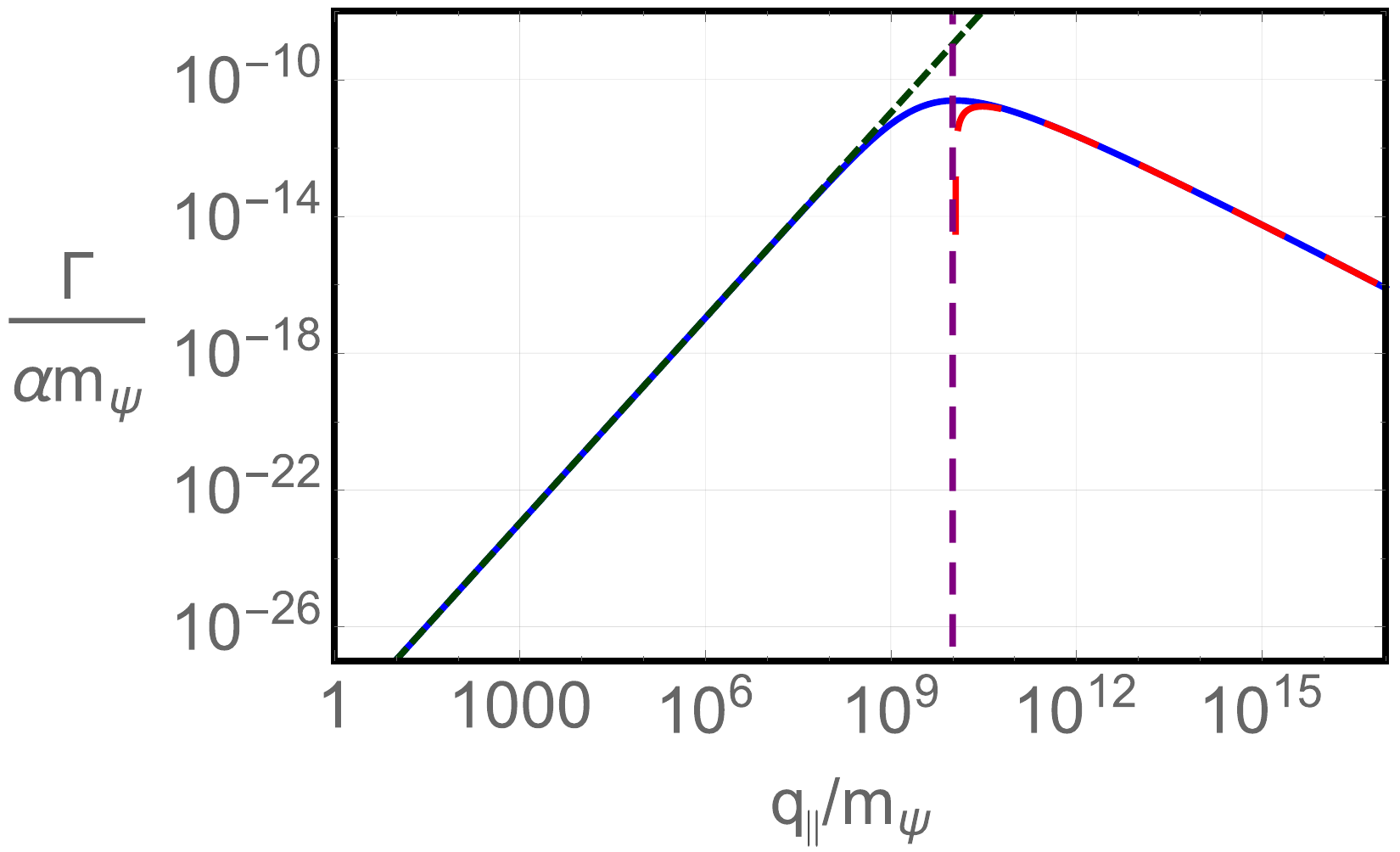}}
\caption{Decay rates of the helicity decay $\oplus\rightarrow \upgamma+\ominus$ for $\ring{b}/m_{\psi}=10^{-10}$ \protect\subref{fig:helicity-decay-b-isotropic}
and for $\bar{b}/m_{\psi}=10^{-10}$ \protect\subref{fig:helicity-decay-b-anisotropic}. The position of the maximum and the asymptotic behaviors on the left-hand
and the right-hand side of the maximum are indicated in both plots. With the naked eye, differences in the two plots cannot be spotted.}
\label{fig:helicity-decays-b-coefficients}
\end{figure}

In \secref{sec:b-coefficients-isotropic} it was shown that vacuum Cherenkov radiation is not possible for the isotropic $b$ coefficients --- as long as the spin projection
of the incoming fermion remains unchanged. However, as described at the beginning, spin-flip processes must be taken into account, as well. Let us first of
all consider the process $\oplus\rightarrow \upgamma+\ominus$. The kinematics differs crucially from the kinematics studied so far. The final polar angle is
given by
\begin{equation}
\cos\vartheta_0=\frac{1}{qk}\left\{kE^{(+)}_{\ring{b}}(\mathbf{q})-\ring{b}\left[q+\ring{b}+\sqrt{\left(k-E^{(+)}_{\ring{b}}(\mathbf{q})\,\right)^2-m_{\psi}^2}\,\right]\right\}\,,
\end{equation}
which allows for restricting the magnitude of the photon momentum to the interval $[k_{\mathrm{min}},k_{\mathrm{max}}]$ where
\begin{equation}
\label{eq:spin-flip-b-coefficients-photon-momentum}
k_{\mathrm{min},\mathrm{max}}=\frac{2q\ring{b}}{4q\ring{b}+m_{\psi}^2}\left[\mp(q-\ring{b})+E^{(+)}_{\ring{b}}(\mathbf{q})\,\right]\,.
\end{equation}
Several remarks are in order. First, the minimal photon momentum is larger than zero in contrast to all of the processes studied without spin flips. Second,
the expression in square brackets in \eqref{eq:spin-flip-b-coefficients-photon-momentum} reveals that both the minimum and the maximum photon momentum
are positive independently of the incoming fermion momentum $q$. Therefore, no restriction is put on the latter, which is why there is no threshold. For the
isotropic $b$ coefficients, a fermion can radiate photons when changing its spin projection, no matter how small its momentum is. Last but not least, the phase
space factor can be cast into a simple form:
\begin{equation}
\Pi(k)=\frac{1}{q}\frac{1}{1-\ring{b}/|\mathbf{q}-\mathbf{k}|}\bigg|_{\vartheta=\vartheta_0}\,.
\end{equation}
It is convenient to perform the phase space integration numerically due to the complicated form of the matrix element squared. A graph of the numerical
result can be found in \figref{fig:helicity-decay-b-isotropic} for a wide range of incoming fermion momenta. This figure allows for making additional
interesting observations. First, as was already pointed out above, the process does not have a threshold, which is why in a double-logarithmic plot, the
decay rate does not have a vertical asymptote. Instead, the decay rate has a maximum, which is a characteristic not observed for spin-conserving processes. An
approximation for the corresponding momentum was found from the plot to be $q_{\mathrm{max}}\approx m_{\psi}^2/\ring{b}$. Second, the decay rates have
asymptotic behaviors on the left- and right-hand sides of the maximum that we denote as $\Gamma_{<,>}$. In a double-logarithmic plot, the decay rate
$\Gamma_<$ has an oblique asymptote that corresponds to a polynomial behavior. Its form is found analytically:
\begin{equation}
\label{eq:decay-rate-isotropic-b-left}
\Gamma_{<}\sim\frac{32\alpha}{3}\left(\frac{\ring{b}}{m_{\psi}}\right)^3\frac{q^2}{m_{\psi}}\,.
\end{equation}
In contrast to the asymptotic behavior just encountered, the decay rate on the right-hand side of the maximum does not behave polynomially,
but it involves a logarithmic dependence:
\begin{equation}
\label{eq:decay-rate-isotropic-b-right}
\Gamma_{>}\sim\frac{\alpha}{2}\left[\ln\left(\frac{4q\ring{b}}{m_{\psi}^2}\right)-\frac{3}{2}\right]\frac{m_{\psi}^2}{q}\,.
\end{equation}
Third, from the asymptotic behaviors and the absolute numbers in the plot we see that the decay rate is highly suppressed --- either by the smallness of
the Lorentz-violating coefficient to the third power or by the ratio $m_{\psi}/q$ that is much smaller than one for energies much larger than the fermion mass.
Finally, asymptotic behaviors for the radiated-energy rates are found to be
\begin{equation}
\label{eq:radiated-energy-rates-isotropic-b}
\left(\frac{\mathrm{d}W}{\mathrm{d}t}\right)_<\sim 32\alpha\left(\frac{q\ring{b}}{m_{\psi}^2}\right)^4m_{\psi}^2\,,\quad \left(\frac{\mathrm{d}W}{\mathrm{d}t}\right)_>\sim \frac{\alpha}{2}\left[\ln\left(\frac{4q\ring{b}}{m_{\psi}^2}\right)-\frac{11}{6}\right]m_{\psi}^2\,.
\end{equation}
These two regimes are also separated by $q_{\mathrm{max}}$ to a good approximation.

Considering the opposite process $\ominus\rightarrow \upgamma+\oplus$,
we find quickly that the energy balance equation does not provide an accessible phase space region. Recalling that the spin-conserving process is forbidden,
this behavior is not totally surprising, as only one particular of the two possible spin-flip processes is expected to improve the energy balance. The
process $\ominus\rightarrow\upgamma+\oplus$ is forbidden for all of the sectors investigated subsequently. The situation changes when switching the sign
of the controlling coefficient, though.

\subsection{Anisotropic $\boldsymbol{b}$ coefficients}

The calculation for an anisotropic case of minimal $b$ coefficients with the preferred direction pointing along the third axis of the coordinate
system has many similarities with that carried out previously. We consider the process $\oplus\rightarrow \upgamma+\ominus$. The
perpendicular momentum component of the photon in the final state can again be expressed in terms of the K\"{a}ll\'{e}n function:
\begin{subequations}
\begin{align}
k_{\bot,0}&=\frac{\sqrt{\Omega(f^{(+)}(q),k_{\parallel}^2,f^{(-)}(q-k_{\parallel}))}}{2E^{(+)}(q)}\,, \displaybreak[0]\\[2ex]
f^{(\pm)}(p_3)&=\left(\sqrt{p_3^2+m_{\psi}^2}\pm \bar{b}\right)^2\,.
\end{align}
\end{subequations}
From the condition that $k_{\bot,0}=0$, we obtain $k_{\parallel}\in [k_{\parallel,\mathrm{min}},k_{\parallel,\mathrm{max}}]$ where both limits are complicated
functions of the incoming fermion momentum, the fermion mass, and the controlling coefficient under consideration. Since these functions do not contribute to
a better understanding, they will be skipped. Due to $k_{\parallel,\mathrm{min}}\sim -\bar{b}$, a photon with a small momentum can, in principle, be emitted in
a direction opposite to the direction of the initial fermion. The phase space factor is given by the inverse of the dispersion relation $E^{(+)}$, as usual for
anisotropic cases characterized by the generic dispersion relation of \eqref{eq:generic-dispersion-relation-anisotropic}. The phase space integration
is carried out numerically leading to a decay rate that shares many characteristics with the decay rate for the isotropic $b$ coefficient,
cf.~\figref{fig:helicity-decay-b-anisotropic}. There is a maximum at $q_{\mathrm{max}}\approx m_{\psi}^2/\bar{b}$. The asymptotic behavior of the decay rate
on its left-hand side is
\begin{equation}
\Gamma_<\sim\frac{32\alpha}{3}\left(\frac{\bar{b}}{m_{\psi}}\right)^3\frac{q^2}{m_{\psi}}\,,
\end{equation}
whereas the asymptotic behavior on the right-hand side again involves a logarithm:
\begin{equation}
\Gamma_>\sim\frac{\alpha}{2}\left[\ln\left(\frac{4q\bar{b}}{m_{\psi}^2}\right)-\frac{3}{2}\right]\frac{m_{\psi}^2}{q}\,.
\end{equation}
Note the similarities to the previous Eqs.~(\ref{eq:decay-rate-isotropic-b-left}), (\ref{eq:decay-rate-isotropic-b-right}) for the isotropic framework. The
opposite process $\ominus\rightarrow\upgamma+\oplus$ can again be shown to be energetically forbidden. The corresponding radiated-energy rates read
\begin{equation}
\left(\frac{\mathrm{d}W}{\mathrm{d}t}\right)_<\sim 32\alpha\left(\frac{q\bar{b}}{m_{\psi}^2}\right)^4m_{\psi}^2\,,\quad \left(\frac{\mathrm{d}W}{\mathrm{d}t}\right)_>\sim \frac{\alpha}{2}\left[\ln\left(\frac{4q\bar{b}}{m_{\psi}^2}\right)-\frac{11}{6}\right]m_{\psi}^2\,,
\end{equation}
and they are completely analogous to \eqref{eq:radiated-energy-rates-isotropic-b}.

\subsection{$\boldsymbol{H}$ coefficients}

The $H$ coefficients do not have an isotropic part. We will consider the representative example of a nonvanishing coefficient $H^{(3)03}\equiv \bar{H}$
where $H^{(3)\mu\nu}=\bar{H}(\ring{\lambda}^{\mu}\bar{\lambda}^{\nu}-\bar{\lambda}^{\mu}\ring{\lambda}^{\nu})$. The kinematics
behaves differently in comparison to the kinematics of the anisotropic $b$ and $d$ coefficients, which is inferred from the dispersion relations:
\begin{subequations}
\begin{align}
E^{(\pm)}(\mathbf{p})&=\sqrt{p_{\parallel}^2+(p_{\bot}\pm \bar{H})^2+m_{\psi}^2}\,, \\[2ex]
p_{\parallel}&\equiv \mathbf{p}\cdot\bar{\boldsymbol{\lambda}}\,,\quad p_{\bot}\equiv\sqrt{\mathbf{p}^2-p_{\parallel}^2}\,.
\end{align}
\end{subequations}
Note that in the dispersion laws given above, it is the perpendicular momentum component that is connected to the controlling coefficient, whereas the parallel
momentum component can be separated from the rest. In comparison to the dispersion relations of the anisotropic $b$, $d$ coefficients, the roles of
$p_{\bot}$ and $p_{\parallel}$ have interchanged, i.e, the decay rate is not expected to depend on the parallel momentum component, but on the
perpendicular one only. Therefore, the process is kinematically possible for the initial fermion traveling perpendicularly to $\bar{\boldsymbol{\lambda}}$
where the observer frame is chosen such that it travels along the $x$ axis specifically, i.e., $\mathbf{q}=(q,0,0)$. With the final photon reasonably parameterized according to
\eqref{eq:momentum-polarization-anisotropic}, the energy balance equation for the process $\oplus\rightarrow\upgamma+\ominus$ can be evaluated neatly.
It is now the parallel momentum component that is expressed via the K\"{a}ll\'{e}n function:
\begin{subequations}
\label{eq:parallel-momentum-component-h3-helicity-decay}
\begin{align}
k_{\parallel,0}^{(\pm)}&=\pm\frac{\sqrt{\Omega(f^{(+)}(q,0),k_{\bot}^2,f^{(-)}(q-k_{\bot}\cos\varphi,-k_{\bot}\sin\varphi))}}{2E^{(+)}(q)}\,, \\[2ex]
f^{(\pm)}(p_1,p_2)&=\left(\sqrt{p_1^2+p_2^2}\pm \bar{H}\right)^2+m_{\psi}^2\,.
\end{align}
\end{subequations}
\begin{figure}
\centering
\subfloat[]{\label{fig:helicity-decay-H}\includegraphics[scale=0.45]{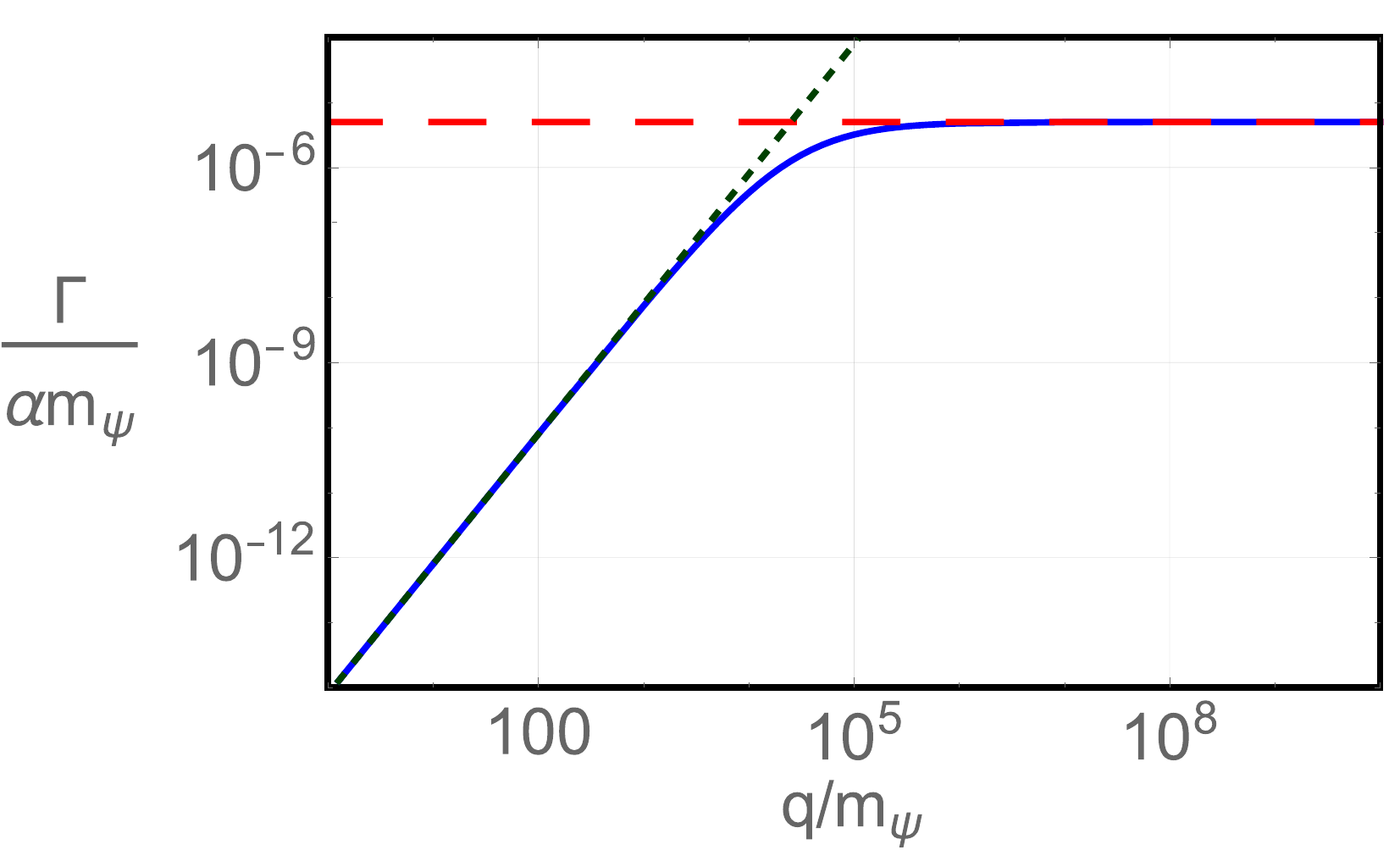}}\qquad
\subfloat[]{\label{fig:helicity-decay-d-isotropic}\includegraphics[scale=0.45]{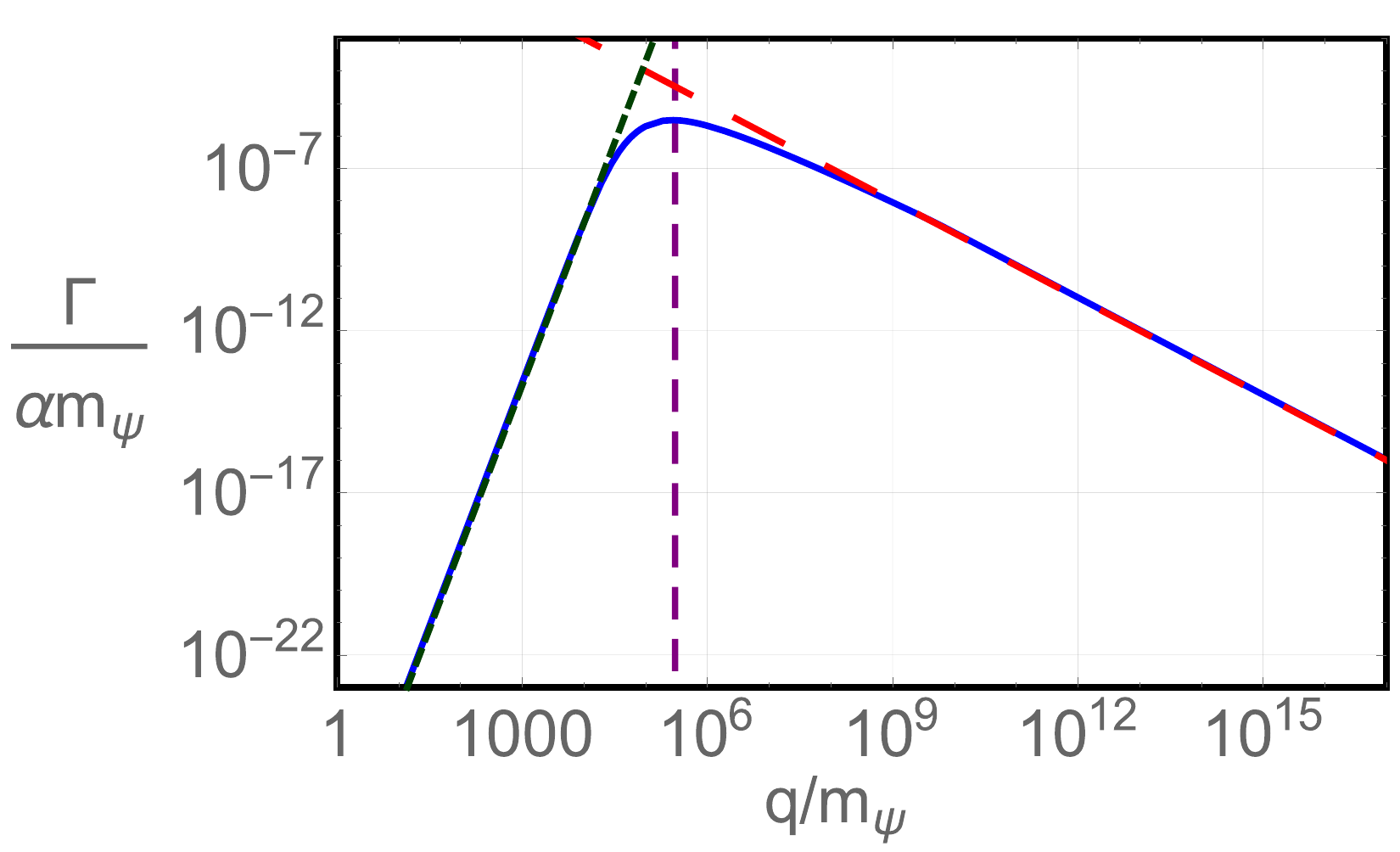}}
\caption{Decay rates of the process $\oplus\rightarrow \upgamma+\ominus$ for $\bar{H}/m_{\psi}=10^{-5}$ \protect\subref{fig:helicity-decay-H}
and for $\ring{d}=10^{-10}$ \protect\subref{fig:helicity-decay-d-isotropic}. The horizontal asymptote in \protect\subref{fig:helicity-decay-H}
is a particular property of this case.}
\label{fig:helicity-decay-H-and-d}
\end{figure}
Note that $k_{\parallel}\in\mathbb{R}$, which is why both the positive and the negative solution have to be taken into account.
From the limit $k_{\parallel,0}^{(\pm)}=0$ we obtain $k_{\bot}\in [0,k_{\bot,\mathrm{max}}]$. Again, the upper limit on $k_{\bot}$ is a complicated function that
will not be stated, because nothing is to be gained from it. The phase space factor is given by $\Pi(k)=k_{\bot}/(|k_{\parallel,0}^{(\pm)}|E^{(+)}(\mathbf{q}))$.
The latter differs from the phase space factor of most of the other anisotropic cases under consideration, as the roles of $k_{\bot}$, $k_{\parallel}$
have interchanged. Since the phase space factor depends on the magnitude of $k_{\parallel}$ only and due to $|\mathcal{M}|^2(-k_{\parallel})=|\mathcal{M}|^2(k_{\parallel})$
for this case, the negative solution $k_{\parallel,0}^{(-)}$ delivers the same contribution to the decay rate or radiated-energy rate as the positive one.
Therefore, the calculation can be restricted to $k_{\parallel,0}^{(+)}$, whereby the final result must simply be doubled.

Another crucial difference from the kinematics of the $b$ and $d$ coefficients is that there is no azimuthal symmetry, i.e., $k_{\parallel,0}^{(+)}$,
$k_{\bot,\mathrm{max}}$, and the matrix element squared depend on $\varphi\in [0,2\pi]$ where the interval is not restricted by the kinematics.
Therefore, the numerical integration is two-dimensional and has to be carried out over $\varphi\in [0,2\pi]$ and
$k_{\bot}\in [0,k_{\bot,\mathrm{max}}(\varphi)]$. Numerical instabilities were encountered for $\bar{H}/m_{\psi}\ll 1$ making the corresponding
result of the decay rate unreliable. The problem is that $k_{\bot,\mathrm{max}}$ strongly depends on $\varphi$. Large contributions lie in the direct
vicinities of $\varphi\in \{0,2\pi\}$ and the function drops extremely quickly outside of the neighborhoods of these two values such that
$k_{\bot,\mathrm{max}}(\pi)\sim\bar{H}$. This behavior makes the integration over $\varphi$ cumbersome. Hence, the \textit{Mathematica} routine is only used for integrating over
$k_{\bot}$ and the integration over $\varphi$ is carried out manually by applying Simpson's rule over subsequent intervals. This procedure has proven to be more stable for a larger
value of the controlling coefficient $\bar{H}/m_{\psi}=10^{-5}$ and the corresponding graph is presented in \figref{fig:helicity-decay-H}. Starting with low
momenta, the decay rate is heavily
suppressed by the smallness of the controlling coefficient. Besides, there is one oblique and one (presumably) horizontal asymptote in the
double-logarithmic plot, which is why the curve does not have a maximum. Due to the complexities of the computation,
it is only possible to determine approximate expressions for the asymptotes from the numerical data points:
\begin{subequations}
\begin{align}
\Gamma_<&\approx 8\alpha\left(\frac{\bar{H}}{m_{\psi}}\right)^3\frac{q^2}{m_{\psi}}\,,\quad \Gamma_>\approx\frac{\alpha}{2}\bar{H}\,, \\[2ex] \left(\frac{\mathrm{d}W}{\mathrm{d}t}\right)_<&\approx 20\alpha \left(\frac{q\bar{H}}{m_{\psi}^2}\right)^4m_{\psi}^2\,,\quad \left(\frac{\mathrm{d}W}{\mathrm{d}t}\right)_>\approx\frac{\alpha}{3}q\bar{H}\,.
\end{align}
\end{subequations}
It seems that in contrast to all of the cases studied previously, both $\Gamma_<$ and $\Gamma_>$ have a polynomial behavior. The existence of a horizontal
asymptote makes this framework different from the other cases investigated.

\subsection{Isotropic $\boldsymbol{d}$ coefficients}

The decay rate of the helicity decay for the $d$ coefficients is also complicated to compute due to the involved structure of the modified dispersion
relations. Considering the decay $\oplus\rightarrow\upgamma+\ominus$, we can proceed formally just as before. The important point is that computer algebra
delivers two functions for the polar angle $\vartheta$ in the final state where only one of these satisfies the energy balance equation. Respecting $\vartheta\in [0,\pi]$
further restricts the magnitude of the photon momentum to an interval whose limits are complicated polynomials involving third roots. Therefore, here it is
more necessary than ever to perform the calculation numerically where the result for the decay rate can be found in \figref{fig:helicity-decay-d-isotropic}.
It is evident that the characteristics of the curve are similar to the properties of the decay rates for the $b$ coefficients. There is a maximum at
$q_{\mathrm{max}}\approx 3/\sqrt{\ring{d}}$ and the asymptotic behavior on the left-hand side seems to be polynomial, whereas the decay rate on the right-hand
side possibly has a polynomial dependence, as well. It is challenging to determine analytical functions describing these behaviors. However, the following
numerical approximations are directly obtained from the plot:
\begin{equation}
\Gamma_<\approx 20\alpha \left(\frac{q\ring{d}}{m_{\psi}}\right)^3\frac{q^2}{m_{\psi}}\,,\quad \Gamma_>\approx 10\alpha\frac{m_{\psi}^2}{q}\,.
\end{equation}
What is significant here is the steep increase of the decay rate for the region on the left-hand side of the maximum. However, note that the decay rate is then still suppressed
by the third power of the small dimensionless controlling coefficient. For the radiated-energy rates we find the approximations
\begin{equation}
\left(\frac{\mathrm{d}W}{\mathrm{d}t}\right)_<\approx 100\alpha \ring{d}^4\left(\frac{q}{m_{\psi}}\right)^8m_{\psi}^2\,,\quad \left(\frac{\mathrm{d}W}{\mathrm{d}t}\right)_>\approx 10.4\alpha m_{\psi}^2\,.
\end{equation}
On the left-hand side of the maximum, $\mathrm{d}W/\mathrm{d}t$ grows very steeply as a function of the momentum, as well, but it is suppressed by the fourth
power of the controlling coefficient.

\subsection{Isotropic $\boldsymbol{g}$ coefficients}

\begin{figure}
\centering
\includegraphics[scale=0.45]{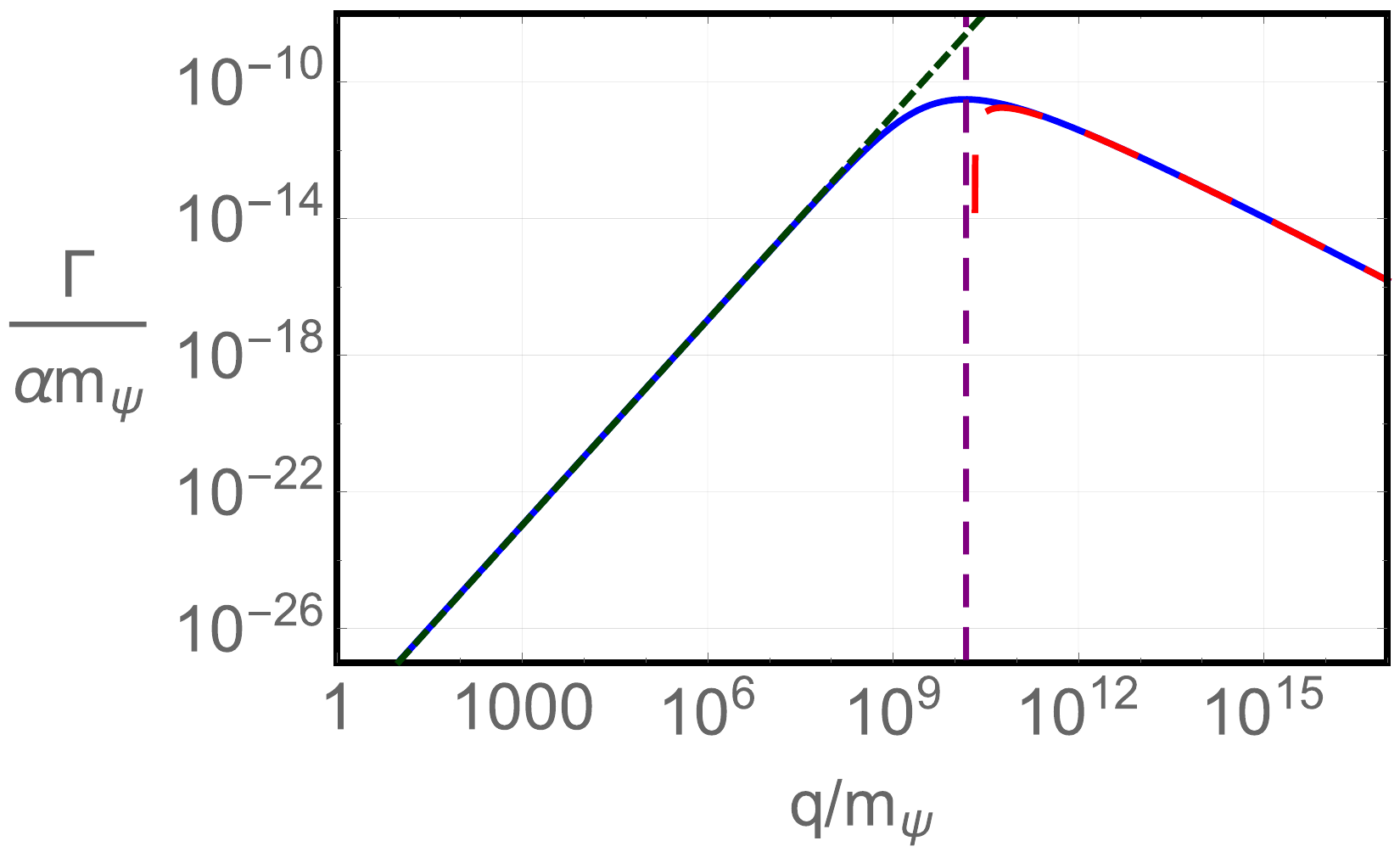}
\caption{Decay rate of the process $\oplus\rightarrow \upgamma+\ominus$ for $\ring{g}=10^{-10}$.}
\label{fig:helicity-decay-g-isotropic}
\end{figure}

The decay rate for the helicity decay $\oplus\rightarrow \upgamma+\ominus$ can be calculated in the same way as that for the $d$ coefficients.
From a technical viewpoint, the calculation is even simpler in comparison. The numerical result is shown in \figref{fig:helicity-decay-g-isotropic}. For
increasing momenta the rate rises as
\begin{equation}
\label{eq:decay-rate-isotropic-g-left}
\Gamma_<\sim \frac{32\alpha}{3}\ring{g}^3\frac{q^2}{m_{\psi}}\,,
\end{equation}
until it reaches the maximum at $q_{\mathrm{max}}\approx (3/2)m_{\psi}/\ring{g}$. Note that the increase is far from as steep as for the isotropic $d$ coefficients.
Here, the decay rate just grows quadratically but it is still suppressed by the third power of the controlling coefficient. After reaching the maximum, the function
decreases again. This decrease cannot be described by a polynomial, but it has a logarithmic behavior. Performing several approximations
of both the integrand and the integration limits, which were checked to be consistent with the numerical result, allows for obtaining the asymptotic decay rate for
very large momenta:
\begin{equation}
\Gamma_>\sim \alpha\left[\ln\left(\frac{q\ring{g}}{m_{\psi}}\right)-\frac{3}{4}\right]\frac{m_{\psi}^2}{q}\,.
\end{equation}
Last but not least, asymptotic behaviors for the radiated-energy rates read
\begin{equation}
\left(\frac{\mathrm{d}W}{\mathrm{d}t}\right)_<\sim 32\alpha\left(\frac{q\ring{g}}{m_{\psi}}\right)^4m_{\psi}^2\,,\quad\left(\frac{\mathrm{d}W}{\mathrm{d}t}\right)_>\sim\alpha\left[\ln\left(\frac{q\ring{g}}{m_{\psi}}\right)-\frac{11}{12}\right]m_{\psi}^2\,.
\end{equation}
Here we point out the great similarities to the results for the isotropic $b$ coefficient, cf.~Eqs.~(\ref{eq:decay-rate-isotropic-b-left}),
(\ref{eq:decay-rate-isotropic-b-right}), and (\ref{eq:radiated-energy-rates-isotropic-b}).

\subsection{Discussion}

A simplified treatment of helicity decays in the context of Lorentz-violating fermions can be found in \cite{Jacobson:2005bg}. In the latter paper,
a collection of Lorentz-violating dimension-5 operators is considered that were initially classified by Myers and Pospelov \cite{Myers:2003fd}. Based on a simplified theoretical analysis,
the decay rate for various unusual processes such as vacuum Cherenkov radiation is obtained for the asymptotic regimes. Although their theory differs from the minimal
frameworks considered here, some of their and our results do have certain characteristics in common. Their decay rate for vacuum Cherenkov radiation in the asymptotic
region of small momenta is given by Eq.~(C30). The common characteristic with Eqs.~(\ref{eq:decay-rate-isotropic-b-left}), (\ref{eq:decay-rate-isotropic-g-left}) is a
suppression by the third power of the controlling coefficient. The momentum dependence is, of course, different, as they consider a higher-dimensional operator including
additional momenta. The decay rate was also obtained in the region of asymptotically large momenta, cf. their Eq.~(C31). The latter shares some properties with
\eqref{eq:decay-rate-isotropic-b-right}, e.g., the decay rate is suppressed by $m_{\psi}^2/q$ but not by the controlling coefficient. The logarithmic dependence is
missing in their result, which is most likely related to the approximations that the authors of \cite{Jacobson:2005bg} use.

To understand the decay rates for processes with and without spin flip a bit better, we consider a linear decay with all particle momenta aligned. In the
process, a photon with a certain helicity is emitted. Without spin flip, the final-state electron will move along a direction opposite to that of the
initial-state electron. The reason for this behavior is that the final-state electron has to change helicity such that the overall helicity in the process is conserved.
Since the electron does not perform a spin flip by definition, its helicity can only change by reversing its momentum. For momenta directly above threshold, the kinematics
does not allow the electron momentum to reverse, which is why the decay rate is highly suppressed in this regime. For momenta much larger than the threshold, a high-energy
photon can be emitted such that the final-state electron moves in a direction opposite to the initial-state electron automatically changing helicity even without flipping
the electron spin. This behavior results in a strong increase of the decay rate. The situation is different for a helicity decay. As the spin of the final-state electron
now flips, its momentum direction does not have to be reversed, which means that the final-state electron can be emitted in forward direction. This behavior is nicely
seen within the anisotropic cases based on $\bar{b}$ and $\bar{H}$ where photons with momenta approximately proportional to the controlling coefficients can be emitted
in directions opposite to the direction of the incoming fermion. Hence, the appearance of such momentum configurations demonstrates that it is possible for the final-state
fermion to continue propagating in forward direction.

The decay rate for helicity decays was found to rise with increasing momentum of the initial
fermion, which can be explained from the growing accessible phase space volume. However, for momenta beyond a certain value, the decay rate decreases again. The initial
fermions are then almost chiral, which renders a helicity-flip more and more unlikely \cite{Jacobson:2005bg}. The situation is different for the coefficient $\bar{H}$ only, as a spin-up
fermion propagating along the $x$-axis is not in a helicity eigenstate. Hence, such a fermion must be in a superposition of states with positive and negative helicity
independently of the particle energy, which is why the probability for photon emission should not be suppressed. This property explains the horizontal asymptote of the decay rate for large momenta.

The authors of~\cite{Jacobson:2005bg} call the maximum of the decay rate, which occurs for helicity decays, an ``effective threshold.'' This notion is reasonable,
since the decay rate below the maximum is strongly suppressed by Lorentz violation. For the $d$ and $g$ coefficients, the approximate values of $q_{\mathrm{max}}$
found from the plots are related to the threshold momenta of the corresponding spin-conserving decays, e.g., $q_{\mathrm{max}}\sim m_{\psi}/\sqrt{\ring{d}}$ for the
isotropic $d$ coefficient. This fact explains the strong increase of the radiated-energy rate of the helicity decay for the isotropic $d$ coefficient, which is
proportional to the eighth power of the momentum. In contrast to that, the radiated-energy rates only rise with the fourth power of the momentum within the remaining sectors
studied. Note also that each asymptote $(\mathrm{d}W/\mathrm{d}t)_<$ is suppressed by the fourth power of the corresponding controlling coefficient, i.e., this property
is common to all of the coefficients studied. Last but not least, there is a direct correspondence between $\ring{g}$ and $\ring{b}/m_{\psi}$ via Eqs.~(27) of
\cite{Kostelecky:2013rta} that explains the very similar results for $\Gamma_<$ and $(\mathrm{d}W/\mathrm{d}t)_<$ for these coefficients. However, this correspondence
is valid at leading order in Lorentz violation only, i.e., deviations of the decay rates for larger momenta emerge when higher-order terms in Lorentz violation become
important.

\section{Constraints}
\setcounter{equation}{0}
\label{sec:constraints}

The most significant theoretical findings are presented in \tabref{tab:phenomenological-results} and they will be discussed
before proceeding to experimental constraints. Vacuum Cherenkov radiation conserving the spin direction of the fermion was found to be only
possible in frameworks with nonzero $c$, $d$, $e$, $f$, and $g$ coefficients, which, interestingly, are the dimensionless ones in the fermion
sector. For these processes, the threshold momentum is generically given by $q^{\mathrm{th}}=\rho m_{\psi}/X_{\subset}^{\sigma}$ at leading order
in the controlling coefficients $X_{\subset}\in X$ with $X$ of \eqref{eq:set-controlling-coefficients}. This expression is valid for both
isotropic and anisotropic frameworks. The parameters $\rho$ and $\sigma$ are dimensionless, i.e., $\rho$ is a global prefactor and $\sigma$ is
the power of the coefficient. The decay rate for momenta that are much larger than the particle mass can be written in the form
$\Gamma^{\infty}/\alpha=rX_{\subset}^sq$ for isotropic sectors and $\Gamma^{\infty}/\alpha=rX_{\subset}^sq_{\parallel}$ for anisotropic ones.
The radiation rate reads $(\mathrm{d}W/\mathrm{d}t)^{\infty}/\alpha=uX_{\subset}^vq^2$ and
$(\mathrm{d}W/\mathrm{d}t)^{\infty}/\alpha=uX_{\subset}^vq_{\parallel}^2$, respectively. Here $q$ is the momentum of the incoming fermion
and $q_{\parallel}$ is the component parallel to the preferred spacelike direction. The remaining dimensionless parameters $r$, $s$, $u$,
and $v$ describe the dynamical properties of the vacuum Cherenkov process for the different frameworks under consideration.

The results for the $c$ and $d$ coefficients are very similar, so are those for the $e$, $f$, and $g$ coefficients. For
the first two, the threshold momenta are inversely proportional to the square-root of the nonzero controlling coefficient, and the decay rates for
large momenta depend linearly on the controlling coefficient and the momentum itself. This behavior differs crucially from the outcomes
obtained for the $e$, $f$, and $g$ coefficients. For the latter, the threshold momenta are inversely proportional to the controlling coefficients
directly where the decay rates for large momenta are suppressed by the squares of the controlling coefficients. Note that the ratio between the
asymptotic decay rates for the $f$ and $c$ coefficients amounts to 3/8, as expected from the existence of the transformation found in \cite{Altschul:2006ts},
cf.~\secref{sec:f-coefficients-isotropic}. The characteristic dimensionless numbers presented for the isotropic and anisotropic frameworks
correspond to each other except for the decay rate for the $g$ coefficients. For the anisotropic coefficient $g^{(4)033}$, the matrix
element squared depends on the azimuthal angle $\varphi$, which has to be integrated over. This integration produces
a numerical constant for the decay rate that is approximately 1/5 and it differs from the 1/3 of the isotropic case.

\begin{table}
\begin{tabular}{c|ccccc|ccccc}
\toprule
 & $\ring{c}$ & $\ring{d}$ & $\ring{e}$ & $\ring{f}$ & $\ring{g}$ & $c^{(4)33}$ & $d^{(4)33}$ & $|\mathbf{e}^{(4)}|$ & $|\mathbf{f}^{(4)}|$ & $g^{(4)033}$ \\
\colrule
$\rho$ & $(1/2)\sqrt{3/2}$ & $(1/2)\sqrt{3/2}$ & 1 & 1 & 1 & $(1/2)\sqrt{3/2}$ & $(1/2)\sqrt{3/2}$ & 1 & 1 & 1 \\
$\sigma$ & 1/2 & 1/2 & 1 & 1 & 1 & 1/2 & 1/2 & 1 & 1 & 1 \\
$r$ & $16/9$ & $16/9$ & 2/3 & 2/3 & 2/3 & $16/9$ & $16/9$ & $2/3$ & $2/3$ & $\approx 0.40$ \\
$s$ & 1 & 1 & 2 & 2 & 2 & 1 & 1 & 2 & 2 & 2 \\
$u$ & $7/9$ & $7/9$ & $7/24$ & $7/24$ & $7/24$ & $7/9$ & $7/9$ & $7/24$ & $7/24$ & $\approx 0.20$ \\
$v$ & 1 & 1 & 2 & 2 & 2 & 1 & 1 & 2 & 2 & 2 \\
\botrule
\end{tabular}
\caption{Summary of the crucial results for vacuum Cherenkov radiation conserving the fermion spin. The definition of the
dimensionless parameters in the first column can be found at the beginning of \secref{sec:constraints}. The first two characterize the
threshold energy and the middle two describe the asymptotic decay rate for large momenta compared to the fermion mass. The final
two are contained in the asymptotic expressions for the radiated-energy rate. The first set of five columns states values for the
isotropic coefficients studied. The second set lists values for the corresponding anisotropic coefficients under consideration.}
\label{tab:phenomenological-results}
\end{table}
Constraints on the controlling coefficients can be obtained from data of UHECR. We assume the energy of
a cosmic ray detected on Earth to be smaller than the threshold energy of vacuum Cherenkov radiation. This will be justified below. So we
consider the condition $E_{\mathrm{prim}}\equiv E<q^{\mathrm{th}}$ with the primary energy $E_{\mathrm{prim}}$ and the generic expression
$q^{\mathrm{th}}$ for the threshold momentum introduced above. The latter equation is solved for the controlling coefficient leading to
$X_{\subset}=(\varrho m_{\psi}/E)^{1/\sigma}$. To obtain a constraint on the generic coefficient $X_{\subset}$ at $2\sigma$ level we add
twice the uncertainty that is linked to the experimental error $\Delta E$ of the energy value measured:
\begin{equation}
\label{eq:constraint-equation}
X_{\subset}>\left(\frac{\varrho m_{\psi}}{E}\right)^{1/\sigma}+2\Delta E\left|\frac{\partial}{\partial E}\left(\frac{\varrho m_{\psi}}{E}\right)^{1/\sigma}\right|\,,
\end{equation}
where $\varrho$, $\sigma$ are the dimensionless parameters in the threshold energy, which can be found in \tabref{tab:phenomenological-results}.
In principle, the constraints can be placed directly on the coefficients discussed in the paper, so far. Note that it is possible to generalize at least
some of the bounds obtained with \eqref{eq:constraint-equation}. For the isotropic coefficients, we use the notation that was introduced in Eq.~(95) of \cite{Kostelecky:2013rta}. This notation
is based on the effective coefficients of Eqs.~(27) in the latter reference. The effective $c$ coefficients additionally involve the $m$ coefficients.
However, $c^{(4)}_{\mathrm{eff}}$ only contains $c^{(4)}$, as $m^{(3)}$ does not exist. Furthermore, there is a coordinate transformation that transforms
between the $c$ coefficients in the fermion sector and the nonbirefringent {\em CPT}-even coefficients in the photon sector at first order in Lorentz
violation~\cite{Altschul:2006zz} (see \secref{sec:c-coefficients-isotropic}, as well). Therefore, any bound on the isotropic $c$ coefficient involves
the isotropic coefficient $\widetilde{\kappa}_{\mathrm{tr}}$ of the photon sector. According to Eqs.~(27) in \cite{Kostelecky:2013rta}, the $d^{(4)}$
coefficients are comprised by the effective coefficients $\widetilde{H}_{\mathrm{eff}}^{(5)}$ where those also contain the dimension-5 $H$ coefficients.
Since the latter have not been considered in the current article, we can constrain the isotropic $d$ coefficients only. The dimension-4 $e$ coefficients
are contained in $a^{(5)}_{\mathrm{eff}}$ together with the dimension-5 $a$ coefficients, which are not taken into account here. So the bound applies to
$e^{(4)}$ only. Last but not least, the coefficients $\widetilde{g}^{(4)}_{\mathrm{eff}}$ involve the dimension-4 $g$ coefficients and the dimension-3
$b$ coefficients. The $b^{(3)}$ do not contribute to the decay rate of the spin-conserving process, though, which is why $\ring{g}_1^{(4)}$ can be
bound directly. Note that in Eq.~(95) of \cite{Kostelecky:2013rta}, the latter is defined with an additional minus sign. There are no effective coefficients
involving $f^{(4)}$ since these contribute to observables at second order in Lorentz violation only. Following the observation of \cite{Altschul:2006ts},
we put the $f$ into the $c$ coefficients.

The isotropic constraints are computed by using the energy $E$ of a primary detected by the Pierre-Auger observatory. The composition of this primary
is unknown. A conservative choice is to take an iron nucleus with $N=56$ nucleons leading to a nucleon energy of $E/N$. The Cherenkov photon
is assumed to be emitted from one of the up or down-type quarks in the nucleons where we take the masses $m_{\mathrm{u}}\approx \unit[2.3\times 10^{-3}]{GeV}$
and $m_{\mathrm{d}}\approx \unit[4.8\times 10^{-3}]{GeV}$. Another conservative assumption is that each of these quarks carries a fraction
$r=0.1(\overset{\wedge}{=}10\%)$ of the nucleon energy (cf.~\cite{Moore:2001bv} and references therein). Hence, we use quark energies of
$E_{\mathrm{u}}=E_{\mathrm{d}}=rE/N$. The anisotropic constraints are based on a different event, but the procedure employed is the same.

A compilation of the constraints based on these values and \eqref{eq:constraint-equation} is presented in
\tabref{tab:constraints}. Several remarks are in order. First, a subset of the bounds is one-sided and the remaining ones are two-sided.
The vacuum Cherenkov process is rendered possible if either the velocity of the photon is decreased with respect to the maximum velocity
of the fermion or if the fermion velocity is increased with respect to the photon. Having a one-sided bound means that this condition is only
fulfilled for a particular sign of the controlling coefficients involved. Another reason for a one-sided bound is that the constraint may
apply to a combination of coefficients that has a fixed sign such as $|\mathbf{e}^{(4)}|$. Second, most of the one-sided bounds are lower ones,
as vacuum Cherenkov radiation only occurs when the corresponding controlling coefficients are negative.  An exception are again the bounds on
the nonnegative $|\mathbf{e}^{(4)}|$. However, considering only one of the three coefficients $e^{(4)}_i$ to be nonzero at a time, results in an
upper bound on $|e^{(4)}_i|$. In principle, the latter is equal to a two-sided constraint on the particular $e^{(4)}_i$.
\begin{table}[t]
\centering
\begin{tabular}{cccc}
\toprule
Sector & Lower constraint & Coefficients & Upper constraint \\
\colrule
up quark & $-3\times 10^{-23}<$ & $\ring{c}^{\mathrm{u}}-(3/4)\widetilde{\kappa}_{\mathrm{tr}}-(3/8)(\ring{f}^{\mathrm{u}})^2$ & \\
         & $-3\times 10^{-23}<$ & $\ring{d}^{\mathrm{u}}$ & $<3\times 10^{-23}$ \\
         & $-9\times 10^{-12}<$ & $\ring{e}^{\mathrm{u}}$ & $<9\times 10^{-12}$ \\
         & $-9\times 10^{-12}<$ & $\ring{g}^{\mathrm{u}}$ & $<9\times 10^{-12}$ \\
         & $-6\times 10^{-22}<$ & $c^{(4)\mathrm{u}}_{33}-(3/8)(f^{(4)\mathrm{u}}_3)^2$ & \\
         & $-6\times 10^{-22}<$ & $d^{(4)\mathrm{u}}_{33}$ & $<6\times 10^{-22}$ \\
         & & $\sqrt{\sum_i (e^{(4)\mathrm{u}}_i)^2}$ & $<4\times 10^{-11}$ \\
         & $-4\times 10^{-11}<$ & $g^{(4)\mathrm{u}}_{033}$ & $<4\times 10^{-11}$ \\
\hline
down quark & $-1\times 10^{-22}<$ & $\ring{c}^{\mathrm{d}}-(3/4)\widetilde{\kappa}_{\mathrm{tr}}-(3/8)(\ring{f}^{\mathrm{d}})^2$ & \\
           & $-1\times 10^{-22}<$ & $\ring{d}^{\mathrm{d}}$ & $<1\times 10^{-22}$ \\
           & $-2\times 10^{-11}<$ & $\ring{e}^{\mathrm{d}}$ & $<2\times 10^{-11}$ \\
           & $-2\times 10^{-11}<$ & $\ring{g}^{\mathrm{d}}$ & $<2\times 10^{-11}$ \\
           & $-3\times 10^{-21}<$ & $c^{(4)\mathrm{d}}_{33}-(3/8)(f^{(4)\mathrm{d}}_3)^2$ & \\
           & $-3\times 10^{-21}<$ & $d^{(4)\mathrm{d}}_{33}$ & $<3\times 10^{-21}$ \\
           & & $\sqrt{\sum_i (e^{(4)\mathrm{d}}_i)^2}$ & $<8\times 10^{-11}$ \\
           & $-8\times 10^{-11}<$ & $g^{(4)\mathrm{d}}_{033}$ & $<8\times 10^{-11}$ \\
\botrule
\end{tabular}
\caption{Constraints on controlling coefficients in the up and down-quark sector at $2\sigma$-level obtained with \eqref{eq:constraint-equation}.
The isotropic bounds are based on the event 737165 detected by the Pierre-Auger observatory and published in \cite{Abraham:2006ar}. In
\cite{Klinkhamer:2008ky} the event energy
was corrected by 5\%. This was done to take into account the missing energy of a hadronic primary, since the energy of \cite{Abraham:2006ar} was based
on the assumption of a photon primary. However, a photon primary can be ruled out to a significance of $3\sigma$ when the shower-maximum atmospheric
depth of the event is considered, cf.~\cite{Klinkhamer:2008ky}. Thus, the corrected event energy is $E=\unit[212\times 10^{18}]{eV}$ where the
experimental error $\Delta E$ constitutes 25\% of the cosmic-ray energy. To obtain the anisotropic constraints, we chose the event
81 of the SUGAR catalog on \cite{CR-Catalog:2017} with an energy of $E=\unit[38.9\times 10^{18}]{eV}$ and a declination of $-88^{\circ}$. The latter
value means that the primary was propagating approximately along the third axes of the coordinate system, i.e., along the preferred spacetime direction
$\bar{\lambda}^{\mu}$ of the anisotropic frameworks. It was hard to find an official statement on the energy uncertainty of SUGAR. Therefore, an estimate
of 10\% was obtained from the averaged differences between the underlying conversion models from the vertical muon number to the primary energy, cf.~Fig.~1
and the first two unnumbered equations in Sec.~2 of \cite{Winn:1986up}.}
\label{tab:constraints}
\end{table}

Third, the absolute numbers appearing in the constraints for the $c$, $d$ and the $e$, $g$ coefficients
are the same because, for large particle energies, the thresholds correspond to each other. The kinematics of the process plays a primary role in
obtaining the constraints where the decay rate itself is secondary. Fourth, the
bounds on the $e$ and $g$ coefficients are weak compared to the constraints for the $c$ and $d$ coefficients, as the thresholds depend on the
controlling coefficients in a different way. Last but not least, we deduce that the vacuum Cherenkov process is highly efficient for
the $c$ and $d$ coefficients. This conclusion follows from either the radiated-energy rate or the decay rate whose inverse corresponds to the lifetime of the
high-energetic particle. Hence, a cosmic ray whose energy lies above the threshold loses this energy excess very rapidly. UHECRs originate
definitely from outside of our solar system, probably even from outside of the Milky Way. So it is granted that a cosmic ray that is detected on
Earth has an energy below the threshold, which is why the assumption made at the very beginning is justified. The only caveat is that for the $e$
and $g$ coefficients, the process is less efficient as it is suppressed by the square of the Lorentz-violating coefficient. This behavior is mirrored
in the threshold momenta leading to much weaker bounds on those coefficients.

\subsection{Energy loss via helicity processes}

In \secref{sec:helicity-processes} we found that helicity processes are devoid of a threshold. Although the corresponding decay rates are
heavily suppressed by the smallness of the controlling coefficients, a charged particle can steadily radiate and lose energy when traveling from its source
to the Earth. Since the particle will radiate over distances that lie in the order of magnitude of many lightyears, these decays can nevertheless play a crucial
role as long as the corresponding controlling coefficients are large enough. This argument could allow us to place additional constraints on the coefficients involved.
Since possible constraints cannot follow from a threshold, we adopt the approach used in \cite{Kostelecky:2015dpa} to derive them. In the latter reference,
the radiated-energy rate for a gravitational Cherenkov process is employed to compute the propagation length that a UHECR travels before its energy
drops below a certain value that has already been observed on Earth. Taking into account that the energy loss $\mathrm{d}E/\mathrm{d}t$ is the negative of the
radiated-energy rate, the following differential equation must be solved:
\begin{equation}
\label{eq:deq-general}
\frac{\mathrm{d}E}{\mathrm{d}t}=-\frac{\mathrm{d}W}{\mathrm{d}t}\,,
\end{equation}
where expressions for $\mathrm{d}W/\mathrm{d}t$ can be found in \secref{sec:helicity-processes} for the cases studied. The asymptotic behavior of the
radiated-energy rate was obtained for both momenta smaller and larger than $E_{\mathrm{max}}\approx q_{\mathrm{max}}$ where the decay rate has a maximum.
Hence, the differential equation can be solved for these cases. In what follows, all particle momenta will be replaced by particle energies, which is an
excellent approximation in the ultrahigh-energetic regime.
\begin{table}
\centering
\begin{tabular}{ccccccccccc}
\toprule
           & \multicolumn{2}{c}{Polynomial $(<)$}          & \multicolumn{2}{c}{Polynomial $(>)$}        & \multicolumn{3}{c}{Logarithmic $(>)$}          & \\
           & $\overline{s}$                          & $n$ & $\overline{s}$                        & $n$ & $s$                  & $\tilde{s}$ & $\hat{s}$ & $E_{\mathrm{max}}$ ($\approx$) \\
\colrule
$\ring{b}$ & $32\alpha(\ring{b}/m_{\psi})^4$         & 4   &                                       &     & $4\ring{b}/m_{\psi}$ & $\alpha/2$  & $11/6$    & $m_{\psi}^2/\ring{b}$ \\
$\bar{b}$  & $32\alpha(\bar{b}/m_{\psi})^4$          & 4   &                                       &     & $4\bar{b}/m_{\psi}$  & $\alpha/2$  & $11/6$    & $m_{\psi}^2/\bar{b}$ \\
$\bar{H}$  & $\approx 20\alpha(\bar{H}/m_{\psi})^4$  & 4   & $\approx(\alpha/3)\bar{H}/m_{\psi}$   & 1   &                      &             &           & --- \\
$\ring{d}$ & $\approx 100\alpha\ring{d}^4$           & 8   & $\approx 10.4\alpha$                  & 0   &                      &             &           & $3m_{\psi}/\sqrt{\ring{d}}$ \\
$\ring{g}$ & $32\alpha\ring{g}^4$                    & 4   &                                       &     & $\ring{g}$           & $\alpha$    & $11/12$   & $(3/2)m_{\psi}/\ring{g}$ \\
\botrule
\end{tabular}
\caption{Parameters used in Eqs.~(\ref{eq:deq-linear}), (\ref{eq:deq-logarithmic}). The first column states the
controlling coefficient under consideration. The next two columns give the parameters for an asymptotic polynomial behavior of the radiated-energy
rate for $E\ll E_{\mathrm{max}}$. In the fourth and fifth column the corresponding parameters can be found that are valid in the regime
$E\gg E_{\mathrm{max}}$. The subsequent three columns list the parameters for a logarithmic behavior again for momenta beyond $E_{\mathrm{max}}$.
The approximate expressions for $E_{\mathrm{max}}$, which separates the different regimes, appear in the final column. The symbol $\approx$ indicates that the
corresponding results are understood to have been obtained from plots or numerically instead of by analytical methods.}
\label{tab:parameters-helicity-decays}
\end{table}
Note that in natural units, time $t$ can be replaced by the path length $l$ traveled by the particle. To obtain conservative bounds, we could
consider an iron nucleus with initial energy $E_i\equiv E(l=0)$ propagating from its source to the Earth. Its final energy when detected on Earth
shall be $E_f\equiv E(l=L)$. Furthermore, we assume that such nuclei survive distances of around $L\simeq \unit[10]{Mpc}$ despite possible scattering
processes with the cosmic microwave background. The latter value corresponds to $\unit[10^{39}]{GeV^{-1}}$ in natural units and was used in
\cite{Kostelecky:2015dpa}, as well. The initial energy is supposed to lie in the regime $E\ll E_{\mathrm{max}}$, which can be justified at the end
when we will be aware of the limiting value of the controlling coefficients. The radiated-energy rate is then characterized by a polynomial behavior
that we write generically as
\begin{equation}
\label{eq:deq-linear}
\frac{\mathrm{d}W}{\mathrm{d}t}=\overline{s}\left(\frac{q}{m_{\psi}}\right)^nm_{\psi}^2\,,
\end{equation}
where $n\geq 0$ and $\overline{s}=\overline{s}(X_{\subset},m_{\psi})$ is a dimensionless number the may include a suitable dimensionless ratio of the
controlling coefficient and the particle mass. For such a polynomial behavior, \eqref{eq:deq-general} can be solved analytically:
\begin{equation}
\label{eq:energy-polynomial-decay}
E(l)=\left\{\begin{array}{ccl}
\left[E_i^{1-n}+(n-1)\overline{s}lm_{\psi}^{2-n}\right]^{1/(1-n)} & \text{for} & n\neq 1 \\
E_i\exp(-\overline{s}lm_{\psi}) & \text{for} & n=1\,. \\
\end{array}
\right.
\end{equation}
Explicit values and expressions for the parameters used in Eqs.~(\ref{eq:deq-logarithmic}), (\ref{eq:deq-linear}) are listed in \tabref{tab:parameters-helicity-decays}.
The question now is how large the corresponding Lorentz-violating coefficients can be chosen such that an iron nucleus with a certain
initial energy $E_i$ can travel a length of $L=\unit[10]{Mpc}$ without its energy dropping below $E_f=\unit[10^{11}]{GeV}$. After all, the latter lies
in the order of magnitude of primary energies that have already been detected on Earth. The only real quarks in the proton and the neutron are the up and
down quarks. When each of these quarks carries a fraction of $r=0.1$ of the kinetic energy of the nucleon, the real quarks would carry 30\% of the
primary energy.\footnote{These numbers are in good accordance with Table I of \cite{Gagnon:2004xh}.} It is safe to assume that statistically half of the
up and down quarks are in a spin-up state, whereas the other half is in a spin-down state. So only 50\% of the quarks will be affected by vacuum Cherenkov
radiation in case there is Lorentz violation. Hence, even if the quarks constantly radiate over a large distance, the nucleon will lose 15\% of its kinetic
energy at a maximum. As the initial energy of the nucleus directly after production at the source is unclear, this energy loss is not sufficient to
obtain reliable constraints on controlling coefficients.

Finally, we would like to make a couple of remarks about the scenario with the energy of the iron nucleus lying in the regime
$E\gg E_{\mathrm{max}}$. For the isotropic and anisotropic $b$ coefficients and the isotropic $g$ coefficients, the asymptotic behavior of the
radiated-energy rate for energies beyond $E_{\mathrm{max}}$ is logarithmic. The corresponding differential equation then generically reads
\begin{equation}
\label{eq:deq-logarithmic}
\frac{\mathrm{d}E}{\mathrm{d}t}=-\tilde{s}\left[\ln\left(s\frac{E}{m_{\psi}}\right)-\hat{s}\right]m_{\psi}^2\,,
\end{equation}
with the dimensionless numbers $\tilde{s}$, $\hat{s}$, and $s=s(X_{\subset},m_{\psi})$. The latter $s$ additionally involves a suitable dimensionless
combination of the controlling coefficient and the particle mass. Integrating both sides delivers
\begin{subequations}
\begin{align}
\label{eq:deq-logarithmic-solution}
t&=\left.\frac{\exp(\hat{s})}{s\tilde{s}m_{\psi}}\mathrm{Ei}\left[\ln\left(s\frac{E}{m_{\psi}}\right)-\hat{s}\right]\right|^{E_f}_{E_i}\sim \frac{E}{\tilde{s}m_{\psi}^2}\left[\ln\left(s\frac{E}{m_{\psi}}\right)-\hat{s}\right]^{-1}\bigg|^{E_f}_{E_i}\,, \\[2ex]
\mathrm{Ei}(x)&\equiv-\int_{-x}^{\infty} \frac{\exp(-t)}{t}\,\mathrm{d}t\,,
\end{align}
\end{subequations}
with the exponential integral $\mathrm{Ei}(x)$. In the second line of \eqref{eq:deq-logarithmic-solution}, the asymptotic behavior is stated for
the argument of the logarithm to be much larger than 1. Unfortunately, it is not possible to solve the latter equation for $E_f$ by analytical
means. A numerical study reveals some properties of the solution, though. We choose energy values $E_i\gg E_{\mathrm{max}}$ where $E_{\mathrm{max}}$
is determined based on the formulas in the last column of \tabref{tab:parameters-helicity-decays} for a particular coefficient choice and, e.g., the
constraints listed in \tabref{tab:constraints} when possible. For these values we found that the particle energy drops radically below $E_i$ long before the particle
reaches a distance of $\unit[10]{Mpc}$ from its source. This result is largely insensitive of the exact value of $E_i$ and the controlling coefficient,
which can be traced back to the logarithmic dependence of $\mathrm{d}W/\mathrm{d}t$. For the $d$ and $H$ coefficients, the radiated-energy rate for
$E\gg E_{\mathrm{max}}$ stays constant and grows linearly with the energy, respectively. The behavior of the propagating nucleus was found to be
similar. Its energy drops below $E_i$ at distances that are much smaller than $\unit[10]{Mpc}$. However, as statistically just 50\% of all particles
will lose energy by radiating electromagnetic waves, this process cannot be interpreted as an alternative to the GZK mechanism.

\subsection{Generalization to more general observer frames}
\label{eq:generalization-observer-frames}

The bounds compiled in \tabref{tab:constraints} hold in a special observer frame, which is identified with the sun-centered equatorial frame.
In general, Lorentz violation is isotropic in a single observer frame only where any observer transformation to a different frame produces
anisotropies. Since the bounds are obtained based on a cosmic ray whose energy was measured in the frame of the moving Earth, the bounds are
valid in this frame. When transforming to an observer frame that is at rest with respect to the cosmic microwave background, the
velocity of the Earth would have to be considered. This procedure would produce further nonzero controlling coefficients that are additionally
suppressed by this velocity. To be able to take those into account, the equations for the threshold momenta that were derived in the previous
sections will, at least, be generalized partially.

In principle, these characteristic momenta can be obtained classically. The procedure was already used for the isotropic $e$ coefficients
in \secref{sec:e-coefficients-isotropic} as a cross check. It takes the modified group velocity of a massive particle as a basis. The limiting
condition for vacuum Cherenkov radiation to occur is $v_{\mathrm{gr}}=1$, i.e., the incoming particle has to propagate faster than light.
Neglecting higher-order terms in Lorentz violation, the momentum components $\mathbf{q}$ are replaced by the particle velocity $\mathbf{v}$
via $\mathbf{q}/|\mathbf{q}|=\widehat{\mathbf{v}}$ with the normalized velocity $\widehat{\mathbf{v}}\equiv\mathbf{v}/|\mathbf{v}|$. Then $v_{\mathrm{gr}}=1$ must be
solved for $|\mathbf{q}|$ rendering the threshold.  For the $c$ and $d$ coefficients, the calculation can be performed in an elegant way by
following the lines of \cite{Altschul:2006zz}. However, for the $e$ and $g$ coefficients the latter approach did not seem to work. Instead,
we tried to solve the group velocity equation directly for $|\mathbf{q}|$. This works well for frameworks with a single coefficient but for
multiple nonzero coefficients, the equation becomes too involved to be solved practically. There is the possibility that deviations
between the group and the phase velocity $v_{\mathrm{ph}}\equiv E(\mathbf{q})/|\mathbf{q}|$ are of higher order in Lorentz violation. Hence,
solving the technically much simpler equation $v_{\mathrm{ph}}=1$ may lead to a momentum $\overline{q}$ that corresponds to the correct
threshold at first order in Lorentz violation: $\overline{q}=q^{\mathrm{th}}+\mathcal{O}(X_{\subset})$. To check whether this is the case,
indeed, both equations were solved numerically with the results $\overline{q}$ and $q^{\mathrm{th}}$ compared subsequently. The quantity
$\overline{q}$ obtained from the condition $v_{\mathrm{ph}}=1$ was only taken as a reliable approximation for the real threshold $q^{\mathrm{th}}$
when the numerical deviations were supposed to be traced back to higher-order terms in Lorentz violation. For the $c$, $d$, $e$, and $g$
coefficients we then obtain (where the threshold momenta for only one of the two modes are shown):
\begin{subequations}
\begin{align}
q^{\mathrm{th}}_{c^{(4)}}&=\frac{m_{\psi}}{\sqrt{(-2c^{(4)\mu\nu}+\overline{k}^{(4)\mu\nu}_F+f^{(4)\mu}f^{(4)\nu})u_{\mu}u_{\nu}}}+\dots\,, \displaybreak[0]\\[2ex]
q^{\mathrm{th}}_{d^{(4)}}&=\frac{m_{\psi}}{\sqrt{-2d^{(4)\mu\nu}u_{\mu}u_{\nu}}}+\dots\,, \displaybreak[0]\\[2ex]
q^{\mathrm{th}}_{e^{(4)}}&=\frac{m_{\psi}}{e^{(4)\mu}u_{\mu}}+\dots\,, \displaybreak[0]\\[2ex]
\label{eq:generalized-threshold-g-coefficients}
q^{\mathrm{th}}_{g^{(4)}}&=\frac{m_{\psi}}{\sqrt{(1/2)g^{(4)\mu\nu\varrho}g^{(4)}_{\mu\nu\sigma}u_{\varrho}u^{\sigma}+2\sqrt{-g^{(4)\mu\alpha\beta}g^{(4)}_{\mu\gamma\delta}u_{\alpha}u_{\beta}u^{\gamma}u^{\delta}}}}\Bigg|_{M}+\dots\,.
\end{align}
\end{subequations}
Here $u^{\mu}\equiv (1,\widehat{\mathbf{v}})^{\mu}$ is the (normalized) four-velocity of the incoming massive particle and $\overline{k}_F^{(4)\mu\nu}$
is the nonbirefringent part of the {\em CPT}-even, minimal photon sector coefficients $k_F^{(4)\mu\nu\varrho\sigma}$ that appear in the
nonbirefringent \textit{Ansatz} of \eqref{eq:nonbirefringent-ansatz} (cf.~also \cite{Altschul:2006zz}).
Note that both the latter photon coefficients and the $f$ coefficients can be put into the $c$ coefficients by suitable transformations, as
described before in Secs.~\ref{sec:c-coefficients-isotropic} and \ref{sec:f-coefficients-isotropic}. The threshold momenta for the $c$, $d$,
and $e$ sectors include all of the minimal coefficients. Apart from that, it seems to be challenging to generalize the threshold momentum for the $g$
coefficients, i.e., the equation given holds for the $g$ coefficients restricted to the set $M\equiv\{g^{(4)0\mu\nu}=g^{(4)\mu\nu0}=g^{(4)121}=0\}$
only. Although the term in front of the second square root in \eqref{eq:generalized-threshold-g-coefficients} is of higher order in the $g$
coefficients, it cannot be discarded, because there are sectors that do not contribute to the double square-root term. For example, this is the
case for the isotropic $g$ coefficients.

\subsection{Comparison to existing bounds}

According to Table D26 of the data tables \cite{Kostelecky:2008ts}, the number of constraints in the quark sector is moderate. Most of the
bounds have been set on minimal $a$ coefficients and those result from various types of meson oscillation experiments. Vacuum Cherenkov radiation cannot bound
the $a$ coefficients, as we have seen, since both the process with and without a spin flip is forbidden for this type of coefficients. The second group of constraints
refers to the minimal $c$ coefficients where the strictest lower ones result from astrophysical data and they lie in the order of magnitude of $-10^{-23}$, which is
in accordance with the corresponding values in \tabref{tab:constraints}. Further
bounds on the minimal $c$ coefficients have been placed on those of the top-quark sector. They approximately cover the range $\{-10^{-1},10^{-1}\}$, i.e., these coefficients
are only weakly bounded. Last but not least, there is a small number of constraints on the minimal $d$ coefficients, again for the top-quark sector.
These lie in the same order of magnitude as the bounds on the $c$ coefficients mentioned before.

To summarize, the majority of already existing constraints in the quark sector were obtained on the minimal $a$ and $c$ coefficients where a small number
exists for the $d$ coefficients. So the current bounds compiled in \tabref{tab:constraints} on the $d$, $e$, and $g$ coefficients obtained from the
absence of vacuum Cherenkov radiation are justified. Some of the remaining new bounds such as those on $d$ improve the currently existing ones by several
orders of magnitude. However, it must also be taken into account that the constraints presented here were computed based on a very simple analysis without taking into
account the nucleon structure appropriately, i.e., the parton distribution functions. In this context, the recent paper~\cite{Kostelecky:2016pyx} shall be
mentioned where upper constraints on the magnitudes of up and down-quark coefficients were derived from data on deep inelastic scattering covering the
orders of magnitude $\{10^{-6},10^{-4}\}$.

Cherenkov-type processes may play a role for neutrinos, as well. There are several processes that can be considered to contribute to this class, e.g.,
$\upnu_{\upmu}\rightarrow \upnu_{\upmu}+\upnu_{\mathrm{e}}+\overline{\upnu}_{\mathrm{e}}$ or $\upnu_{\upmu}\rightarrow \upnu_{\upmu}+\upgamma$ at
one-loop level. Neutrinos would lose energy by such processes occurring, which would distort the energy spectrum of a neutrino beam. This distortion
depends on the traveling length, and its nonobservation allows for estimating neutrino coefficients that are known as oscillation-free. The associated
particles are left-handed Dirac fermions with a single flavor; so it makes sense to refer to them in this context. The estimated values are compiled in
Tab.~XV of \cite{Kostelecky:2011gq}. The value for the oscillation-free isotropic $c$ coefficient $\ring{c}^{(4)\upnu}$ for a neutrino energy of $\unit[100]{TeV}$
amounts to $-3\times 10^{-11}$. The absence of the Cherenkov process in combination with the first PeV neutrino events detected by IceCube led to remarkable
bounds on oscillation-free $c$ coefficients of different mass dimension~\cite{Diaz:2013wia}. For example, a lower bound of  $-4\times 10^{-19}$ was placed on
$(c_{\mathrm{of}}^{(4)\upnu})_{00}$.

\section{Conclusions}
\setcounter{equation}{0}
\label{eq:conclusions}

In the current work, vacuum Cherenkov radiation was studied where Lorentz violation was assumed to be situated in the minimal fermion sector. First of
all, the process was analyzed without a spin flip of the fermion occurring. It was found that such a process is not allowed for the minimal $a$, $b$, and $H$
coefficients, whereas it can occur for subsets of
the $c$, $d$, $e$, $f$, and $g$ sectors. A charged particle can radiate photons when its energy exceeds a certain threshold that depends on some inverse
power of the Lorentz-violating coefficients. Furthermore, both the decay rates and radiated-energy rates were computed based on formal results on the
modified particle spinors developed earlier in~\cite{Reis:2016hzu}. For large energies, the decay rates were found to grow linearly with the
momentum. For the $c$ and $d$ coefficients, the rate is suppressed by the nonvanishing controlling coefficient where for the $e$, $f$, and $g$
coefficients, this suppression is even quadratic. Last but not least, when the momentum approaches the threshold, the rates tend to zero, as expected.

On the other hand, the scenario including a fermion spin flip was examined. This process is crucial for the spin-nondegenerate
$b$, $d$, $H$, and $g$ coefficients, as the fermion energy jumps from one branch of the dispersion relation to the other when the spin projection changes. Because of this
behavior, the process has several peculiar properties. First of all, this realization of vacuum Cherenkov radiation is possible for the dimensionful $b$
and $H$ coefficients in contrast to the spin-conserving process. The helicity decay does not have any threshold, i.e., it occurs for arbitrary initial fermion momentum.
Furthermore, the decay rate grows for increasing momentum until it reaches a maximum at which point it decreases again. The increase and decrease were either
found to depend polynomially or logarithmically on the initial fermion momentum.

Spin-conserving vacuum Cherenkov radiation works very efficiently, and the radiating fermion loses its energy excess on short time scales. Hence, if a
cosmic ray is detected on Earth, we can deduce that its energy is smaller than the threshold energy. From this condition, several new constraints on
isotropic and anisotropic Lorentz violation in the up and down-quark sectors have been obtained. For the $c$ and $d$ coefficients, their
magnitudes range from $3\times 10^{-23}$ to $3\times 10^{-21}$. For the $e$ and $g$ coefficients, weaker bounds have been computed in the range of
$9\times 10^{-12}$ to $2\times 10^{-11}$, as the threshold has a different dependence on the controlling coefficients. Some bounds are even two-sided
due to the fact that vacuum Cherenkov radiation is not tied to a particular sign of the related coefficients.

Although the decay rates of the helicity processes are heavily suppressed by the smallness of the controlling coefficients, a fermion can radiate energy
steadily when traveling light years. However, as a nucleon consists to 30\% of real quarks and as statistically only half of those radiate, there
is no possibility of obtaining a reliable set of constraints from helicity decays at the moment. Nevertheless, studying helicity processes in the context
of spin-nondegenerate operators has led to several fruitful theoretical insights.

Bounds on the quark sector are rare and the current paper contributes to closing this gap. Furthermore, it shows two issues.
First, it demonstrates the power of cosmic-ray data to search for Lorentz violation in the
fermion sector where hitherto the main focus had been on the photon sector. Second, it illustrates the usefulness of the recently developed theoretical results
of~\cite{Reis:2016hzu} for phenomenological studies related to Lorentz violation in the fermion sector.

\section{Acknowledgments}

It is a pleasure to thank V.A.~Kosteleck\'{y} for useful remarks, especially on transforming the $f$ onto the $c$ coefficients and with regards to more
general observer frames. Also, the author is grateful for the opportunity of giving two seminars at Indiana University. Subsequent discussions with
V.A.~Kosteleck\'{y}, R.~Lehnert, and the local group have been very fruitful with respect to obtaining the constraints from experimental data. Additionally,
the author thanks the anonymous referee for a considerable amount of remarks that helped to broaden the scope of the paper. Pointing out the possibility of
helicity decays has led to an increased number of interesting results. This work was partially funded by the Brazilian foundation FAPEMA.

\newpage
\begin{appendix}
\numberwithin{equation}{section}

\section{Propagators}
\label{sec:propagators}
\setcounter{equation}{0}

Here we give the propagators for the Lorentz-violating frameworks that are characterized by additional time derivatives in the Lagrange density.
For such sectors, the asymptotic states possess an unconventional time evolution, which is why both the spinor solutions and the propagator
have to be modified. We give the propagators for these sectors, because they cannot simply be derived from Eqs.~(2.6), (2.7) in \cite{Reis:2016hzu}.

\subsection{Pseudovector $\boldsymbol{d}$ coefficients}
\label{sec:pseudovector-d-coefficients}

For the framework of isotropic coefficients, the Dirac operator has to be modified with the matrix $A_{\ring{d}}$ where for the anisotropic case
the analogous procedure must be carried out with $A_{\bar{d}}$. Both matrices can be chosen diagonal and are given by
\begin{subequations}
\begin{align}
\label{eq:transformation-matrix-isotropic}
A_{\ring{d}}&=\mathrm{diag}\left(\frac{1}{\sqrt{1+\ring{d}}},\frac{1}{\sqrt{1+\ring{d}}},\frac{1}{\sqrt{1-\ring{d}}},\frac{1}{\sqrt{1-\ring{d}}}\right)\,, \\[2ex]
\label{eq:transformation-matrix-anisotropic}
A_{\bar{d}}&=\mathrm{diag}\left(\frac{1}{\sqrt{1+\bar{d}/3}},\frac{1}{\sqrt{1+\bar{d}/3}},\frac{1}{\sqrt{1-\bar{d}/3}},\frac{1}{\sqrt{1-\bar{d}/3}}\right)\,.
\end{align}
\end{subequations}
The propagator of the modified Dirac operator for the isotropic sector has the form
\begin{subequations}
\label{eq:propagator-d-coefficients-isotropic}
\begin{align}
S_{\ring{d}}&=\frac{1}{\Delta_{\ring{d}}}\left(\widehat{\xi}_{\ring{d}}^{\mu}\gamma_{\mu}+\widehat{\Xi}_{\ring{d}}\mathds{1}_4+\widehat{\zeta}_{\ring{d}}^{\mu}\gamma^5\gamma_{\mu}+\widehat{\psi}_{\ring{d}}^{\mu\nu}\sigma_{\mu\nu}\right)\,, \displaybreak[0]\\[2ex]
\widehat{\xi}_{\ring{d}}^{\mu}&=\frac{1}{27}\left\{4\ring{d}^2p_0\left[9m_{\psi}^2-\ring{\mathfrak{y}}p_0^2-(5\ring{\mathfrak{x}}-4\ring{d}^2)\mathbf{p}^2\right]\ring{\lambda}^{\mu}-\ring{\mathfrak{x}}(\ring{\mathfrak{z}}\mathbf{p}^2+9m_{\psi}^2-\ring{\mathfrak{y}}p_0^2)p^{\mu}\right\}\,, \displaybreak[0]\\[2ex]
\widehat{\Xi}_{\ring{d}}&=-\frac{\sqrt{\ring{\mathfrak{y}}}}{27}m_{\psi}(\ring{\mathfrak{z}}\mathbf{p}^2+9m_{\psi}^2-\ring{\mathfrak{y}}p_0^2)\,, \displaybreak[0]\\[2ex]
\widehat{\zeta}_{\ring{d}}^{\mu}&=\frac{4\ring{d}}{27}\left\{(\ring{\mathfrak{z}}\mathbf{p}^2+9m_{\psi}^2+\ring{\mathfrak{y}}p_0^2)p^{\mu}+p_0\left[(\ring{\mathfrak{z}}+8\ring{d}^2)\mathbf{p}^2-(9m_{\psi}^2+\ring{\mathfrak{y}}p_0^2)\right]\ring{\lambda}^{\mu}\right\}\,, \displaybreak[0]\\[2ex]
\widehat{\psi}_{\ring{d}}^{\mu\nu}&=\frac{4\sqrt{\ring{\mathfrak{y}}}}{9}\ring{d}m_{\psi}p_0\varepsilon^{\mu\nu\varrho\sigma}p_{\varrho}\ring{\lambda}_{\sigma}\,, \displaybreak[0]\\[2ex]
\Delta_{\ring{d}}&=\left(\frac{\ring{\mathfrak{y}}}{9}\right)^2(p_0-E_{\ring{d}}^{(+)})(p_0-E_{\ring{d}}^{(-)})(p_0+E_{\ring{d}}^{(+)})(p_0+E_{\ring{d}}^{(-)})\,,
\end{align}
\end{subequations}
with the particle energies given by \eqref{eq:dispersion-relations-isotropic-d} and $\ring{\mathfrak{x}}$, $\ring{\mathfrak{y}}$, $\ring{\mathfrak{z}}$ defined in
Eqs.~(\ref{eq:definitions-isotropic-x-y}), (\ref{eq:definitions-isotropic-z}).
For the anisotropic framework we obtain:
\begin{subequations}
\label{eq:propagator-d-coefficients-anisotropic}
\begin{align}
S_{\bar{d}}&=\frac{1}{\Delta_{\bar{d}}}\left(\widehat{\xi}_{\bar{d}}^{\mu}\gamma_{\mu}+\widehat{\Xi}_{\bar{d}}\mathds{1}_4+\widehat{\zeta}_{\bar{d}}^{\mu}\gamma^5\gamma_{\mu}+\widehat{\psi}_{\bar{d}}^{\mu\nu}\sigma_{\mu\nu}\right)\,, \displaybreak[0]\\[2ex]
\widehat{\xi}_{\bar{d}}^{\mu}&=\frac{1}{81}\left\{\left[\bar{\mathfrak{z}}(\bar{\mathfrak{z}}\mathfrak{P}^2-9m_{\psi}^2)-9\bar{\mathfrak{X}}p_3^2\right]\mathfrak{P}^{\mu}+3\bar{\mathfrak{x}}p_3\left[\bar{\mathfrak{z}}\mathfrak{P}^2-9m_{\psi}^2-\bar{\mathfrak{y}}p_3^2\right]\bar{\lambda}^{\mu}\right\}\,, \displaybreak[0]\\[2ex]
\widehat{\Xi}_{\bar{d}}&=\frac{\sqrt{\bar{\mathfrak{z}}}}{27}m_{\psi}\left[\bar{\mathfrak{z}}\mathfrak{P}^2-9m_{\psi}^2-\bar{\mathfrak{y}}p_3^2\right]\,, \displaybreak[0]\\[2ex]
\widehat{\zeta}_{\bar{d}}^{\mu}&=\frac{4\bar{d}}{27}\left\{6\bar{\mathfrak{x}}p_3^2p^{\mu}+p_3\left[\bar{\mathfrak{z}}\mathfrak{P}^2+9m_{\psi}^2-(\bar{\mathfrak{z}}+16\bar{d}^2)p_3^2\right]\bar{\lambda}^{\mu}\right\}\,, \displaybreak[0]\\[2ex]
\widehat{\psi}_{\bar{d}}^{\mu\nu}&=-\frac{4\sqrt{\bar{\mathfrak{z}}}}{9}\bar{d}m_{\psi}p_3\varepsilon^{\mu\nu\varrho\sigma}p_{\varrho}\bar{\lambda}_{\sigma}\,, \displaybreak[0]\\[2ex]
\Delta_{\bar{d}}&=\left(\frac{\bar{\mathfrak{z}}}{9}\right)^2(p_0-E_{\bar{d}}^{(+)})(p_0-E_{\bar{d}}^{(-)})(p_0+E_{\bar{d}}^{(+)})(p_0+E_{\bar{d}}^{(-)})\,,
\end{align}
\end{subequations}
with the fermion energies of \eqref{eq:dispersion-relations-anisotropic-d} and the quantities $\bar{\mathfrak{X}}$, $\bar{\mathfrak{x}}$, $\bar{\mathfrak{y}}$
defined in Eqs.~(\ref{eq:anisotropic-d-quantity-X}), (\ref{eq:anisotropic-d-quantities-x-y}). Furthermore, the ``reduced momentum''
$\mathfrak{P}^{\mu}\equiv p^{\mu}-p_3\bar{\lambda}^{\mu}$ has been introduced to express the result in a convenient way.

\subsection{Vector $\boldsymbol{c}$ coefficients}
\label{sec:vector-c-coefficients}

For the $c$ coefficients, which are spin-degenerate, the results are much simpler. The Dirac operator for the isotropic case must be modified
with $A_{\ring{c}}$. For the anisotropic case, $A_{\bar{c}}$ is introduced. Each matrix is given by the identity matrix in spinor space that
is multiplied by a global factor:
\begin{equation}
\label{eq:transformation-matrices-a-coefficients}
A_{\ring{c}}=\frac{1}{\sqrt{1+\ring{c}}}\mathds{1}_4\,,\quad A_{\bar{c}}=\frac{1}{\sqrt{1-\bar{c}/3}}\mathds{1}_4\,.
\end{equation}
The propagator for the isotropic framework is then given by
\begin{subequations}
\label{eq:propagator-c-coefficients-isotropic}
\begin{align}
S_{\ring{c}}&=\frac{1}{\Delta_{\ring{c}}}(\widehat{\xi}_{\ring{c}}^{\mu}\gamma_{\mu}+\widehat{\Xi}_{\ring{c}}\mathds{1}_4)\,, \displaybreak[0]\\[2ex]
\widehat{\xi}_{\ring{c}}^{\mu}&=\ring{\mathfrak{a}}\ring{\mathfrak{b}}p^{\mu}+\frac{4}{3}\ring{c}\ring{\mathfrak{a}}p_0\ring{\lambda}^{\mu}\,, \displaybreak[0]\\[2ex]
\widehat{\Xi}_{\ring{c}}&=\ring{\mathfrak{a}}m_{\psi}\,, \displaybreak[0]\\[2ex]
\Delta_{\ring{c}}&=\ring{\mathfrak{a}}^2(p_0-E_{\ring{c}})(p_0+E_{\ring{c}})\,,
\end{align}
\end{subequations}
with $\ring{\mathfrak{a}}$, $\ring{\mathfrak{b}}$ of \eqref{eq:isotropic-c-quantities-a-b}.
For the anisotropic sector we obtain
\begin{subequations}
\label{eq:propagator-c-coefficients-anisotropic}
\begin{align}
S_{\bar{c}}&=\frac{1}{\Delta_{\bar{c}}}(\widehat{\xi}_{\bar{c}}^{\mu}\gamma_{\mu}+\widehat{\Xi}_{\bar{c}}\mathds{1}_4)\,, \displaybreak[0]\\[2ex]
\widehat{\xi}^{\mu}_{\bar{c}}&=\bar{\mathfrak{b}}^2p^{\mu}-\frac{4}{3}\bar{c}\bar{\mathfrak{b}}p_3\bar{\lambda}^{\mu}\,, \displaybreak[0]\\[2ex]
\widehat{\Xi}_{\bar{c}}&=\bar{\mathfrak{b}}m_{\psi}\,, \displaybreak[0]\\[2ex]
\Delta_{\bar{c}}&=\bar{\mathfrak{b}}^2(p_0-E_{\bar{c}})(p_0+E_{\bar{c}})\,,
\end{align}
\end{subequations}
with $\bar{\mathfrak{a}}$, $\bar{\mathfrak{b}}$ given by \eqref{eq:anisotropic-c-quantities-a-b}.

\subsection{Scalar $\boldsymbol{e}$ coefficients}
\label{sec:scalar-e-coefficients}

The Dirac operator for the isotropic sector must be transformed with the matrix
\begin{equation}
\label{eq:transformation-matrix-e-coefficients}
A_{\ring{e}}=-a\gamma^5+b\gamma^0\gamma^5\,,\quad a=\frac{\ring{e}}{\sqrt{2\mathfrak{r}_{\ring{e}}^2(1-\mathfrak{r}_{\ring{e}})}}\,,\quad b=\frac{\ring{e}}{\sqrt{2\mathfrak{r}_{\ring{e}}^2(1+\mathfrak{r}_{\ring{e}})}}\,,
\end{equation}
to remove the additional time derivatives in the Lagrange density. Note that the latter matrix is not diagonal in contrast to those encountered
previously. The propagator is then
\begin{subequations}
\label{eq:propagator-e-coefficients-isotropic}
\begin{align}
S_{\ring{e}}&=\frac{1}{\Delta_{\ring{e}}}\left(\widehat{\xi}^{\mu}_{\ring{e}}\gamma_{\mu}+\widehat{\Xi}_{\ring{e}}\mathds{1}_4\right)\,, \displaybreak[0]\\[2ex]
\widehat{\xi}^{\mu}_{\ring{e}}&=\mathfrak{r}_{\ring{e}}p^{\mu}+\ring{e}\left(m_{\psi}-\frac{\ring{e}p_0}{1+\mathfrak{r}_{\ring{e}}^{-1}}\right)\ring{\lambda}^{\mu}\,, \displaybreak[0]\\[2ex]
\widehat{\Xi}_{\ring{e}}&=-m_{\psi}\,, \displaybreak[0]\\[2ex]
\Delta_{\ring{e}}&=\mathfrak{r}_{\ring{e}}^2(p_0-E_{\ring{e}})(p_0+E_{-\ring{e}})\,,\quad E_{-\ring{e}}\equiv E_{\ring{e}}|_{\ring{e}\mapsto -\ring{e}}\,.
\end{align}
\end{subequations}
This propagator does produce the standard textbook result for vanishing Lorentz violation. The reason is that the Dirac operator was
transformed with the matrix $A_{\ring{e}}$ of \eqref{eq:transformation-matrix-e-coefficients} whereby the latter does not correspond to the identity
matrix for $\ring{e}=0$. This behavior is different from that of the cases previously analyzed.

\subsection{Scalar $\boldsymbol{f}$ coefficients}
\label{sec:scalar-f-coefficients}

For the isotropic sector of the minimal $f$ coefficients, the additional time derivative can be removed by a transformation in spinor space such
as before. Here, two distinct transformation matrices have been found that are given as follows:
\begin{subequations}
\begin{align}
\label{eq:transformation-matrix-f-coefficients}
A_{\ring{f}}&=a\mathds{1}_4+\mathrm{i}b\gamma^5\gamma^0\,,\quad a=\frac{1}{\sqrt{\mathfrak{r}_{\ring{f}}}}\cosh\left[\frac{1}{2}\mathrm{artanh}(\ring{f})\right]\,,\quad b=\frac{1}{\sqrt{\mathfrak{r}_{\ring{f}}}}\sinh\left[\frac{1}{2}\mathrm{artanh}(\ring{f})\right]\,, \\[2ex]
\label{eq:transformation-matrix-f-coefficients-2}
A'_{\ring{f}}&=a'\sigma^{02}-b'\gamma^5\gamma^2\,,\quad a'=\frac{\ring{f}}{\sqrt{2\mathfrak{r}_{\ring{f}}^2(1-\mathfrak{r}_{\ring{f}})}}\,,\quad b'=\frac{\ring{f}}{\sqrt{2\mathfrak{r}_{\ring{f}}^2(1+\mathfrak{r}_{\ring{f}})}}\,,
\end{align}
\end{subequations}
with the quantity $\mathfrak{r}_{\ring{f}}$ of \eqref{eq:quantity-rf}. Note the similarities of the matrix coefficients $a$, $b$ in
\eqref{eq:transformation-matrix-e-coefficients} and $a'$, $b'$ in \eqref{eq:transformation-matrix-f-coefficients-2}. The first matrix was found by
making an \textit{Ansatz} in terms of particular Dirac bilinears and solving the resulting equations. The second matrix is based on the observations
in \cite{Altschul:2006ts}. In the latter reference, a transformation was constructed mapping the $f$ coefficients onto the $c$ coefficients. For the
timelike case, the corresponding transformation reads
\begin{subequations}
\begin{align}
\chi&=\exp\left[\frac{\mathrm{i}}{2}\gamma^0\gamma^5\mathrm{artanh}(\ring{f})\right]\psi=\left[\cosh(x)\mathds{1}_4-\mathrm{i}\sinh(x)\gamma^5\gamma^0\right]\psi\,, \\[2ex]
x&=\frac{1}{2}\mathrm{artanh}(\ring{f})\,.
\end{align}
\end{subequations}
This transformation generates an isotropic $c$ coefficient that is of the form stated in \eqref{eq:transformation-c-f-timelike}.
Then \eqref{eq:transformation-matrices-a-coefficients}
can be applied to remove the additional time derivative from the Lagrange density. This procedure results in the matrix $A_{\ring{f}}$ given in
\eqref{eq:transformation-matrix-f-coefficients} above.\footnote{The existence of two distinct transformation matrices shows explicitly that the
coordinate transformation removing the additional
time derivatives from the Lagrange density is not unique. Such a behavior is reminiscent of the observation made in Sec.~III~A of
\cite{Kostelecky:2010ze}. In the latter paper, a Hamiltonian is obtained by two different procedures: a field-redefinition method and a method
named after Parker. The resulting Hamiltonians are different, but they produce analog physical results.}
Since the matrix $A_{\ring{f}}$ has a simpler structure compared to $A'_{\ring{f}}$, the first will be used. The propagator then reads
\begin{subequations}
\label{eq:propagator-e-coefficients-isotropic}
\begin{align}
S_{\ring{f}}&=\frac{1}{\Delta_{\ring{f}}}\left(\widehat{\xi}^{\mu}_{\ring{f}}\gamma_{\mu}+\widehat{\Xi}_{\ring{f}}\mathds{1}_4\right)\,, \displaybreak[0]\\[2ex]
\widehat{\xi}^{\mu}_{\ring{f}}&=\mathfrak{r}_{\ring{f}}\left[p^{\mu}+p_0(\mathfrak{r}_{\ring{f}}-1)\lambda^{\mu}\right]\,, \\[2ex]
\widehat{\Xi}_{\ring{f}}&=\mathfrak{r}_{\ring{f}}m_{\psi}\,, \\[2ex]
\Delta_{\ring{f}}&=\mathfrak{r}_{\ring{f}}^2(p_0-E_{\ring{f}})(p_0+E_{\ring{f}})\,.
\end{align}
\end{subequations}
Note that when $A'_{\ring{f}}$ is employed, the standard textbook result for the propagator is not obtained for a vanishing controlling coefficient, which
is a behavior analog to that of the isotropic $e$ coefficient. After all, $A'_{\ring{f}}$ does not correspond to the identity matrix for vanishing Lorentz
violation. This is another reason for why the transformation mediated by $A_{\ring{f}}$ is the preferable one to be used for calculations of the decay
rate.

\section{Sum over spinor matrices for spin-degenerate operators}
\label{sec:spinor-matrices-spin-degenerate}

In the current section, the sum over the spinor matrices $u^{(s)}\overline{u}^{(s)}$ shall be obtained for a spin-degenerate operator in the SME
fermion sector. These comprise the $c$, $e$, and $f$ coefficients. The derivation works in analogy to the proof based on the optical theorem that
was carried out in~\cite{Reis:2016hzu}. Our investigations are restricted to minimal coefficients. Therefore, the propagator for a spin-degenerate
operator has two poles only. In general, it can be written as
\begin{subequations}
\begin{align}
\mathrm{i}S&=\frac{\mathrm{i}}{\Delta}\left(\widehat{\xi}_{\mu}\gamma^{\mu}+\widehat{\Xi}\mathds{1}_4+\widehat{\Upsilon}\gamma^5+\widehat{\zeta}_{\mu}\gamma^5\gamma^{\mu}+\widehat{\psi}_{\mu\nu}\sigma^{\mu\nu}\right)\,, \\[2ex]
\Delta&=\mathcal{Z}(p_0-E_u)(p_0-E_{<})\,,
\end{align}
\end{subequations}
with the denominator $\Delta$. The latter is decomposed into its roots where $\mathcal{Z}$ is a momentum-independent constant, $E_u$
is the fermion energy, and $E_{<}$ the corresponding negative propagator pole. The structure in spinor space is controlled by the
parameters $\{\widehat{\xi}_{\mu},\widehat{\Xi},\widehat{\Upsilon},\widehat{\zeta}_{\mu},\widehat{\psi}_{\mu\nu}\}$. For the spin-degenerate
cases, $\widehat{\zeta}_{\mu}$ and $\widehat{\psi}_{\mu\nu}$ do possibly not contribute. Based on the validity of the optical theorem at tree-level,
the spin sum over the spinor matrices $u^{(s)}\overline{u}^{(s)}$ is directly linked to the propagator. The general proof for the spin-nondegenerate
operators is shown in \cite{Reis:2016hzu}. Adapting the latter to the spin-degenerate sectors, produces
\begin{equation}
\label{eq:sum-spinor-matrices}
\sum_{s=\pm} u^{(s)}\overline{u}^{(s)}=\frac{2E_u}{\mathcal{Z}(E_u-E_{<})}\left(\widehat{\xi}_{\mu}\gamma^{\mu}+\widehat{\Xi}\mathds{1}_4+\widehat{\Upsilon}\gamma^5+\widehat{\zeta}_{\mu}\gamma^5\gamma^{\mu}+\widehat{\psi}_{\mu\nu}\sigma^{\mu\nu}\right)\Big|_{p_0=E_u}\,.
\end{equation}
Hence, the spinor space structure of the propagator is directly taken over to the sum over the spinor matrices and all $p_0$ have to be replaced by
the fermion energy. After all, the latter correspondence follows from cutting a propagator resulting in on-shell fermions. The prefactor is the analog
of the function $\mathscr{C}$ in Eq.~(4.11a) of \cite{Reis:2016hzu}, which is the remainder of the propagator denominator.

\end{appendix}

\end{fmffile}

\newpage


\end{document}